\documentclass[onecolumn]{aastex631}
\usepackage[caption=false]{subfig}
\usepackage{graphicx}
\usepackage{amsmath}
\usepackage{multirow}
\usepackage{graphicx}
\usepackage{booktabs}
\usepackage{amsmath,amssymb}
\usepackage[T1]{fontenc}
\graphicspath{{./}{figures/}}
\usepackage{longtable}
\usepackage{epstopdf}
\usepackage[]{amsmath}
\usepackage{threeparttablex}
\usepackage[caption=false]{subfig}
\usepackage{graphicx}
\usepackage{amsmath}
\usepackage{multirow}
\usepackage{graphicx}
\usepackage{booktabs}
\usepackage{amsmath,amssymb}
\usepackage[T1]{fontenc}
\graphicspath{{./}{figures/}}
\usepackage{longtable}
\usepackage{epstopdf}
\usepackage[]{amsmath}
\usepackage{threeparttablex}

\begin{document}

\title{Searching for Associations Between Short Gamma-ray Bursts and Fast Radio Bursts}

\correspondingauthor{Xiang-Gao Wang}
\email{E-mail: wangxg@gxu.edu.cn}

\author{Ming-Xuan Lu}
\affiliation{Guangxi Key Laboratory for Relativistic Astrophysics,
	School of Physical Science and Technology, Guangxi University, Nanning 530004,
	China}

\author{Long Li}
\affiliation{Department of Astronomy, School of Physical Sciences, University of
Science and Technology of China, Hefei 230026, China}

\author{Xiang-Gao Wang}
\affiliation{Guangxi Key Laboratory for Relativistic Astrophysics,
	School of Physical Science and Technology, Guangxi University, Nanning 530004,
	China}

\author{Can-Min Deng}
\affiliation{Guangxi Key Laboratory for Relativistic Astrophysics,
	School of Physical Science and Technology, Guangxi University, Nanning 530004,
	China}

\author{Yun-Feng Liang}
\affiliation{Guangxi Key Laboratory for Relativistic Astrophysics,
	School of Physical Science and Technology, Guangxi University, Nanning 530004,
	China}

\author{Da-Bin Lin}
\affiliation{Guangxi Key Laboratory for Relativistic Astrophysics,
	School of Physical Science and Technology, Guangxi University, Nanning 530004,
	China}

\author{En-Wei Liang}
\affiliation{Guangxi Key Laboratory for Relativistic Astrophysics,
	School of Physical Science and Technology, Guangxi University, Nanning 530004,
	China}

\begin{abstract}
The physical origin of fast radio bursts (FRBs) is still unclear. However, young magnetars associated with short-duration gamma-ray bursts (SGRBs) have been thought to be possible central engines for some FRBs. In this paper, we perform a systematic search for SGRBs that are associated with FRBs in a sample including 623 FRBs (601 one-off bursts and 22 repeaters) and 168 SGRBs with precise localizations. We find that FRB 190309A is spatially associated with GRB 060502B, with a chance probability of 0.05 when temporal and redshift information is taken into account. 
Considering the high chance probability (the statistical significance is $<3\sigma$), we examine other observational properties such as the host galaxy, the dispersion measure, and the energy budget of the central engine to check the possibility of their association.
 Although the available observational information is insufficient to determine whether they are physically associated, it does not rule out such a possibility. As the only pair of FRB and GRB that are spatially associated, it remains an interesting case worthy of further attention.
\end{abstract}

\keywords{Radio transient sources (2008) --- Gamma-ray bursts (629) --- Magnetars (992)}

%
\section{Introduction}           
\label{sect:intro}

Fast radio bursts (FRBs) are mysterious radio transients with millisecond durations \citep{2007Sci...318..777L,  2013Sci...341...53T,2019PhR...821....1P}. FRBs have extremely high brightness temperatures \citep[][]{2019ARA&A..57..417C,2019A&ARv..27....4P}. The dispersion measures (DMs) of FRBs exceed the Milky Way's contribution and some of them have been reliably determined to be located in the extragalactic systems \citep{2017Natur.541...58C,2019Sci...365..565B,2019Sci...366..231P,2019Natur.572..352R,2020Natur.577..190M}.
Up to now, except for the X-ray burst from the Galactic soft gamma repeater (SGR) SGR1935+2154 coincident with the fast radio burst FRB200428 \citep{2020Natur.587...59B,2020Natur.587...54C}, no other multiwavelength/multimessenger transients associated with FRBs are definitely observed  \citep{2016MNRAS.460.2875Y,2016ApJ...832L...1D,2017MNRAS.472.2800H,2017ApJ...843L..13Z,2017ApJ...842L...8X,2019MNRAS.489L..75J,2019ApJ...879...40C,2020ApJ...898L..29M,2020ATel13446....1T,2021ApJ...922...78X,2021NatAs...5..378L,2022arXiv220312038T,2022arXiv220317222W}, though there have been some works claiming tentative associations of FRBs with multiwavelength transient counterparts \citep{2020ApJ...894L..22W,2022arXiv220306994L}.
Current FRB models can be divided into two categories: catastrophic models \citep[e.g.,][]{2013ApJ...776L..39K,2013PASJ...65L..12T,2014A&A...562A.137F,2014ApJ...780L..21Z,2016ApJ...827L..31Z,2016ApJ...826...82L,2016ApJ...822L...7W} and non-catastrophic models \citep[e.g.,][]{2016ApJ...829...27D,2016MNRAS.461.1498M,2017ApJ...841...14M,2018ApJ...868L...4M,2020ApJ...890L..24Z,2020ApJ...893L..26I,2020ApJ...891...72W,2021Innov...200152G,2021ApJ...922...98D}.
The former (e.g. NS-NS merger or NS-BH merger model with NS and BH denoting neutron star and black hole, respectively) is invoked to explain one-off FRBs.
The latter usually invokes a NS born in a catastrophic event as the central engine of the FRB.

The localization of FRB sources and the identification of their host galaxies are important for understanding their physical origins, and several host galaxies of FRBs have been identified.
The host galaxy of the first repeat source FRB 121102 is found to be similar to the host galaxies of superluminous supernovae (SLSNe) and long gamma-ray bursts (LGRBs) \citep{2017ApJ...834L...7T,2017ApJ...843...84N,2017ApJ...841...14M,2017ApJ...834L...7T,2017Natur.541...58C,2017ApJ...834L...8M,2019MNRAS.487.3672Z}. However, the host environment of another well-localized repeater FRB 20180916B is completely different to FRB 121102 \citep{2021ApJ...908L..12T}.
The one-off bursts FRB 190523A and FRB 180924B 
are found to occur in massive galaxies with low star formation rates and have large offsets from the centers of their hosts \citep{2019Natur.572..352R,2019Sci...365..565B}. These two FRBs share similar properties with the host environments of short gamma-ray bursts (SGRBs) and binary neutron star (BNS) mergers \citep{2019ApJ...886..110M, 2020MNRAS.497.3131G}.
In addition, \citet{2019ApJ...884L..26L} also suggested that some FRB host candidates have low star formation and large offsets, and the corresponding FRBs may be driven by NSs that were born in BNS mergers.

Gamma-ray bursts (GRBs) are the most luminous explosions in the universe. Based on the duration distribution, they can be divided into LGRBs and SGRBs with a separation line of about 2 seconds \citep{1993ApJ...413L.101K,2015PhR...561....1K,2018pgrb.book.....Z}. For SGRBs, the NS-NS or NS-BH merger is the preferred progenitor model. The joint detection of an SGRB and the gravitational wave event confirms the NS-NS merger origin for at least some SGRBs \citep{2017ApJ...848L..12A,2017ApJ...848L..14G}.  Although the event rate of SGRBs is much smaller than that of FRBs, we believe that some FRBs may be related to SGRBs for the following reasons:
1.\, Several theoretical models predict FRB emissions are also in connection with neutron star mergers \citep{2013PASJ...65L..12T,2016ApJ...822L...7W,2020ApJ...890L..24Z}.
2.\, The BNS merger may leave behind a massive, stable, rapidly-spinning magnetar that could serve as the central engine of SGRB, which could also power FRBs after an uncertain time delay \citep{2013arXiv1307.4924P,2014MNRAS.442L...9L,2014ApJ...797...70K,2016PhRvD..93d4065G,2016ApJ...826..226K,2018ApJ...868...31Y,2018MNRAS.477.2470L,2019ApJ...886..110M,2020ApJ...891...72W,2020MNRAS.498.2384L,2020ApJ...896..142B}. 
3.\, For most host galaxies of one-off FRBs, their properties are similar to the host galaxies of SGRBs, which have low star formation rates and large offsets from the centers of the galaxies \citep{2005Natur.438..994B,2005Natur.437..851G,2005Natur.437..845F,2006ApJ...638..354B,2013ApJ...769...56F,2019Sci...365..565B,2019Natur.572..352R,2020Natur.577..190M,2020ApJ...891...72W}.

Searches for continuous radio emission and/or FRBs associated with SGRBs have been conducted in many works but have not revealed a reliable association \citep{1996MNRAS.281..977D,2012ApJ...757...38B,2014ApJ...785...27O,2014ApJ...790...63P,2015ApJ...814L..25K,2018ApJ...864...22A,2019MNRAS.490.3483R,2021PASA...38...26A,2021MNRAS.506.5268R,2022PASA...39....3T}. 
It is valuable to study the connection between SGRBs and FRBs with large FRB and SGRB samples \citep{2022arXiv220800803C}.
Recently,
the Canadian Hydrogen Intensity Mapping Experiment (CHIME) released a large sample of FRBs \citep{2021arXiv210604352T}.
In this work, we perform a systematic search for SGRBs that may be associated with FRBs in the SGRB and FRB samples which include all publicly reported FRBs and precisely-localized SGRBs before September 2022, making our work the one considering the most complete samples to date.

\section{Search for FRBs Associated with SGRBs}
\label{sect:Obs}

\begin{figure}[tbph]
    \begin{center}
        \includegraphics[width=0.9\textwidth]{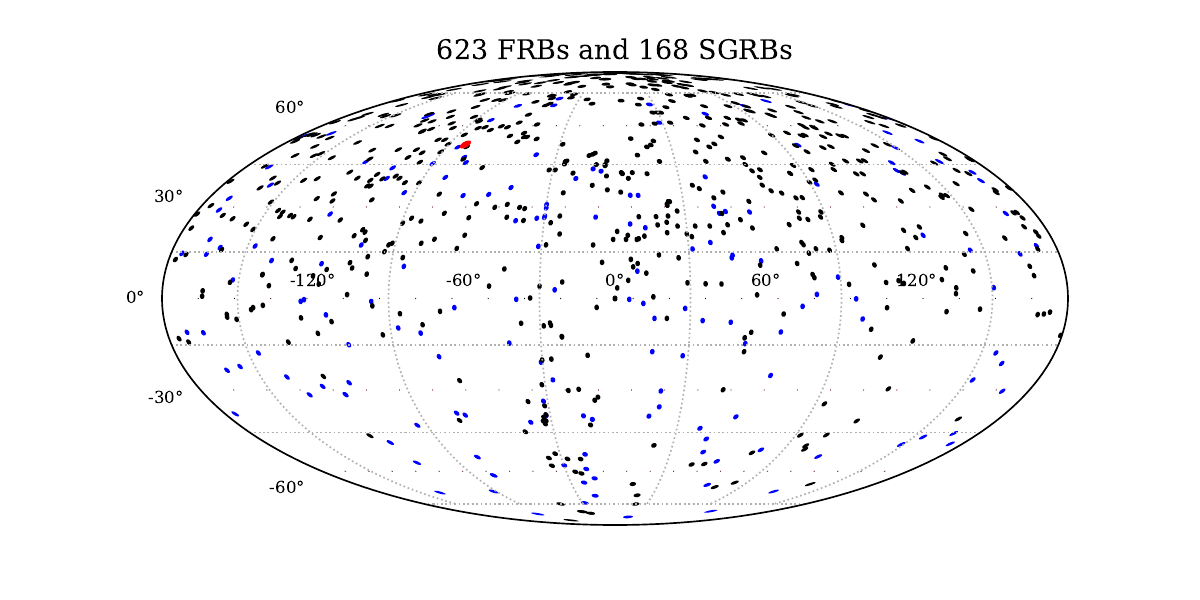}
    \end{center}
    \caption{Sky distributions in celestial coordinates of the 623 FRBs and 168 SGRBs considered in this work. The black points represent the positions of the FRBs. The blue points represent the locations of SGRBs. The region enclosing GRB 060502B and FRB 190309A is marked by a red circle.}
    \label{distribution}     
\end{figure}

The first FRB was reported in 2007 \citep{2007Sci...318..777L}. Since then, 807 FRBs have been reported in the literature until September 2022, including 601 one-off bursts and 204 repeat bursts from 22 repeaters\footnote{\href{https://www.wis-tns.org}{https://www.wis-tns.org}} \citep{2020TNSAN.160....1P}.
For the SGRB sample, we consider the GRB catalog\footnote{\href{https://www.mpe.mpg.de/~jcg/grbgen.html\#userconsent\#}{https://www.mpe.mpg.de/$\sim$jcg/grbgen.html\#userconsent\#}} presented by J. Greiner (JG catalog), which compiles GRBs detected by various detectors, e.g., Fermi, Swift, HETE-2, BeppoSAX, BATSE, AGILE, etc. The JG catalog contains thousands of objects and is updated almost every day. Until September 2022, there are more than one thousand GRBs with afterglow detections recorded in the JG catalog.
In general, T$_{90}$ is used to categorize LGRBs (T$_{90}$ $\textgreater$ 2s) and SGRBs (T$_{90}$ $\le$ 2s), and there are 111 SGRBs with afterglow detections in the JG catalog (we have also checked the \textit{Swift}/XRT website \footnote{\url{https://swift.gsfc.nasa.gov/archive/grb\_table.html/}} and confirmed that there are no other SGRBs with afterglow detections by \textit{Swift}/XRT). In addition, some GRBs with the T$_{90}$ larger than 2s are also classified as SGRBs because there is evidence that they are of merger origin,
which include 19 SGRBs (T$_{90}$ $\textgreater$ 2s).
In total, there are 130 SGRBs with afterglow detections in our sample.
In addition, considering the excellent localization capability of \textit{Swift}/BAT, we also use the SGRBs detected by {\em Swift}/BAT even without an afterglow detection (38 SGRBs). We list our SGRB sample in Table \ref{table2}. Figure \ref{distribution} shows the sky distribution of our sample, including 168 SGRBs and 623 FRBs (601 one-off bursts and 22 repeaters).

We perform a systematic search of our sample based on the following three criteria \citep{2020ApJ...894L..22W}: I, the SGRB should positionally overlap with the FRB within the localization error circle; II, the SGRB should occur earlier than the FRB; III, the redshift of the SGRB should be compatible with the FRB distance derived from its DM. The criterion III is described in more detail below.

The observed total DMs can usually be composed of
\begin{equation}
\rm{DM}_{\rm{total,obs}} = \rm{DM_{MW}} + \rm{DM_{IGM}} + \frac{DM\rm{_{host}}}{1+z}
\end{equation}
where DM$\rm{_{MW}}$ is the DM contribution from the Milky Way, the ${\rm DM_{IGM}}$ is the DM contribution from intergalactic medium (IGM) and DM$\rm{_{host}}$ is the DM contribution from the FRB host galaxy.
Although for a single FRB the relationship between ${\rm DM_{IGM}}$ and redshift is difficult to be determined, a mean value can be estimated through
\citep{2018ApJ...867L..21Z,2019JHEAp..23....1D}

\vspace{-0.36cm}
\begin{equation}
{\rm \overline{DM}_{IGM}}(z) = \frac {3 c H_0 \Omega_b f_{\rm IGM} f_{\rm e}} {8\pi G m{\rm _p}}\int_{0}^{z} \frac{1+z^{'}}{E(z^{'})} dz^{'}
\label{eq:dmz}
\end{equation}
where $E(z')$ = $H(z')/{H_0}$ with $H_0$ the Hubble constant, $\Omega_b$ is the baryon density, $f_{\rm{IGM}}\sim 0.83$ is the fraction of baryons in IGM \citep {1998ApJ...503..518F}, $m{\rm _p}$ is the proton mass, and the number of free electrons per baryon in the universe is $f_{\rm e} \, \sim$ 7/8 \citep{2014ApJ...783L..35D}. The $\Lambda$CDM cosmological parameters of the Planck results are adopted, i.e., $H_0$=67.74 km s$^{-1}$ Mpc$^{-1}$, $\Omega_{\rm m}$ = 0.3089, $\Omega_{\rm \Lambda}$ = 0.6911, and $\Omega_b$ = 0.0486 \citep{2016A&A...594A..13P}.
Because Eq.~(\ref{eq:dmz}) assumes that the matter in the IGM is uniformly distributed, the effects of inhomogeneities should also be considered, leading to an uncertainty $\sigma \rm _{IGM}$ $(z)$ of the $\rm{\overline{DM}_{IGM}}$ $(z)$
\citep{2014ApJ...780L..33M}.
Thus, we have to take into account the $\sigma \rm _{IGM} (z)$ when converting a redshift into $\rm{DM}_{IGM}$.
This means that for a given redshift, a DM range of $\rm{\overline{DM}_{IGM}}$ $(z)$ $\pm$ $\sigma \rm _{IGM} (z)$ is acceptable. The criterion III requires this range covers the $\rm{DM_{IGM}}=\rm{DM}_{\rm{total,obs}}-\rm{DM_{MW}}- {DM\rm{_{host}}}/{(1+z)}$. We use the results in \citet{2014ApJ...780L..33M} to obtain the value of $\sigma \rm _{IGM}$ and adopt the baryon distribution model that considers the details of the accretion rates of baryons into dark matter halos.

The DM$\rm{_{MW}}$ values have been provided in the FRB catalog \citep{2020TNSAN.160....1P}, which are derived based on two Galactic electron models \citep{2002astro.ph..7156C,2017ApJ...835...29Y}.
However, only a handful of FRBs have the measurements of DM$\rm{_{host}}$ to date (e.g., FRB 121102 \citep{2017ApJ...834L...7T}, FRB 171020 \citep{2018ApJ...867L..10M}, FRB 200120E \citep{2021ApJ...910L..18B}).
Therefore, we adopt the $\rm{DM_{obs}} - \rm{DM_{MW}}$ as a proxy / an upper limit of the DM$\rm{_{IGM}}$ (i.e., ignoring the DM contribution from the host galaxy). We have tested that assuming a typical value of 50 of the ${\rm DM_{host}}$ \citep{2019JHEAp..23....1D} would not change our main conclusions. By requiring $\rm{\overline{DM}_{IGM}}$ $(z)$ $\pm$ $\sigma \rm _{IGM} (z)$ covers the $\rm{DM_{IGM}}=\rm{DM}_{\rm{obs}}-\rm{DM_{MW}}$, we estimate the redshift range compatible with the observed DM for each FRB in our sample. 
Only one pair of SGRB and FRB satisfies the three criteria (in fact, only this pair satisfies criterion I): GRB 060502B and FRB 190309A.
The position of GRB 060502B located by {\em Swift}/XRT is (RA, Dec) = $(18\,35\,45.74,+52\,37\,52.47)$ with an 90\% error radius of $4''.4$ \citep{2006GCN..5093....1T}.
The redshift of GRB 060502B is $z$ = 0.287 \citep{2007ApJ...654..878B}.
FRB 190309A was detected by CHIME with a position of (RA, Dec) = $(278.96\pm0.23,52.41\pm0.24)$\footnote{\url{http://www.chime-frb.ca/catalog/FRB20190309A}} \citep{2021arXiv210604352T}. The angular separation between the two sources is $0.22^\circ$, less than the localization uncertainty of FRB 190309A by CHIME.
FRB 190309A was detected 4694 days ($\sim$12.8 yr) after the GRB 060502B trigger. The observed total DM is 356.9 $\rm pc\,cm^{-3}$ and the extragalactic contribution ${\rm DM_E}={\rm DM_{IGM}} + \rm{DM_{host}}/(1+z)$ is 298.3 $\rm pc \, cm^{-3}$ \citep{2021arXiv210604352T} when using the NE2001 model\citep{2002astro.ph..7156C}.
We estimate its redshift range to be $z \sim (0.23 \sim 0.54)$ (assuming a $\rm DM_{host}$ of 50 gives $z \sim (0.20 \sim 0.48)$), which covers the redshift of GRB 060502B.

To derive the significance of the association, we calculate the chance possibility of the putative GRB 060502B-FRB 190309A association by Monte Carlo (MC) simulation. The CHIME sky coverage is $2\pi(1-\cos101^\circ)=7.48\,{\rm sr}$, where $101^{\circ}$ is the latitude range of the CHIME observation (RA: $0-360^{\circ}$, Dec: $-11^\circ-90^\circ$) \citep{2021arXiv210604352T}. Due to the large sky coverage, the present FRB sample is dominated by the CHIME FRBs with 503 (483 one-off FRBs and 20 repeaters) out of 623 FRBs in the whole sample. Therefore, the MC simulation could be performed considering only the CHIME FRBs and the SGRBs within the CHIME sky coverage. There are $103/168 \approx 61.3\%$ SGRBs in our SGRB sample within the CHIME sky coverage.
Finally, 503 CHIME FRBs and 103 SGRBs are used in this MC simulation.

We perform the MC simulation as follows: Based on the sky distributions of the observed 103 SGRBs and 503 FRBs, we generate 103 and 503 pseudo- SGRBs and FRBs within the CHIME sky coverage. The SGRBs are randomly sampled isotropically, while the FRBs follow the CHIME FRB distribution in the sky, as shown in Figure 2. For the simulated FRBs and SGRBs, we randomly assign each of them a localization error, observation time, and redshift taken from the actual data of the real FRBs and SGRBs\footnote{Only 36 SGRBs have redshift measurements, and we randomly draw one from these 36 redshifts for each pseudo-SGRB.}. It's important to note that the localization errors of CHIME FRBs are not uniform across the sky; they are notably better at lower declinations than at higher declinations. Therefore, we divide the sky into 3-degree declination bins and randomly select a real localization error from the CHIME FRBs within the corresponding bin based on the declination of the simulated FRB.

We perform $10^5$ MC simulations (for each simulation 103 SGRBs and 503 FRBs are sampled) and calculate the chance probabilities as follows.
For each simulation, we apply the three criteria mentioned above to select association candidates. We record the number of simulations in which there is at least one pair of pseudo-sources satisfying the criteria, denoted as $N$. The chance probability is then calculated as $P = N/10^5$.
Considering only criterion I, we find a chance probability of 0.27. When criterion II is included, the chance probability is 0.21. With criterion III included, the chance probability is further reduced to 0.05.

\begin{figure}
\center
\includegraphics[scale=0.5]{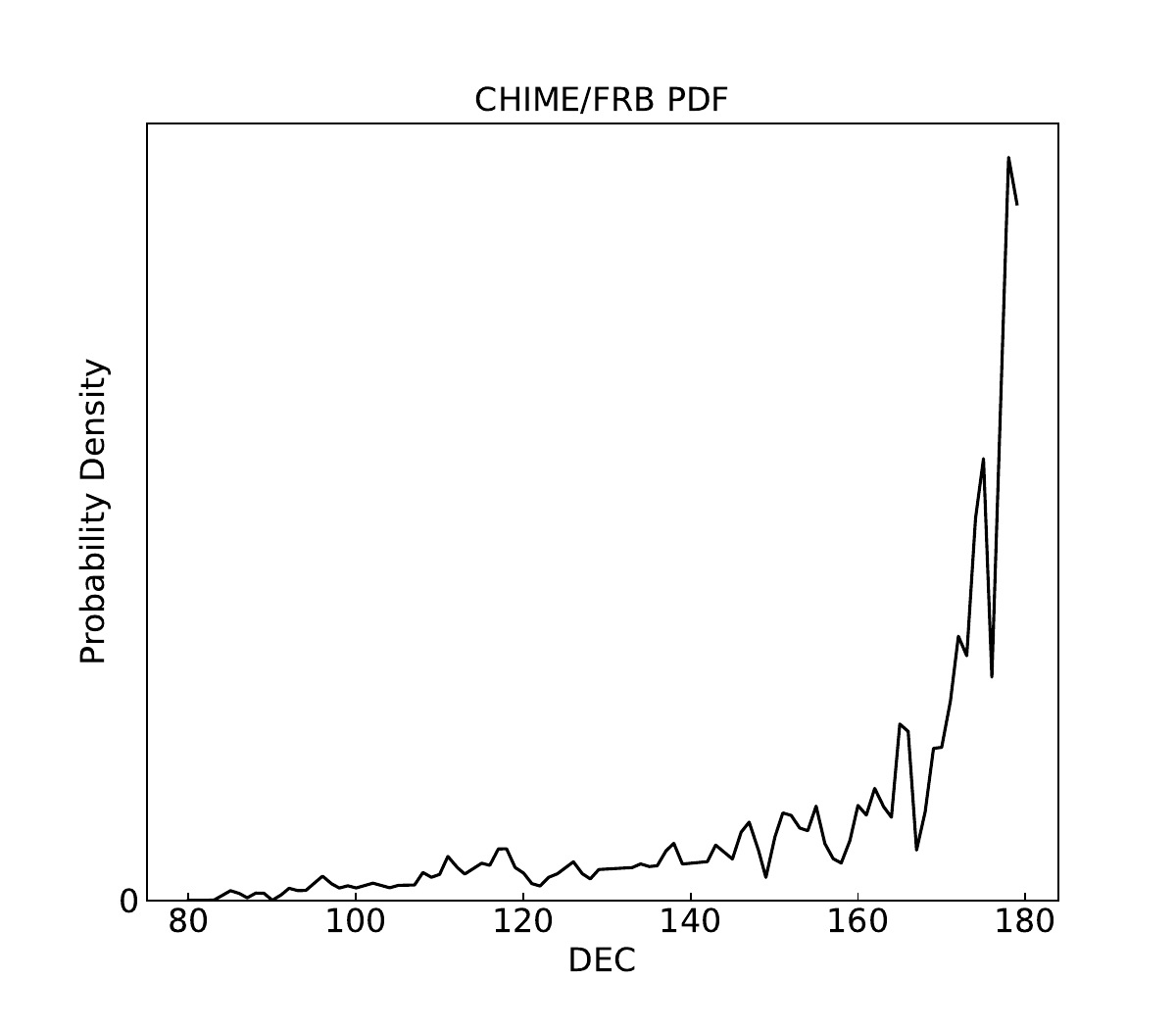}
\caption{CHIME does not have a uniform exposure across the sky \citep{2021arXiv210604352T}. This plot shows the probability density distribution of the 503 CHIME FRBs as a function of declination, which is given by the relative number of FRBs in the DEC $\pm$ 1$^{\circ}$ region divided by the corresponding solid angle.}
\label{PDF}
\end{figure}

\section{Is GRB 060502B Associated with FRB 190309A?}
\label{sect:GRB 060502B}
\subsection{Host Galaxy}

GRB 060502B triggered the {\em Swift}/BAT on 05 June 2006 (UT) at 17:24:41 ($T_{0}$), with $T_{90}=0.131\,{\rm s}$ \citep{2006GCN..5055....1T,2006GCN..5064....1S}. 
The {\em Swift}/XRT and {\em Swift}/UVOT began to observe the X-ray and optical afterglow of GRB 060502B at 76 and 100 s after the {\em Swift}/BAT trigger, respectively \citep{2006GCN..5055....1T,2006GCN..5069....1P}. 
Many optical telescopes have carried out follow-up observations, e.g., MASTER optical telescopes \citep{2006GCN..5056....1L}, Xinglong 0.8m telescope \citep{2006GCN..5057....1Z}, Tautenburg 1.34m Schmidt telescope \citep{2006GCN..5062....1K}, AROMA optical telescope \citep{2006GCN..5065....1T,2006GCN..5073....1T}, MDM 1.3m telescope \citep{2006GCN..5066....1H,2006GCN..5072....1H}, Cassini telescope \citep{2006GCN..5074....1M},
Gemini North telescope \citep{2006GCN..5071....1B,2006GCN..5077....1P} and MAO 1.5m telescope \citep{2006GCN..5184....1R}. 
No optical counterpart was found from several minutes to more than ten hours after the GRB 060502B trigger, and only a faint X-ray afterglow was detected by {\em Swift}/XRT.

The identification of the host galaxy of GRB 060502B has been studied in some previous works \citep{2006GCN..5238....1B,2007ApJ...654..878B,2007ApJ...664.1000B,2011MNRAS.413.2004C,2012MSAIS..21..104C}. \citet{2007ApJ...654..878B} proposed that a bright galaxy (referred to as $G^*$), situated south of the Swift/XRT localization of the GRB with an angular separation of 17.5 arcsec (approximately 73 kpc if assuming a redshift of $z\sim0.287$), is the host galaxy of GRB 060502B (chance probability of 0.03). 
\citet{2011MNRAS.413.2004C} noted that such a large offset is inconsistent with the scenario that GRB 060502B originates from a compact binary merger in the galaxy. 
They suggest that either the precursor of GRB 060502B was a binary neutron star system that formed within a globular cluster inside $G^*$, giving this binary a large initial kick, or $G^*$ is not the actual host galaxy of GRB 060502B.
Except for the $G^*$, within 1 arcmin of the GRB 060502B position, we find no other possible host galaxy candidates for GRB 060502B with redshift measurements\footnote{Here, we search for the candidates from the Extragalactic Database \citep{https://doi.org/10.26132/ned1} and Sloan Digital Sky Survey \href{https://www.sdss.org/dr17/}{https://www.sdss.org/dr17/}.}. Furthermore, the results obtained from the relation between the spectral peak energy ($E_p$) and the isotropic gamma-ray energy ($E_{\gamma,\rm iso}$), namely the \textit{Amati}-relation \citep{2002A&A...390...81A,2018ApJ...859..160W}, indicate that when adopting a redshift of $G^*$ ($z$ = 0.287 \citet{2009ApJ...703.1696Z}) the GRB 060502B well lies within the SGRB's \textit{Amati}-relation. Therefore, we also consider the bright galaxy $G^*$ as the host galaxy for GRB 060502B.

The properties of the host environment (e.g., the offset from the host center, the star-formation rate (SFR), and the total stellar mass of the galaxy ($M_{\rm s,tot}$)) may provide important clues for unveiling the origin of FRBs. 
Although the sample size of FRB host galaxies is still limited, some statistical analyses between FRB hosts and the host galaxies of other transient sources, e.g., SGRBs, core-collapse supernovae (CCSNe), Type Ia supernovae (SNe Ia), LGRBs, SLSNe, have been performed \citep{2020ApJ...905L..30S,2020ApJ...895L..37B,2020ApJ...903..152H,2020ApJ...899L...6L,2021ApJ...917...75M,2021ApJ...907L..31B,2022AJ....163...69B,2023arXiv230703344L}. A majority of these works suggest that the host galaxies of SGRBs and CCSNe are similar to those of FRBs. In particular, \citet{2020ApJ...899L...6L} investigated 9 FRBs' hosts and suggested that the SFRs and the $M_{\rm s,tot}$ of 8 of them are consistent with SGRBs. \citet{2022AJ....163...69B} presented that the global properties (offset, SFR, $M_{\rm s,tot}$) of FRB hosts are indistinguishable from those of the hosts of CCSNe and SGRBs.

\begin{figure*}
\centering
\includegraphics[scale=0.29]{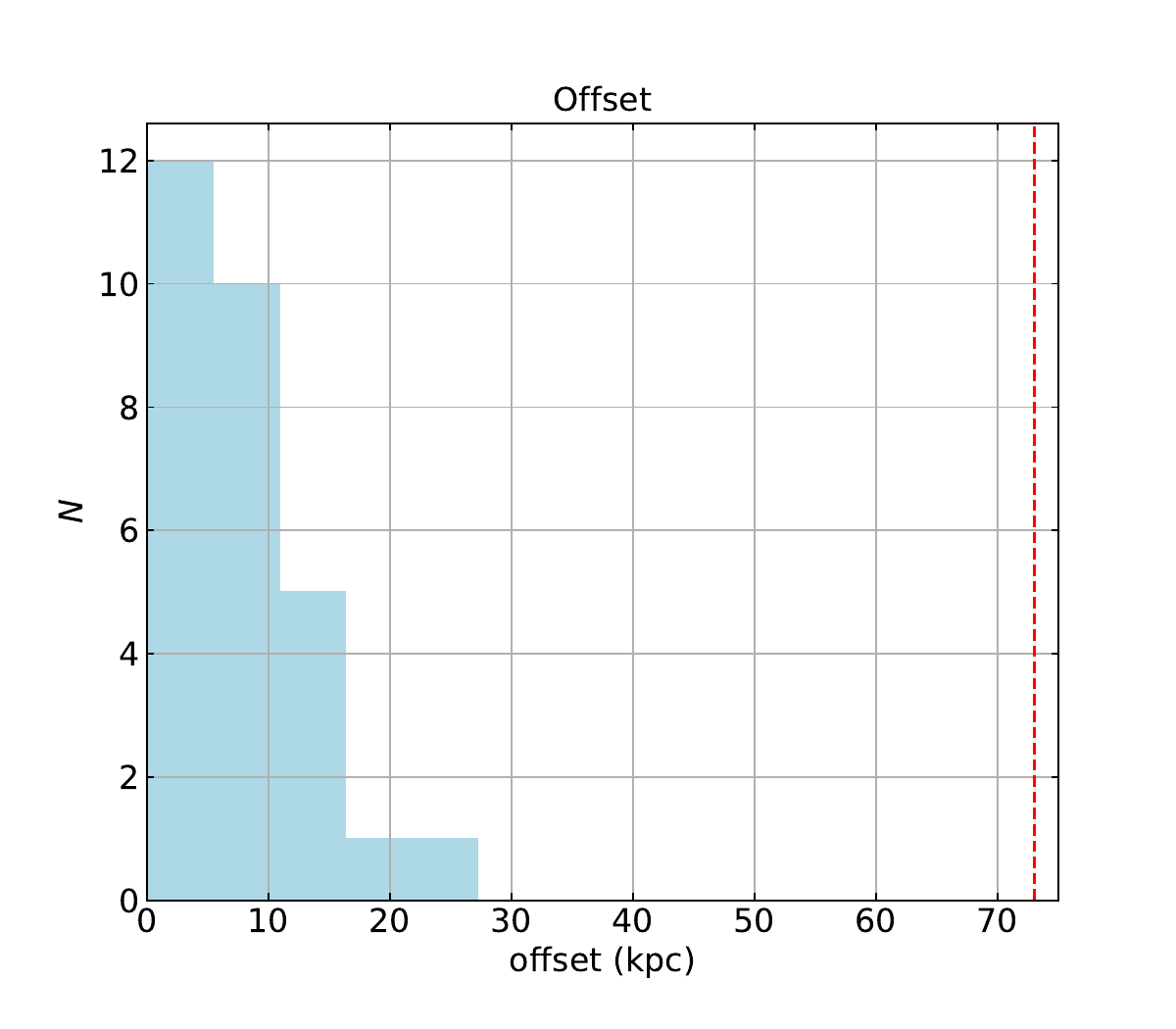}
\includegraphics[scale=0.29]{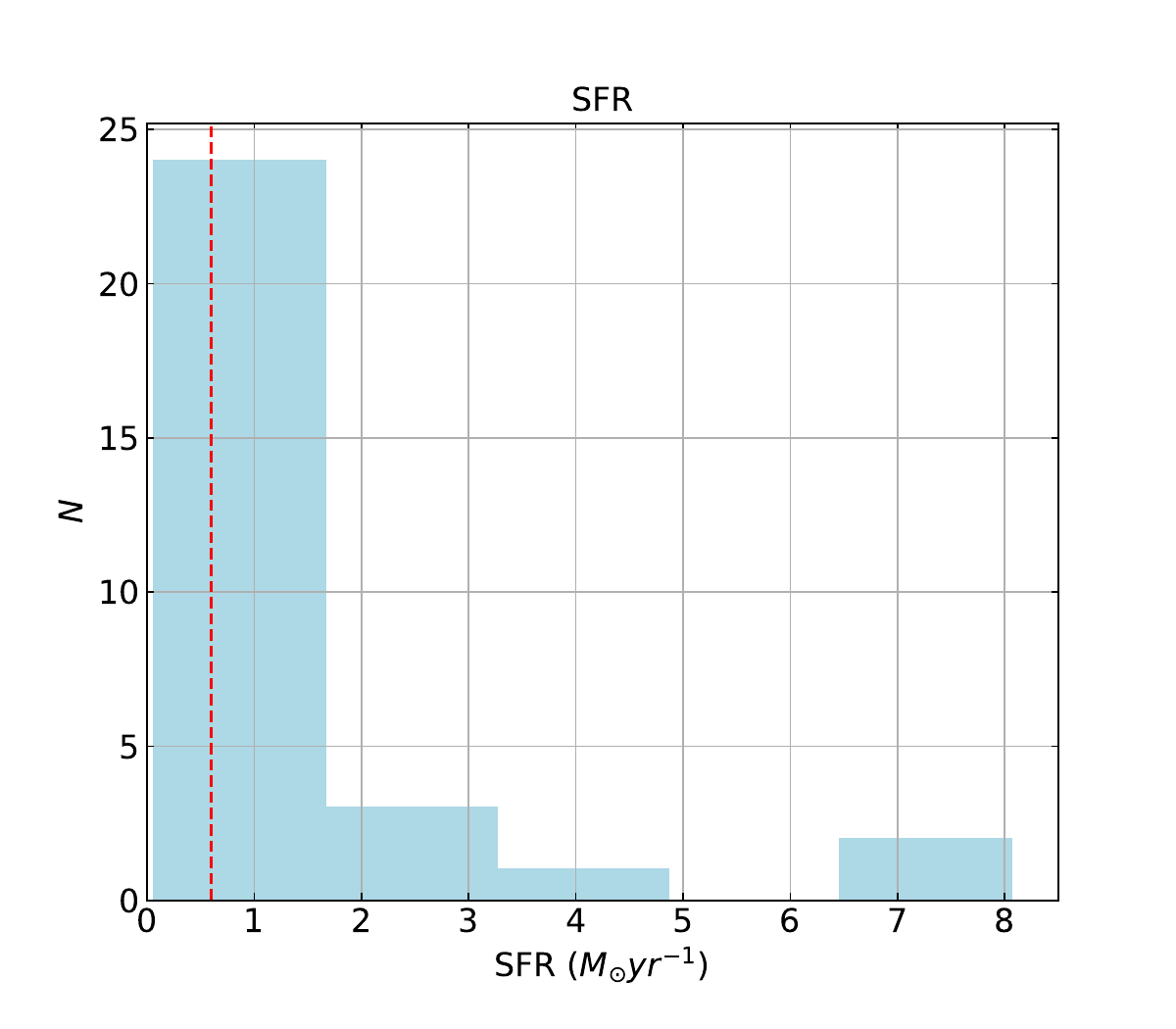}
\includegraphics[scale=0.29]{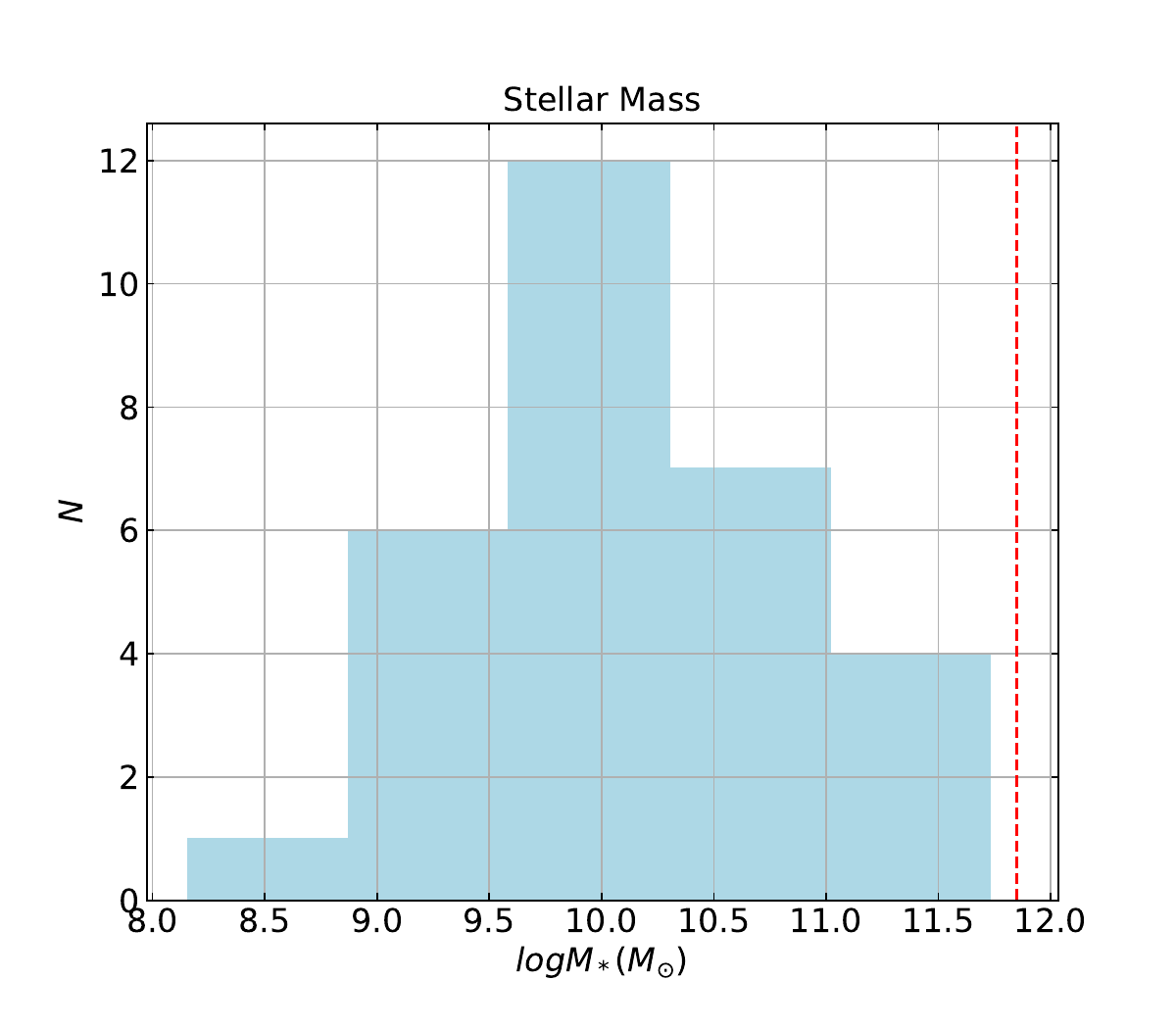}
\caption{Distributions of offsets (left panel), host galaxy SFRs (middle panel), and stellar masses (right panel) for 30 FRBs (light blue region). The dashed red lines represent the values for GRB 060502B.}
\label{Figure2}
\end{figure*}

To compare the host galaxy properties between the GRB 060502B (i.e., $G^*$) and the FRB population, we also demonstrate the offsets, SFRs, and $M_{\rm s,tot}$ for 30  FRB hosts in Table \ref{table1} and Figure~\ref{Figure2}, including 17 FRBs from \citet{2022AJ....163...69B}, 11 FRBs from \citet{2023arXiv230703344L} and two possible hosts from \citet{2021ApJ...919L..24B} (FRB 181030A) and \citet{2022AJ....163...69B} (FRB 181112A).
The red lines denote the values for the $G^*$.
As is shown in the figures, although the SFR of $G^*$ falls well into the range of the FRB host galaxy population, the other two properties (offset and $M_{\rm s,tot}$) are not consistent with those of FRBs, which do not support that $G^*$ is a typical FRB host galaxy.

However, as mentioned above, if $G^*$ is considered as the host galaxy of GRB 060502B, the possible born channel for this GRB is the merger of a BNS system formed in a GC of $G^*$ \citep{2011MNRAS.413.2004C}. 
Therefore, if FRB 190309A is associated with GRB 060502B, the FRB should also arise from this channel, which is not the same as the born channels of most of FRBs.
One FRB that has a similar channel is the burst FRB 200120E.
FRB 200120E could have arisen in a GC and been powered by a magnetar born from the merger of a compact binary system within the GC \citep{2022Natur.602..585K}.
Although the merger position of the progenitor of FRB 190309A is not identical to that of FRB 200120E,
the sparse sample size currently does not allow us to tell whether $G^*$ matches the host properties of this class of FRBs and thus cannot rule out the possibility that $G^*$ serves as the host galaxy of FRB 190309A. 

\subsection{Dispersion Measure}
\label{sect:Associate}
DM${\rm_{host}}$ can be resolved into three components, including DM$_{\rm_{src}}$, DM${\rm_{ISM}}$ and DM${\rm_{halo}}$. They are the DMs contributed by the source environment, the interstellar medium (ISM), and the halo of the host galaxy, respectively. Here, we assume that GRB 060502B and FRB 190309A are associated, and we want to estimate whether the source environment of GRB 060502B can allow the FRB to escape and whether the observational DM${\rm_ {host}}$ of FRB 190309A can be explained by the host galaxy of GRB 060502B.

Since GRB 060502B is supposed to be from a BNS merger, its DM$_{\rm_{src}}$ would be contributed by the ejecta of the BNS merger. The BNS merger ejecta is different from the CCSNe and has a higher velocity $v \sim (0.1 - 0.3)c$ and a lower mass $M \sim (10^{-4} - 10^{-2}) M_{\odot}$.
After the BNS merger, the DM$_{\rm {src}}$ can be derived as follows \citep{2020ApJ...891...72W},
\begin{equation}
{\rm DM_ {src}} = n_e \Delta R \simeq \frac{ \eta Y_e M } { 4 \pi m{ _{ \rm p } } (vt)^2  } \simeq 0.17 {\rm pc \, cm^{-3}}
\\ \times \,\eta \Big(\frac{Y_e}{0.2}\Big) \Big(\frac{M}{10^{-3}M_\odot}\Big) \Big(\frac {v} {0.2\, c}\Big)^{-2}  \Big(\frac{t}{1 \, {\rm yr}}\Big)^{-2}
\end{equation}
where $n_e \simeq { \eta Y_e M }/{[4 \pi m{ _{ \rm p } } (vt)^3]}$ is the free electron density, $\Delta R$ $\sim$ $vt$ is the ejecta thickness, $v$ is the speed of the ejecta, $t$ is the elapsed time after the BNS merger, $\eta$ is the ionization fraction, $Y_e$ is the electron fraction, and $M$ is the mass of the ejecta.
The ${\rm DM_ {src}}$ is derived to be 7.9 $\times$ $10^{-4}$ ${\rm pc \, cm^{-3}}$ using $t=12.8$ yrs (the time difference between GRB 060502B and FRB 190309A).
This indicates a clean source environment, which is thought to be a prerequisite for FRB escape if the magnetar acts as the central engine to power the FRB \citep{2016MNRAS.461.1498M,2017ApJ...841...14M,2018ApJ...868L...4M}.

According to the redshift of GRB 060502B ($z$ = 0.287), the $\rm{\overline{DM}_{IGM}}$ can be estimated from equation (2), i.e. $\rm{\overline{DM}_{IGM}} \simeq 245.2\, \rm pc \,  cm^{-3}$. We can derive that the ${\rm DM_{host}}$ is
\begin{equation}
{\rm DM_{host}} = (1+z)({\rm DM_{obs}-DM_{MW}-\rm {\overline{DM}_{IGM}}) \simeq 68.3 \, \rm pc \, \rm cm^{-3}}
\end{equation}
where ${\rm DM_{MW}} = 58.6\,{\rm pc\,cm^{-3}}$ is the contribution from the Milky Way.
Clearly, the DM$_{\rm {src}}$ ($\sim$ 7.9 $\times$ $10^{-4}$ ${\rm pc \, cm^{-3}}$) can be neglected. 
Meanwhile, due to the large offset between GRB 060502B and $G^*$, the DM$_{\rm {ISM}}$ part should also be neglected.
Thus, the DM$\rm{_{host}}$ is mainly contributed by the DM$_{\rm {halo}}$.

Following \citet{2011MNRAS.413.2004C}, we adopt the below profile to model the dark matter halo of $G^*$ \citep{2009ApJ...691..770T},
\begin{equation}
\rho (r) = \frac{v^2_{\rm{h}}}{4\pi G} \frac{3r^2_{\rm{h}} + r^2}{(r^2_{\rm{h}} + r^2)^2}
\label{eq:profile}
\end{equation}
where $r_{\rm{h}}=20.45\,{\rm kpc}$ is the core radius of the halo, $v_{\rm{h}}=505.31\,{\rm km\,s^{-1}}$ is the circular velocity at infinity and $G=6.67\times10^{-11}\,{\rm m^{3}\,kg^{-1}s^{-2}}$ is the gravitational constant.
We take the halo mass from \citet{2011MNRAS.413.2004C}, i.e. 6.01 $\times 10^{12} M_{\odot}$, which is enclosed within a spherical radius of $R_{\rm max}=105\,{\rm kpc}$ of the halo.
We consider the simplest model for baryons in the halo, which are assumed to trace the underlying dark matter halo distribution \citep{2019MNRAS.485..648P}. The baryon density in the halo is thus, $\rho_{\rm{b}} (r) \propto f_{\rm{b, halo}}\rho (r){\Omega_{\rm{b}}}/{\Omega_{\rm{m}}}$, where the cosmic baryon fraction is ${\Omega_{\rm{b}}}/{\Omega_{\rm{m}}}$ $\thickapprox$ 0.158 and $f_{\rm{b,halo}}$ is the fraction of baryons existing in a halo form (i.e., excluding those portion in the form of ISM and stars). Here, we consider two values: (1) $f_{\rm{b,halo}}$ = 0.75, which assumes that $\thickapprox 25 \%$ of the baryons exist in the ISM, stars and their remnants \citep{1998ApJ...503..518F}. (2) $f_{\rm{b,halo}}$ = 0.40, which is a lower limit for a $\sim$ 10$^{12} M_{\odot}$ halo found by \citet{2019MNRAS.488.1248H}. 

\begin{figure}
\centering
\includegraphics[scale=0.4]{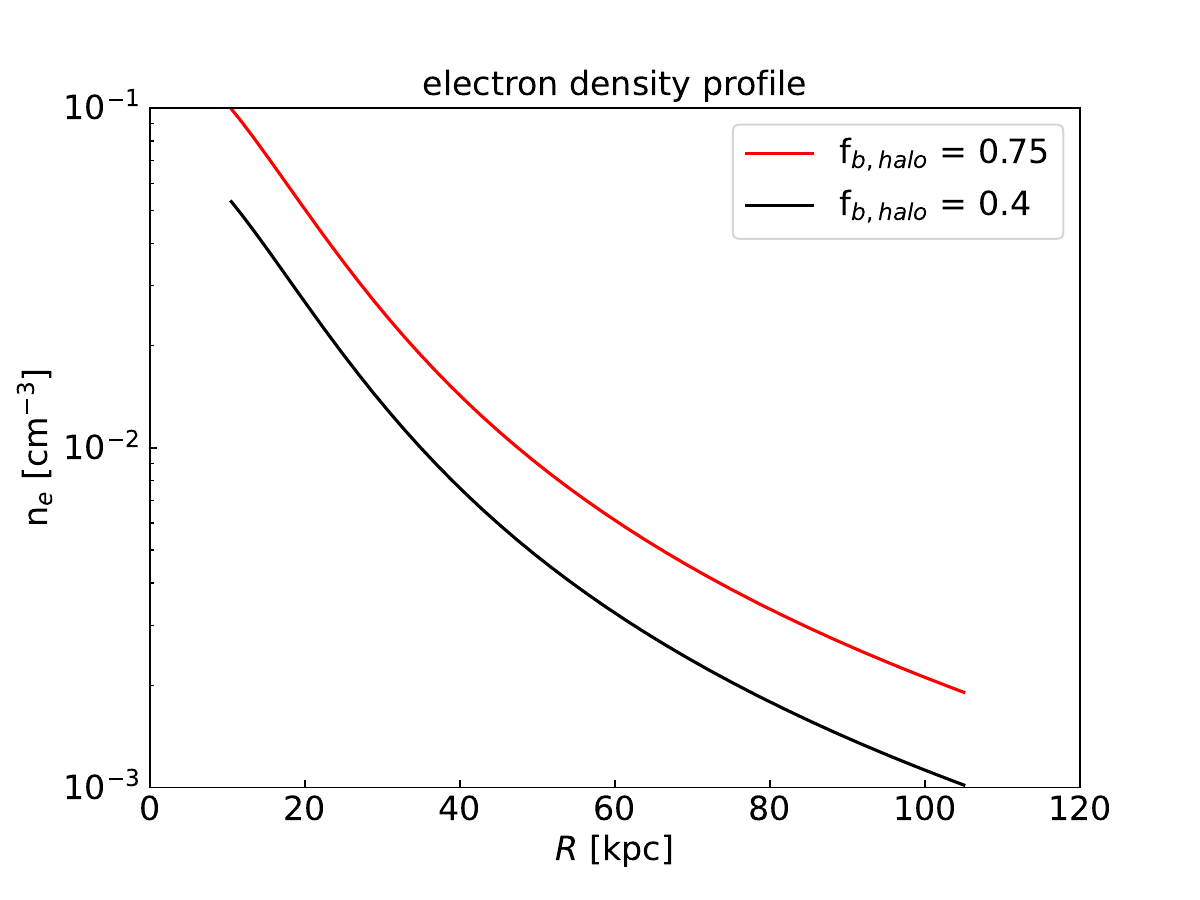}
\caption{The electron number density as a function of halo radius in the halo of $G^*$. 
}
\label{Figure3}
\end{figure}

The line of sight may travel through the $G^*$ halo, and thus the DM contributed by the halo is 
\begin{equation}
{\rm{DM}}(x) = \int_0^x n_{\rm{e}} dl.
\label{eq:dmhalo}
\end{equation}
The $n_e$ is the free-electron number density in the halo, which can be estimated by \citep{2019MNRAS.485..648P}
\begin{equation}
n_e = \mu_{\rm{e}} \frac{\rho_{\rm{b}}}{m_{\rm{p}}\mu_{\rm{H}}}
\end{equation}
where $\mu_{\rm{H}}$ = 1.3 the reduced mass (accounting for Helium) and 
$\mu_{\rm{e}}$ = 1.167 accounts for fully ionized Helium and Hydrogen.
The electron number density distribution is shown in Figure \ref{Figure3}. 
Considering that we don't know exactly the position of the source on the line of sight, the $x$ in Eq.~(\ref{eq:dmhalo}) could be $0 \le x \le 2{\sqrt{{R_{\rm{max}}^2-R_{\perp}^2}}}=151\,{\rm kpc}$ (where $R_\perp=73\,{\rm kpc}$ is the impact parameter) and we can only estimate the DM$\rm{_{halo}}$ range corresponding to \textit{x} varying from 0 to 151 kpc.
Therefore, we obtain the range of DM$\rm{_{halo}}$ of 0 $-$ 470 pc cm$^{-3}$
(for $f_{\rm{b,halo}}$ = 0.75) or 0 $-$ 250 pc cm$^{-3}$ (for $f_{\rm{b,halo}}$ = 0.4).
Although the DM$\rm{_{halo}}$ result is affected by the profile model of the galaxy halo (one can see Figure 1 in
\citet{2019MNRAS.485..648P}, which shows the DM$\rm{_{halo}}$ results for different halo profiles; see also \citet{2023ApJ...946...58C} for a newly developed model of the Galactic electron halo), at least for the profile considered in this paper [Eq.~(\ref{eq:profile})], the DM${\rm_{host}}$ of FRB 190309A (68.3 $\rm pc\,cm^{-3}$) is compatible with the possibility that $G^*$ is the host galaxy.

\subsection{Optical Depth and Time Delay}
In addition, the ejecta surrounding the newborn magnetar could affect the observational signature of FRBs. We here consider the free-free optical depth due to the ejecta of the BNS merger. When the ejecta is transparent to the FRB, the free-free optical depth should be less than 1. We can derive the free-free optical depth with \citep{2020ApJ...891...72W}
\begin{equation}
\begin{split}
\tau{_{\rm ff}}  = \alpha_{\rm ff} \Delta R \simeq (0.018T_{\rm{eje}}^{-3/2}Z^2n_en_i\nu^{-2}\overline{g}_{\rm ff})\Delta R  = 2.7 \times 10^{-8} \\
\times \, \eta ^2 \Big(\frac{Y_{e}}{0.2}\Big)^{2} \Big(\frac{M}{10^{-3}M_\odot}\Big)^{2} \Big(\frac{T_{\rm {eje}}}{10^4{\rm K}}\Big)^{-3/2} \Big(\frac{\nu}{1 \,{\rm GHz}}\Big)^{-2} \Big(\frac{v}{0.2 \, \rm c}\Big)^{-5} \Big(\frac{t}{1 {\rm yr}}\Big)^{-5}
\end{split}
\end{equation}
where $n_e$ and $n_i$ are the number densities of electrons and ions, respectively, and $n_e \sim n_i$ and $Z \sim 1$ are assumed in the ejecta that is a fully ionized hydrogen-dominated composition, $T_{\rm eje}$ is the ejecta temperature, $\overline{g}_{\rm ff} \sim 1$ is the Gaunt factor. The calculated result suggests that the FRB could be detected a few weeks after the BNS merger.
Therefore, the optical depth allows the FRB 190309A to escape the source environment of GRB 060502B.

The FRB emitted from a magnetar has been proposed to be due to magnetospheric 
activities. When the magnetar magnetosphere is triggered by a crust fracturing, the 
magnetic energy in the crust will be released and converted into particle energy 
and radiation, and an FRB is emitted
\citep{2020MNRAS.494.2385K,2019ApJ...879....4W,2020ApJ...902L..32D,2020MNRAS.498.1397L}.
In this picture, as long as the magnetar exists, it is possible to produce FRBs through a crust fracturing, and the time of the crust fracturing is extremely uncertain during the existence of the magnetar.
Here, we estimate the timescale for the existence of the magnetar (i.e., the timescale of magnetic field decay).
The magnetic field in the crust would decay via the Ohmic dissipation and the evolution of the magnetic field can be approximated by $dB/dt\simeq-AB$ \citep{2021ApJ...919...89Y} with $A = 10^{18}\,\rm G^{-1}yr^{-1}$ \citep{2000ApJ...529L..29C}. 
Then the typical timescale of magnetic field decay is \citep{2021ApJ...919...89Y}
\begin{equation}
\tau_B = \frac{1}{AB_0} = 10^4 {\rm yr} \Big(\frac{B_0}{10^{14}\rm G}\Big)^{-1}
\end{equation}
where $B_0$ is the initial magnetic field. 
One can see that the timescale of magnetic field decay (and thus the time delay of the FRB) could be 10$^{4}$ years for $B_0$ = 10$^{14}$ G. 
For example, FRB 200428 may be released by the magnetospheric activities of a Galactic magnetar with an age of approximately 10$^{4}$ yrs \citep{2020Natur.587...59B,2021MNRAS.507.2208W}. 
Therefore, the relatively large time span ($\sim 12.8$ yr) between GRB 060502B and FRB 190309A is also compatible with the model.

\subsection{Energy Budget}
Although the central engine of FRBs is generally considered to be a magnetar, the exact process for the generation of FRBs differs in different models.
For instance,  FRBs could be generated by synchrotron maser in relativistic shocks driven by the magnetar flares \citep{2014MNRAS.442L...9L,2019MNRAS.485.4091M,2020ApJ...896..142B}, by the curvature radiation in the magnetosphere  \citep{2020MNRAS.498.1397L}, by the magnetic reconnection of external magnetosphere \citep{2020ApJ...897....1L}, or by the inverse Compton scattering in the magnetosphere \citep{2022ApJ...925...53Z}.
Here, rather than going into the details of any of the above models, we consider only a conservative estimate of the magnetic field of the magnetar, since for all of these models FRBs are essentially driven by the magnetic energy of the magnetar.

Within this picture,
we try to calculate whether the magnetar could be used as the central engine of FRB 190309A.
FRB 190309A has a duration of
$\Delta t \sim 1.97\,\rm ms$, a flux of $S_{\nu,p} \sim 0.39\,\rm Jy$ and a fluence of $f_\nu \sim0.72\,\rm Jy\,ms$ at $\nu_c \sim 468\,\rm MHz$.
Assuming that FRB 190309A is associated with GRB 060502B, one can
derive the luminosity distance as $D_{\rm L}\sim 1.52\,\rm Gpc$ according to the redshift of GRB 060502B ($z \sim$ 0.287). The luminosity and isotropic energy of FRB 190309A can be calculated as $L_{\rm p} \backsimeq 4 \pi D_{\rm L}^2 S_{\nu,p} \nu_c$ $\simeq$
5.1 $\times$ 10$^{41}$ erg s$^{-1}$ and $E\rm_{FRB}$   $\backsimeq$ ${4\pi D\rm_L^2} f_\nu \nu_c/({1+z})\simeq 7.3 \times$
10$^{38}$ erg.
If this energy is provided by the magnetic energy of the magnetar, one can place a constraint on the strength of the surface polar cap magnetic field of the underlying magnetar. 

The emission radius can be approximately estimated as $r_e \sim c\Delta t \backsimeq 5.91 \times 10^7$ cm, which is consistent with the predicted emission radius of the magnetar that produced the FRB, i.e., a few $\times 10R\rm _{NS} \sim 10^7$ cm \citep{2021arXiv211008435L}. The magnetic field strength at $r_e$ should satisfy
\begin{equation}
\frac{B_e^2}{8\pi} \left(\frac{4\pi}{3} r_e^3\right) \geq E\rm_{FRB},
\end{equation}
where $B_e=B\rm_p$  $(r_e/R)^{-3}$ \citep{2020ApJ...894L..22W} and $B\rm_p$, $R (\sim 10^6$ cm) are the magnetic field strength of the surface polar cap and typical radius of the magnetar, respectively. Therefore, the observation of FRB 190309A demands that the magnetic field strength of the surface polar cap is
\begin{equation}
B{\rm_p} \geq \left(\frac{6Er_e^3}{R^6}\right)^{1/2} \simeq 3 \times 10^{13} \rm\, G,
\end{equation}
which is indeed consistent with a magnetar central engine.
\section{Summary}

In this work, we systematically search for possible associations between SGRBs and FRBs based on a sample of 623 FRBs (601 one-off bursts and 22 repeaters) and 168 SGRBs.
We find that FRB 190309A is spatially coincident with GRB 060502B. Moreover, GRB 060502B occurred earlier than FRB 190309A, and its redshift is consistent with the range of the distance derived from the DM of FRB 190309A. 
Considering the observational information such as spatial location, time of occurrence, and redshift, we obtain a chance probability of the association of $\sim$ 0.05.

Considering the statistical significance is not high enough ($<3\sigma$) to claim a reliable association between GRB 060502B and FRB 190309A, we further investigate if there is any other evidence to support the physical association between them. 
We find that the (candidate) host galaxy of GRB 060502B, $G^*$, has an SFR similar to the hosts of the FRB population, but for another two host properties we have investigated (i.e., the offset from the host center and the total stellar mass of the galaxy), the $G^*$ ones do not coincide with the distributions of 19 FRBs.
However, considering that FRB 190309A may have a different born channel of its underlying magnetar from these FRBs, it may not be enough to exclude the association between  FRB 190309A and GRB 060502B on this basis. 
In addition, adopting the redshift of GRB 060502B, we estimate that FRB 190309A has a DM$\rm_{host}  \simeq $ 68.3 $\rm pc\,cm^{-3}$, which is able to be contributed by the (candidate) host galaxy $G^*$ of GRB 060502B. 
We also derive the free-free optical depth around the source and find it allows the FRB to be detectable. 
Finally, adopting the redshift of GRB 060502B, we obtain an isotropic energy of $E_{\rm FRB}\sim 7.3 \times 10^{38}\,{\rm erg}$ for FRB190309A; accordingly, the required surface magnetic field to power FRB 190309A is $B\rm_p \geq$ 3 $\times$ 10$^{13}$ G, which is also consistent with the typical magnetic field of the SGRB magnetars.
These results indicate that a physical association between GRB 060502B and FRB 190309A is feasible.

Overall, in this paper we do not find a reliable SGRB counterpart that is associated with FRBs; one possible FRB-SGRB association is between GRB 060502B and FRB 190309A, but it has a relatively large chance probability (p$\sim$0.05).
Even though, considering that this is at present the only pair of FRB and GRB that are spatially associated, it is still worthy of our attention. 
We therefore detailedly examine the possibility of their physical association from the aspects of the host galaxy, the DM, the energy budget, etc., 
and find that all of these could not exclude the possibility of their association. 
For this reason, we suggest that GRB 060502B/FRB 190309A pair is still a promising case of FRB-SGRB association that is worth further studying.

\section*{Acknowledgements}
We acknowledge the use of public data from the {\em Swift}, CHIME, GCN, TNS and frbhosts data archive. This work is supported by the National Natural Science Foundation of China (Nos. U1938201, 12133003, U1731239, 12203013), the Guangxi Science Foundation (grants 2018GXNSFGA281007, 2017AD22006, 2021AC19263). 
\label{lastpage}

\bibliography{master}{}

\begin{thebibliography}{}
\expandafter\ifx\csname natexlab\endcsname\relax\def\natexlab#1{#1}\fi
\providecommand{\url}[1]{\href{#1}{#1}}
\providecommand{\dodoi}[1]{doi:~\href{http://doi.org/#1}{\nolinkurl{#1}}}
\providecommand{\doeprint}[1]{\href{http://ascl.net/#1}{\nolinkurl{http://ascl.net/#1}}}
\providecommand{\doarXiv}[1]{\href{https://arxiv.org/abs/#1}{\nolinkurl{https://arxiv.org/abs/#1}}}

\bibitem[{{Abbott} {et~al.}(2017{\natexlab{a}}){Abbott}, {Abbott}, {Abbott},
  {Acernese}, {Ackley}, {Adams}, {Adams}, {Addesso}, {Adhikari}, {Adya},
  {Affeldt}, {Afrough}, {Agarwal}, {Agathos}, {Agatsuma}, {Aggarwal}, {Aguiar},
  {Aiello}, {Ain}, {Ajith}, {Allen}, {Allen}, {Allocca}, {Altin}, {Amato},
  {Ananyeva}, {Anderson}, {Anderson}, {Angelova}, {Antier}, {Appert}, {Arai},
  {Araya}, {Areeda}, {Arnaud}, {Arun}, {Ascenzi}, {Ashton}, {Ast}, {Aston},
  {Astone}, {Atallah}, {Aufmuth}, {Aulbert}, {AultONeal}, {Austin},
  {Avila-Alvarez}, {Babak}, {Bacon}, {Bader}, {Bae}, {Bailes}, {Baker},
  {Baldaccini}, {Ballardin}, {Ballmer}, {Banagiri}, {Barayoga}, {Barclay},
  {Barish}, {Barker}, {Barkett}, {Barone}, {Barr}, {Barsotti}, {Barsuglia},
  {Barta}, {Barthelmy}, {Bartlett}, {Bartos}, {Bassiri}, {Basti}, {Batch},
  {Bawaj}, {Bayley}, {Bazzan}, {B{\'e}csy}, {Beer}, {Bejger}, {Belahcene},
  {Bell}, {Berger}, {Bergmann}, {Bernuzzi}, {Bero}, {Berry}, {Bersanetti},
  {Bertolini}, {Betzwieser}, {Bhagwat}, {Bhandare}, {Bilenko}, {Billingsley},
  {Billman}, {Birch}, {Birney}, {Birnholtz}, {Biscans}, {Biscoveanu}, {Bisht},
  {Bitossi}, {Biwer}, {Bizouard}, {Blackburn}, {Blackman}, {Blair}, {Blair},
  {Blair}, {Bloemen}, {Bock}, {Bode}, {Boer}, {Bogaert}, {Bohe}, {Bondu},
  {Bonilla}, {Bonnand}, {Boom}, {Bork}, {Boschi}, {Bose}, {Bossie},
  {Bouffanais}, {Bozzi}, {Bradaschia}, {Brady}, {Branchesi}, {Brau}, {Briant},
  {Brillet}, {Brinkmann}, {Brisson}, {Brockill}, {Broida}, {Brooks}, {Brown},
  {Brown}, {Brunett}, {Buchanan}, {Buikema}, {Bulik}, {Bulten}, {Buonanno},
  {Buskulic}, {Buy}, {Byer}, {Cabero}, {Cadonati}, {Cagnoli}, {Cahillane},
  {Calder{\'o}n Bustillo}, {Callister}, {Calloni}, {Camp}, {Canepa},
  {Canizares}, {Cannon}, {Cao}, {Cao}, {Capano}, {Capocasa}, {Carbognani},
  {Caride}, {Carney}, {Carullo}, {Casanueva Diaz}, {Casentini}, {Caudill},
  {Cavagli{\`a}}, {Cavalier}, {Cavalieri}, {Cella}, {Cepeda},
  {Cerd{\'a}-Dur{\'a}n}, {Cerretani}, {Cesarini}, {Chamberlin}, {Chan}, {Chao},
  {Charlton}, {Chase}, {Chassande-Mottin}, {Chatterjee}, {Chatziioannou},
  {Cheeseboro}, {Chen}, {Chen}, {Chen}, {Cheng}, {Chia}, {Chincarini},
  {Chiummo}, {Chmiel}, {Cho}, {Cho}, {Chow}, {Christensen}, {Chu}, {Chua},
  {Chua}, {Chung}, {Chung}, {Ciani}, {Ciolfi}, {Cirelli}, {Cirone}, {Clara},
  {Clark}, {Clearwater}, {Cleva}, {Cocchieri}, {Coccia}, {Cohadon}, {Cohen},
  {Colla}, {Collette}, {Cominsky}, {Constancio}, {Conti}, {Cooper}, {Corban},
  {Corbitt}, {Cordero-Carri{\'o}n}, {Corley}, {Cornish}, {Corsi}, {Cortese},
  {Costa}, {Coughlin}, {Coughlin}, {Coulon}, {Countryman}, {Couvares}, {Covas},
  {Cowan}, {Coward}, {Cowart}, {Coyne}, {Coyne}, {Creighton}, {Creighton},
  {Cripe}, {Crowder}, {Cullen}, {Cumming}, {Cunningham}, {Cuoco}, {Dal Canton},
  {D{\'a}lya}, {Danilishin}, {D'Antonio}, {Danzmann}, {Dasgupta}, {Da Silva
  Costa}, {Dattilo}, {Dave}, {Davier}, {Davis}, {Daw}, {Day}, {De}, {DeBra},
  {Degallaix}, {De Laurentis}, {Del{\'e}glise}, {Del Pozzo}, {Demos}, {Denker},
  {Dent}, {De Pietri}, {Dergachev}, {De Rosa}, {DeRosa}, {De Rossi}, {DeSalvo},
  {de Varona}, {Devenson}, {Dhurandhar}, {D{\'\i}az}, {Dietrich}, {Di Fiore},
  {Di Giovanni}, {Di Girolamo}, {Di Lieto}, {Di Pace}, {Di Palma}, {Di Renzo},
  {Doctor}, {Dolique}, {Donovan}, {Dooley}, {Doravari}, {Dorrington},
  {Douglas}, {Dovale {\'A}lvarez}, {Downes}, {Drago}, {Dreissigacker},
  {Driggers}, {Du}, {Ducrot}, {Dudi}, {Dupej}, {Dwyer}, {Edo}, {Edwards},
  {Effler}, {Eggenstein}, {Ehrens}, {Eichholz}, {Eikenberry}, {Eisenstein},
  {Essick}, {Estevez}, {Etienne}, {Etzel}, {Evans}, {Evans}, {Factourovich},
  {Fafone}, {Fair}, {Fairhurst}, {Fan}, {Farinon}, {Farr}, {Farr},
  {Fauchon-Jones}, {Favata}, {Fays}, {Fee}, {Fehrmann}, {Feicht}, {Fejer},
  {Fernandez-Galiana}, {Ferrante}, {Ferreira}, {Ferrini}, {Fidecaro},
  {Finstad}, {Fiori}, {Fiorucci}, {Fishbach}, {Fisher}, {Fitz-Axen},
  {Flaminio}, {Fletcher}, {Fong}, {Font}, {Forsyth}, {Forsyth}, {Fournier},
  {Frasca}, {Frasconi}, {Frei}, {Freise}, {Frey}, {Frey}, {Fries}, {Fritschel},
  {Frolov}, {Fulda}, {Fyffe}, {Gabbard}, {Gadre}, {Gaebel}, {Gair},
  {Gammaitoni}, {Ganija}, {Gaonkar}, {Garcia-Quiros}, {Garufi}, {Gateley},
  {Gaudio}, {Gaur}, {Gayathri}, {Gehrels}, {Gemme}, {Genin}, {Gennai},
  {George}, {George}, {Gergely}, {Germain}, {Ghonge}, {Ghosh}, {Ghosh},
  {Ghosh}, {Giaime}, {Giardina}, {Giazotto}, {Gill}, {Glover}, {Goetz},
  {Goetz}, {Gomes}, {Goncharov}, {Gonz{\'a}lez}, {Gonzalez Castro},
  {Gopakumar}, {Gorodetsky}, {Gossan}, {Gosselin}, {Gouaty}, {Grado}, {Graef},
  {Granata}, {Grant}, {Gras}, {Gray}, {Greco}, {Green}, {Gretarsson}, {Groot},
  {Grote}, {Grunewald}, {Gruning}, {Guidi}, {Guo}, {Gupta}, {Gupta}, {Gushwa},
  {Gustafson}, {Gustafson}, {Halim}, {Hall}, {Hall}, {Hamilton}, {Hammond},
  {Haney}, {Hanke}, {Hanks}, {Hanna}, {Hannam}, {Hannuksela}, {Hanson},
  {Hardwick}, {Harms}, {Harry}, {Harry}, {Hart}, {Haster}, {Haughian}, {Healy},
  {Heidmann}, {Heintze}, {Heitmann}, {Hello}, {Hemming}, {Hendry}, {Heng},
  {Hennig}, {Heptonstall}, {Heurs}, {Hild}, {Hinderer}, {Ho}, {Hoak}, {Hofman},
  {Holt}, {Holz}, {Hopkins}, {Horst}, {Hough}, {Houston}, {Howell}, {Hreibi},
  {Hu}, {Huerta}, {Huet}, {Hughey}, {Husa}, {Huttner}, {Huynh-Dinh}, {Indik},
  {Inta}, {Intini}, {Isa}, {Isac}, {Isi}, {Iyer}, {Izumi}, {Jacqmin}, {Jani},
  {Jaranowski}, {Jawahar}, {Jim{\'e}nez-Forteza}, {Johnson},
  {Johnson-McDaniel}, {Jones}, {Jones}, {Jonker}, {Ju}, {Junker}, {Kalaghatgi},
  {Kalogera}, {Kamai}, {Kandhasamy}, {Kang}, {Kanner}, {Kapadia}, {Karki},
  {Karvinen}, {Kasprzack}, {Kastaun}, {Katolik}, {Katsavounidis}, {Katzman},
  {Kaufer}, {Kawabe}, {K{\'e}f{\'e}lian}, {Keitel}, {Kemball}, {Kennedy},
  {Kent}, {Key}, {Khalili}, {Khan}, {Khan}, {Khan}, {Khazanov}, {Kijbunchoo},
  {Kim}, {Kim}, {Kim}, {Kim}, {Kim}, {Kim}, {Kimbrell}, {King}, {King},
  {Kinley-Hanlon}, {Kirchhoff}, {Kissel}, {Kleybolte}, {Klimenko}, {Knowles},
  {Koch}, {Koehlenbeck}, {Koley}, {Kondrashov}, {Kontos}, {Korobko}, {Korth},
  {Kowalska}, {Kozak}, {Kr{\"a}mer}, {Kringel}, {Krishnan}, {Kr{\'o}lak},
  {Kuehn}, {Kumar}, {Kumar}, {Kumar}, {Kuo}, {Kutynia}, {Kwang}, {Lackey},
  {Lai}, {Landry}, {Lang}, {Lange}, {Lantz}, {Lanza}, {Larson},
  {Lartaux-Vollard}, {Lasky}, {Laxen}, {Lazzarini}, {Lazzaro}, {Leaci},
  {Leavey}, {Lee}, {Lee}, {Lee}, {Lee}, {Lee}, {Lehmann}, {Lenon}, {Leon},
  {Leonardi}, {Leroy}, {Letendre}, {Levin}, {Li}, {Linker}, {Littenberg},
  {Liu}, {Liu}, {Lo}, {Lockerbie}, {London}, {Lord}, {Lorenzini}, {Loriette},
  {Lormand}, {Losurdo}, {Lough}, {Lousto}, {Lovelace}, {L{\"u}ck}, {Lumaca},
  {Lundgren}, {Lynch}, {Ma}, {Macas}, {Macfoy}, {Machenschalk}, {MacInnis},
  {Macleod}, {Maga{\~n}a Hernandez}, {Maga{\~n}a-Sandoval}, {Maga{\~n}a
  Zertuche}, {Magee}, {Majorana}, {Maksimovic}, {Man}, {Mandic}, {Mangano},
  {Mansell}, {Manske}, {Mantovani}, {Marchesoni}, {Marion}, {M{\'a}rka},
  {M{\'a}rka}, {Markakis}, {Markosyan}, {Markowitz}, {Maros}, {Marquina},
  {Marsh}, {Martelli}, {Martellini}, {Martin}, {Martin}, {Martynov}, {Marx},
  {Mason}, {Massera}, {Masserot}, {Massinger}, {Masso-Reid}, {Mastrogiovanni},
  {Matas}, {Matichard}, {Matone}, {Mavalvala}, {Mazumder}, {McCarthy},
  {McClelland}, {McCormick}, {McCuller}, {McGuire}, {McIntyre}, {McIver},
  {McManus}, {McNeill}, {McRae}, {McWilliams}, {Meacher}, {Meadors}, {Mehmet},
  {Meidam}, {Mejuto-Villa}, {Melatos}, {Mendell}, {Mercer}, {Merilh},
  {Merzougui}, {Meshkov}, {Messenger}, {Messick}, {Metzdorff}, {Meyers},
  {Miao}, {Michel}, {Middleton}, {Mikhailov}, {Milano}, {Miller}, {Miller},
  {Miller}, {Millhouse}, {Milovich-Goff}, {Minazzoli}, {Minenkov}, {Ming},
  {Mishra}, {Mitra}, {Mitrofanov}, {Mitselmakher}, {Mittleman}, {Moffa},
  {Moggi}, {Mogushi}, {Mohan}, {Mohapatra}, {Molina}, {Montani}, {Moore},
  {Moraru}, {Moreno}, {Morisaki}, {Morriss}, {Mours}, {Mow-Lowry}, {Mueller},
  {Muir}, {Mukherjee}, {Mukherjee}, {Mukherjee}, {Mukund}, {Mullavey}, {Munch},
  {Mu{\~n}iz}, {Muratore}, {Murray}, {Nagar}, {Napier}, {Nardecchia},
  {Naticchioni}, {Nayak}, {Neilson}, {Nelemans}, {Nelson}, {Nery}, {Neunzert},
  {Nevin}, {Newport}, {Newton}, {Ng}, {Nguyen}, {Nguyen}, {Nichols}, {Nielsen},
  {Nissanke}, {Nitz}, {Noack}, {Nocera}, {Nolting}, {North}, {Nuttall},
  {Oberling}, {O'Dea}, {Ogin}, {Oh}, {Oh}, {Ohme}, {Okada}, {Oliver},
  {Oppermann}, {Oram}, {O'Reilly}, {Ormiston}, {Ortega}, {O'Shaughnessy},
  {Ossokine}, {Ottaway}, {Overmier}, {Owen}, {Pace}, {Page}, {Page}, {Pai},
  {Pai}, {Palamos}, {Palashov}, {Palomba}, {Pal-Singh}, {Pan}, {Pan}, {Pang},
  {Pang}, {Pankow}, {Pannarale}, {Pant}, {Paoletti}, {Paoli}, {Papa}, {Parida},
  {Parker}, {Pascucci}, {Pasqualetti}, {Passaquieti}, {Passuello}, {Patil},
  {Patricelli}, {Pearlstone}, {Pedraza}, {Pedurand}, {Pekowsky}, {Pele},
  {Penn}, {Perez}, {Perreca}, {Perri}, {Pfeiffer}, {Phelps}, {Piccinni},
  {Pichot}, {Piergiovanni}, {Pierro}, {Pillant}, {Pinard}, {Pinto}, {Pirello},
  {Pitkin}, {Poe}, {Poggiani}, {Popolizio}, {Porter}, {Post}, {Powell},
  {Prasad}, {Pratt}, {Pratten}, {Predoi}, {Prestegard}, {Prijatelj},
  {Principe}, {Privitera}, {Prix}, {Prodi}, {Prokhorov}, {Puncken}, {Punturo},
  {Puppo}, {P{\"u}rrer}, {Qi}, {Quetschke}, {Quintero}, {Quitzow-James},
  {Raab}, {Rabeling}, {Radkins}, {Raffai}, {Raja}, {Rajan}, {Rajbhandari},
  {Rakhmanov}, {Ramirez}, {Ramos-Buades}, {Rapagnani}, {Raymond}, {Razzano},
  {Read}, {Regimbau}, {Rei}, {Reid}, {Reitze}, {Ren}, {Reyes}, {Ricci},
  {Ricker}, {Rieger}, {Riles}, {Rizzo}, {Robertson}, {Robie}, {Robinet},
  {Rocchi}, {Rolland}, {Rollins}, {Roma}, {Romano}, {Romano}, {Romel}, {Romie},
  {Rosi{\'n}ska}, {Ross}, {Rowan}, {R{\"u}diger}, {Ruggi}, {Rutins}, {Ryan},
  {Sachdev}, {Sadecki}, {Sadeghian}, {Sakellariadou}, {Salconi}, {Saleem},
  {Salemi}, {Samajdar}, {Sammut}, {Sampson}, {Sanchez}, {Sanchez},
  {Sanchis-Gual}, {Sandberg}, {Sanders}, {Sassolas}, {Sathyaprakash},
  {Saulson}, {Sauter}, {Savage}, {Sawadsky}, {Schale}, {Scheel}, {Scheuer},
  {Schmidt}, {Schmidt}, {Schnabel}, {Schofield}, {Sch{\"o}nbeck}, {Schreiber},
  {Schuette}, {Schulte}, {Schutz}, {Schwalbe}, {Scott}, {Scott}, {Seidel},
  {Sellers}, {Sengupta}, {Sentenac}, {Sequino}, {Sergeev}, {Shaddock},
  {Shaffer}, {Shah}, {Shahriar}, {Shaner}, {Shao}, {Shapiro}, {Shawhan},
  {Sheperd}, {Shoemaker}, {Shoemaker}, {Siellez}, {Siemens}, {Sieniawska},
  {Sigg}, {Silva}, {Singer}, {Singh}, {Singhal}, {Sintes}, {Slagmolen},
  {Smith}, {Smith}, {Smith}, {Somala}, {Son}, {Sonnenberg}, {Sorazu},
  {Sorrentino}, {Souradeep}, {Spencer}, {Srivastava}, {Staats}, {Staley},
  {Steinke}, {Steinlechner}, {Steinlechner}, {Steinmeyer}, {Stevenson},
  {Stone}, {Stops}, {Strain}, {Stratta}, {Strigin}, {Strunk}, {Sturani},
  {Stuver}, {Summerscales}, {Sun}, {Sunil}, {Suresh}, {Sutton}, {Swinkels},
  {Szczepa{\'n}czyk}, {Tacca}, {Tait}, {Talbot}, {Talukder}, {Tanner},
  {T{\'a}pai}, {Taracchini}, {Tasson}, {Taylor}, {Taylor}, {Tewari}, {Theeg},
  {Thies}, {Thomas}, {Thomas}, {Thomas}, {Thorne}, {Thorne}, {Thrane},
  {Tiwari}, {Tiwari}, {Tokmakov}, {Toland}, {Tonelli}, {Tornasi},
  {Torres-Forn{\'e}}, {Torrie}, {T{\"o}yr{\"a}}, {Travasso}, {Traylor},
  {Trinastic}, {Tringali}, {Trozzo}, {Tsang}, {Tse}, {Tso}, {Tsukada}, {Tsuna},
  {Tuyenbayev}, {Ueno}, {Ugolini}, {Unnikrishnan}, {Urban}, {Usman},
  {Vahlbruch}, {Vajente}, {Valdes}, {Vallisneri}, {van Bakel}, {van Beuzekom},
  {van den Brand}, {Van Den Broeck}, {Vander-Hyde}, {van der Schaaf}, {van
  Heijningen}, {van Veggel}, {Vardaro}, {Varma}, {Vass}, {Vas{\'u}th},
  {Vecchio}, {Vedovato}, {Veitch}, {Veitch}, {Venkateswara}, {Venugopalan},
  {Verkindt}, {Vetrano}, {Vicer{\'e}}, {Viets}, {Vinciguerra}, {Vine}, {Vinet},
  {Vitale}, {Vo}, {Vocca}, {Vorvick}, {Vyatchanin}, {Wade}, {Wade}, {Wade},
  {Walet}, {Walker}, {Wallace}, {Walsh}, {Wang}, {Wang}, {Wang}, {Wang},
  {Wang}, {Ward}, {Warner}, {Was}, {Watchi}, {Weaver}, {Wei}, {Weinert},
  {Weinstein}, {Weiss}, {Wen}, {Wessel}, {We{\ss}els}, {Westerweck},
  {Westphal}, {Wette}, {Whelan}, {Whitcomb}, {Whiting}, {Whittle}, {Wilken},
  {Williams}, {Williams}, {Williamson}, {Willis}, {Willke}, {Wimmer},
  {Winkler}, {Wipf}, {Wittel}, {Woan}, {Woehler}, {Wofford}, {Wong}, {Worden},
  {Wright}, {Wu}, {Wysocki}, {Xiao}, {Yamamoto}, {Yancey}, {Yang}, {Yap},
  {Yazback}, {Yu}, {Yu}, {Yvert}, {Zadro{\.Z}ny}, {Zanolin}, {Zelenova},
  {Zendri}, {Zevin}, {Zhang}, {Zhang}, {Zhang}, {Zhang}, {Zhao}, {Zhou},
  {Zhou}, {Zhu}, {Zhu}, {Zimmerman}, {Zucker}, {Zweizig}, {LIGO Scientific
  Collaboration}, \& {Virgo Collaboration}}]{2017PhRvL.119p1101A}
{Abbott}, B.~P., {Abbott}, R., {Abbott}, T.~D., {et~al.} 2017{\natexlab{a}},
  \prl, 119, 161101, \dodoi{10.1103/PhysRevLett.119.161101}

\bibitem[{{Abbott} {et~al.}(2017{\natexlab{b}}){Abbott}, {Abbott}, {Abbott},
  {Acernese}, {Ackley}, {Adams}, {Adams}, {Addesso}, {Adhikari}, {Adya},
  {Affeldt}, {Afrough}, {Agarwal}, {Agathos}, {Agatsuma}, {Aggarwal}, {Aguiar},
  {Aiello}, {Ain}, {Ajith}, {Allen}, {Allen}, {Allocca}, {Altin}, {Amato},
  {Ananyeva}, {Anderson}, {Anderson}, {Angelova}, {Antier}, {Appert}, {Arai},
  {Araya}, {Areeda}, {Arnaud}, {Arun}, {Ascenzi}, {Ashton}, {Ast}, {Aston},
  {Astone}, {Atallah}, {Aufmuth}, {Aulbert}, {AultONeal}, {Austin},
  {Avila-Alvarez}, {Babak}, {Bacon}, {Bader}, {Bae}, {Baker}, {Baldaccini},
  {Ballardin}, {Ballmer}, {Banagiri}, {Barayoga}, {Barclay}, {Barish},
  {Barker}, {Barkett}, {Barone}, {Barr}, {Barsotti}, {Barsuglia}, {Barta},
  {Barthelmy}, {Bartlett}, {Bartos}, {Bassiri}, {Basti}, {Batch}, {Bawaj},
  {Bayley}, {Bazzan}, {B{\'e}csy}, {Beer}, {Bejger}, {Belahcene}, {Bell},
  {Berger}, {Bergmann}, {Bero}, {Berry}, {Bersanetti}, {Bertolini},
  {Betzwieser}, {Bhagwat}, {Bhandare}, {Bilenko}, {Billingsley}, {Billman},
  {Birch}, {Birney}, {Birnholtz}, {Biscans}, {Biscoveanu}, {Bisht}, {Bitossi},
  {Biwer}, {Bizouard}, {Blackburn}, {Blackman}, {Blair}, {Blair}, {Blair},
  {Bloemen}, {Bock}, {Bode}, {Boer}, {Bogaert}, {Bohe}, {Bondu}, {Bonilla},
  {Bonnand}, {Boom}, {Bork}, {Boschi}, {Bose}, {Bossie}, {Bouffanais}, {Bozzi},
  {Bradaschia}, {Brady}, {Branchesi}, {Brau}, {Briant}, {Brillet}, {Brinkmann},
  {Brisson}, {Brockill}, {Broida}, {Brooks}, {Brown}, {Brown}, {Brunett},
  {Buchanan}, {Buikema}, {Bulik}, {Bulten}, {Buonanno}, {Buskulic}, {Buy},
  {Byer}, {Cabero}, {Cadonati}, {Cagnoli}, {Cahillane}, {Calder{\'o}n
  Bustillo}, {Callister}, {Calloni}, {Camp}, {Canepa}, {Canizares}, {Cannon},
  {Cao}, {Cao}, {Capano}, {Capocasa}, {Carbognani}, {Caride}, {Carney},
  {Casanueva Diaz}, {Casentini}, {Caudill}, {Cavagli{\`a}}, {Cavalier},
  {Cavalieri}, {Cella}, {Cepeda}, {Cerd{\'a}-Dur{\'a}n}, {Cerretani},
  {Cesarini}, {Chamberlin}, {Chan}, {Chao}, {Charlton}, {Chase},
  {Chassande-Mottin}, {Chatterjee}, {Chatziioannou}, {Cheeseboro}, {Chen},
  {Chen}, {Chen}, {Cheng}, {Chia}, {Chincarini}, {Chiummo}, {Chmiel}, {Cho},
  {Cho}, {Chow}, {Christensen}, {Chu}, {Chua}, {Chua}, {Chung}, {Chung},
  {Ciani}, {Ciolfi}, {Cirelli}, {Cirone}, {Clara}, {Clark}, {Clearwater},
  {Cleva}, {Cocchieri}, {Coccia}, {Cohadon}, {Cohen}, {Colla}, {Collette},
  {Cominsky}, {Constancio}, {Conti}, {Cooper}, {Corban}, {Corbitt},
  {Cordero-Carri{\'o}n}, {Corley}, {Cornish}, {Corsi}, {Cortese}, {Costa},
  {Coughlin}, {Coughlin}, {Coulon}, {Countryman}, {Couvares}, {Covas}, {Cowan},
  {Coward}, {Cowart}, {Coyne}, {Coyne}, {Creighton}, {Creighton}, {Cripe},
  {Crowder}, {Cullen}, {Cumming}, {Cunningham}, {Cuoco}, {Dal Canton},
  {D{\'a}lya}, {Danilishin}, {D'Antonio}, {Danzmann}, {Dasgupta}, {Da Silva
  Costa}, {Dattilo}, {Dave}, {Davier}, {Davis}, {Daw}, {Day}, {De}, {DeBra},
  {Degallaix}, {De Laurentis}, {Del{\'e}glise}, {Del Pozzo}, {Demos}, {Denker},
  {Dent}, {De Pietri}, {Dergachev}, {De Rosa}, {DeRosa}, {De Rossi}, {DeSalvo},
  {de Varona}, {Devenson}, {Dhurandhar}, {D{\'\i}az}, {Di Fiore}, {Di
  Giovanni}, {Di Girolamo}, {Di Lieto}, {Di Pace}, {Di Palma}, {Di Renzo},
  {Doctor}, {Dolique}, {Donovan}, {Dooley}, {Doravari}, {Dorrington},
  {Douglas}, {Dovale {\'A}lvarez}, {Downes}, {Drago}, {Dreissigacker},
  {Driggers}, {Du}, {Ducrot}, {Dupej}, {Dwyer}, {Edo}, {Edwards}, {Effler},
  {Ehrens}, {Eichholz}, {Eikenberry}, {Eisenstein}, {Essick}, {Estevez},
  {Etienne}, {Etzel}, {Evans}, {Evans}, {Factourovich}, {Fafone}, {Fair},
  {Fairhurst}, {Fan}, {Farinon}, {Farr}, {Farr}, {Fauchon-Jones}, {Favata},
  {Fays}, {Fee}, {Fehrmann}, {Feicht}, {Fejer}, {Fernandez-Galiana},
  {Ferrante}, {Ferreira}, {Ferrini}, {Fidecaro}, {Finstad}, {Fiori},
  {Fiorucci}, {Fishbach}, {Fisher}, {Fitz-Axen}, {Flaminio}, {Fletcher},
  {Fong}, {Font}, {Forsyth}, {Forsyth}, {Fournier}, {Frasca}, {Frasconi},
  {Frei}, {Freise}, {Frey}, {Frey}, {Fries}, {Fritschel}, {Frolov}, {Fulda},
  {Fyffe}, {Gabbard}, {Gadre}, {Gaebel}, {Gair}, {Gammaitoni}, {Ganija},
  {Gaonkar}, {Garcia-Quiros}, {Garufi}, {Gateley}, {Gaudio}, {Gaur},
  {Gayathri}, {Gehrels}, {Gemme}, {Genin}, {Gennai}, {George}, {George},
  {Gergely}, {Germain}, {Ghonge}, {Ghosh}, {Ghosh}, {Ghosh}, {Giaime},
  {Giardina}, {Giazotto}, {Gill}, {Glover}, {Goetz}, {Goetz}, {Gomes},
  {Goncharov}, {Gonz{\'a}lez}, {Gonzalez Castro}, {Gopakumar}, {Gorodetsky},
  {Gossan}, {Gosselin}, {Gouaty}, {Grado}, {Graef}, {Granata}, {Grant}, {Gras},
  {Gray}, {Greco}, {Green}, {Gretarsson}, {Griswold}, {Groot}, {Grote},
  {Grunewald}, {Gruning}, {Guidi}, {Guo}, {Gupta}, {Gupta}, {Gushwa},
  {Gustafson}, {Gustafson}, {Halim}, {Hall}, {Hall}, {Hamilton}, {Hammond},
  {Haney}, {Hanke}, {Hanks}, {Hanna}, {Hannam}, {Hannuksela}, {Hanson},
  {Hardwick}, {Harms}, {Harry}, {Harry}, {Hart}, {Haster}, {Haughian}, {Healy},
  {Heidmann}, {Heintze}, {Heitmann}, {Hello}, {Hemming}, {Hendry}, {Heng},
  {Hennig}, {Heptonstall}, {Heurs}, {Hild}, {Hinderer}, {Hoak}, {Hofman},
  {Holt}, {Holz}, {Hopkins}, {Horst}, {Hough}, {Houston}, {Howell}, {Hreibi},
  {Hu}, {Huerta}, {Huet}, {Hughey}, {Husa}, {Huttner}, {Huynh-Dinh}, {Indik},
  {Inta}, {Intini}, {Isa}, {Isac}, {Isi}, {Iyer}, {Izumi}, {Jacqmin}, {Jani},
  {Jaranowski}, {Jawahar}, {Jim{\'e}nez-Forteza}, {Johnson}, {Jones}, {Jones},
  {Jonker}, {Ju}, {Junker}, {Kalaghatgi}, {Kalogera}, {Kamai}, {Kandhasamy},
  {Kang}, {Kanner}, {Kapadia}, {Karki}, {Karvinen}, {Kasprzack}, {Katolik},
  {Katsavounidis}, {Katzman}, {Kaufer}, {Kawabe}, {K{\'e}f{\'e}lian}, {Keitel},
  {Kemball}, {Kennedy}, {Kent}, {Key}, {Khalili}, {Khan}, {Khan}, {Khan},
  {Khazanov}, {Kijbunchoo}, {Kim}, {Kim}, {Kim}, {Kim}, {Kim}, {Kim},
  {Kimbrell}, {King}, {King}, {Kinley-Hanlon}, {Kirchhoff}, {Kissel},
  {Kleybolte}, {Klimenko}, {Knowles}, {Koch}, {Koehlenbeck}, {Koley},
  {Kondrashov}, {Kontos}, {Korobko}, {Korth}, {Kowalska}, {Kozak},
  {Kr{\"a}mer}, {Kringel}, {Krishnan}, {Kr{\'o}lak}, {Kuehn}, {Kumar}, {Kumar},
  {Kumar}, {Kuo}, {Kutynia}, {Kwang}, {Lackey}, {Lai}, {Landry}, {Lang},
  {Lange}, {Lantz}, {Lanza}, {Larson}, {Lartaux-Vollard}, {Lasky}, {Laxen},
  {Lazzarini}, {Lazzaro}, {Leaci}, {Leavey}, {Lee}, {Lee}, {Lee}, {Lee}, {Lee},
  {Lehmann}, {Lenon}, {Leonardi}, {Leroy}, {Letendre}, {Levin}, {Li}, {Linker},
  {Littenberg}, {Liu}, {Lo}, {Lockerbie}, {London}, {Lord}, {Lorenzini},
  {Loriette}, {Lormand}, {Losurdo}, {Lough}, {Lousto}, {Lovelace}, {L{\"u}ck},
  {Lumaca}, {Lundgren}, {Lynch}, {Ma}, {Macas}, {Macfoy}, {Machenschalk},
  {MacInnis}, {Macleod}, {Maga{\~n}a Hernandez}, {Maga{\~n}a-Sandoval},
  {Maga{\~n}a Zertuche}, {Magee}, {Majorana}, {Maksimovic}, {Man}, {Mandic},
  {Mangano}, {Mansell}, {Manske}, {Mantovani}, {Marchesoni}, {Marion},
  {M{\'a}rka}, {M{\'a}rka}, {Markakis}, {Markosyan}, {Markowitz}, {Maros},
  {Marquina}, {Marsh}, {Martelli}, {Martellini}, {Martin}, {Martin},
  {Martynov}, {Mason}, {Massera}, {Masserot}, {Massinger}, {Masso-Reid},
  {Mastrogiovanni}, {Matas}, {Matichard}, {Matone}, {Mavalvala}, {Mazumder},
  {McCarthy}, {McClelland}, {McCormick}, {McCuller}, {McGuire}, {McIntyre},
  {McIver}, {McManus}, {McNeill}, {McRae}, {McWilliams}, {Meacher}, {Meadors},
  {Mehmet}, {Meidam}, {Mejuto-Villa}, {Melatos}, {Mendell}, {Mercer}, {Merilh},
  {Merzougui}, {Meshkov}, {Messenger}, {Messick}, {Metzdorff}, {Meyers},
  {Miao}, {Michel}, {Middleton}, {Mikhailov}, {Milano}, {Miller}, {Miller},
  {Miller}, {Millhouse}, {Milovich-Goff}, {Minazzoli}, {Minenkov}, {Ming},
  {Mishra}, {Mitra}, {Mitrofanov}, {Mitselmakher}, {Mittleman}, {Moffa},
  {Moggi}, {Mogushi}, {Mohan}, {Mohapatra}, {Montani}, {Moore}, {Moraru},
  {Moreno}, {Morriss}, {Mours}, {Mow-Lowry}, {Mueller}, {Muir}, {Mukherjee},
  {Mukherjee}, {Mukherjee}, {Mukund}, {Mullavey}, {Munch}, {Mu{\~n}iz},
  {Muratore}, {Murray}, {Napier}, {Nardecchia}, {Naticchioni}, {Nayak},
  {Neilson}, {Nelemans}, {Nelson}, {Nery}, {Neunzert}, {Nevin}, {Newport},
  {Newton}, {Ng}, {Nguyen}, {Nguyen}, {Nichols}, {Nielsen}, {Nissanke}, {Nitz},
  {Noack}, {Nocera}, {Nolting}, {North}, {Nuttall}, {Oberling}, {O'Dea},
  {Ogin}, {Oh}, {Oh}, {Ohme}, {Okada}, {Oliver}, {Oppermann}, {Oram},
  {O'Reilly}, {Ormiston}, {Ortega}, {O'Shaughnessy}, {Ossokine}, {Ottaway},
  {Overmier}, {Owen}, {Pace}, {Page}, {Page}, {Pai}, {Pai}, {Palamos},
  {Palashov}, {Palomba}, {Pal-Singh}, {Pan}, {Pan}, {Pang}, {Pang}, {Pankow},
  {Pannarale}, {Pant}, {Paoletti}, {Paoli}, {Papa}, {Parida}, {Parker},
  {Pascucci}, {Pasqualetti}, {Passaquieti}, {Passuello}, {Patil}, {Patricelli},
  {Pearlstone}, {Pedraza}, {Pedurand}, {Pekowsky}, {Pele}, {Penn}, {Perez},
  {Perreca}, {Perri}, {Pfeiffer}, {Phelps}, {Piccinni}, {Pichot},
  {Piergiovanni}, {Pierro}, {Pillant}, {Pinard}, {Pinto}, {Pirello}, {Pitkin},
  {Poe}, {Poggiani}, {Popolizio}, {Porter}, {Post}, {Powell}, {Prasad},
  {Pratt}, {Pratten}, {Predoi}, {Prestegard}, {Price}, {Prijatelj}, {Principe},
  {Privitera}, {Prodi}, {Prokhorov}, {Puncken}, {Punturo}, {Puppo},
  {P{\"u}rrer}, {Qi}, {Quetschke}, {Quintero}, {Quitzow-James}, {Raab},
  {Rabeling}, {Radkins}, {Raffai}, {Raja}, {Rajan}, {Rajbhandari}, {Rakhmanov},
  {Ramirez}, {Ramos-Buades}, {Rapagnani}, {Raymond}, {Razzano}, {Read},
  {Regimbau}, {Rei}, {Reid}, {Reitze}, {Ren}, {Reyes}, {Ricci}, {Ricker},
  {Rieger}, {Riles}, {Rizzo}, {Robertson}, {Robie}, {Robinet}, {Rocchi},
  {Rolland}, {Rollins}, {Roma}, {Romano}, {Romel}, {Romie}, {Rosi{\'n}ska},
  {Ross}, {Rowan}, {R{\"u}diger}, {Ruggi}, {Rutins}, {Ryan}, {Sachdev},
  {Sadecki}, {Sadeghian}, {Sakellariadou}, {Salconi}, {Saleem}, {Salemi},
  {Samajdar}, {Sammut}, {Sampson}, {Sanchez}, {Sanchez}, {Sanchis-Gual},
  {Sandberg}, {Sanders}, {Sassolas}, {Sathyaprakash}, {Saulson}, {Sauter},
  {Savage}, {Sawadsky}, {Schale}, {Scheel}, {Scheuer}, {Schmidt}, {Schmidt},
  {Schnabel}, {Schofield}, {Sch{\"o}nbeck}, {Schreiber}, {Schuette}, {Schulte},
  {Schutz}, {Schwalbe}, {Scott}, {Scott}, {Seidel}, {Sellers}, {Sengupta},
  {Sentenac}, {Sequino}, {Sergeev}, {Shaddock}, {Shaffer}, {Shah}, {Shahriar},
  {Shaner}, {Shao}, {Shapiro}, {Shawhan}, {Sheperd}, {Shoemaker}, {Shoemaker},
  {Siellez}, {Siemens}, {Sieniawska}, {Sigg}, {Silva}, {Singer}, {Singh},
  {Singhal}, {Sintes}, {Slagmolen}, {Smith}, {Smith}, {Smith}, {Somala}, {Son},
  {Sonnenberg}, {Sorazu}, {Sorrentino}, {Souradeep}, {Spencer}, {Srivastava},
  {Staats}, {Staley}, {Steinke}, {Steinlechner}, {Steinlechner}, {Steinmeyer},
  {Stevenson}, {Stone}, {Stops}, {Strain}, {Stratta}, {Strigin}, {Strunk},
  {Sturani}, {Stuver}, {Summerscales}, {Sun}, {Sunil}, {Suresh}, {Sutton},
  {Swinkels}, {Szczepa{\'n}czyk}, {Tacca}, {Tait}, {Talbot}, {Talukder},
  {Tanner}, {T{\'a}pai}, {Taracchini}, {Tasson}, {Taylor}, {Taylor}, {Tewari},
  {Theeg}, {Thies}, {Thomas}, {Thomas}, {Thomas}, {Thorne}, {Thorne}, {Thrane},
  {Tiwari}, {Tiwari}, {Tokmakov}, {Toland}, {Tonelli}, {Tornasi},
  {Torres-Forn{\'e}}, {Torrie}, {T{\"o}yr{\"a}}, {Travasso}, {Traylor},
  {Trinastic}, {Tringali}, {Trozzo}, {Tsang}, {Tse}, {Tso}, {Tsukada}, {Tsuna},
  {Tuyenbayev}, {Ueno}, {Ugolini}, {Unnikrishnan}, {Urban}, {Usman},
  {Vahlbruch}, {Vajente}, {Valdes}, {van Bakel}, {van Beuzekom}, {van den
  Brand}, {Van Den Broeck}, {Vander-Hyde}, {van der Schaaf}, {van Heijningen},
  {van Veggel}, {Vardaro}, {Varma}, {Vass}, {Vas{\'u}th}, {Vecchio},
  {Vedovato}, {Veitch}, {Veitch}, {Venkateswara}, {Venugopalan}, {Verkindt},
  {Vetrano}, {Vicer{\'e}}, {Viets}, {Vinciguerra}, {Vine}, {Vinet}, {Vitale},
  {Vo}, {Vocca}, {Vorvick}, {Vyatchanin}, {Wade}, {Wade}, {Wade}, {Walet},
  {Walker}, {Wallace}, {Walsh}, {Wang}, {Wang}, {Wang}, {Wang}, {Wang}, {Ward},
  {Warner}, {Was}, {Watchi}, {Weaver}, {Wei}, {Weinert}, {Weinstein}, {Weiss},
  {Wen}, {Wessel}, {Wessels}, {Westerweck}, {Westphal}, {Wette}, {Whelan},
  {Whitcomb}, {Whiting}, {Whittle}, {Wilken}, {Williams}, {Williams},
  {Williamson}, {Willis}, {Willke}, {Wimmer}, {Winkler}, {Wipf}, {Wittel},
  {Woan}, {Woehler}, {Wofford}, {Wong}, {Worden}, {Wright}, {Wu}, {Wysocki},
  {Xiao}, {Yamamoto}, {Yancey}, {Yang}, {Yap}, {Yazback}, {Yu}, {Yu}, {Yvert},
  {Zadro{\.z}ny}, {Zanolin}, {Zelenova}, {Zendri}, {Zevin}, {Zhang}, {Zhang},
  {Zhang}, {Zhang}, {Zhao}, {Zhou}, {Zhou}, {Zhu}, {Zhu}, {Zimmerman},
  {Zucker}, {Zweizig}, {LIGO Scientific Collaboration}, {Virgo Collaboration},
  {Wilson-Hodge}, {Bissaldi}, {Blackburn}, {Briggs}, {Burns}, {Cleveland},
  {Connaughton}, {Gibby}, {Giles}, {Goldstein}, {Hamburg}, {Jenke}, {Hui},
  {Kippen}, {Kocevski}, {McBreen}, {Meegan}, {Paciesas}, {Poolakkil}, {Preece},
  {Racusin}, {Roberts}, {Stanbro}, {Veres}, {von Kienlin}, {GBM}, {Savchenko},
  {Ferrigno}, {Kuulkers}, {Bazzano}, {Bozzo}, {Brandt}, {Chenevez},
  {Courvoisier}, {Diehl}, {Domingo}, {Hanlon}, {Jourdain}, {Laurent}, {Lebrun},
  {Lutovinov}, {Martin-Carrillo}, {Mereghetti}, {Natalucci}, {Rodi}, {Roques},
  {Sunyaev}, {Ubertini}, {INTEGRAL}, {Aartsen}, {Ackermann}, {Adams},
  {Aguilar}, {Ahlers}, {Ahrens}, {Samarai}, {Altmann}, {Andeen}, {Anderson},
  {Ansseau}, {Anton}, {Arg{\"u}elles}, {Auffenberg}, {Axani}, {Bagherpour},
  {Bai}, {Barron}, {Barwick}, {Baum}, {Bay}, {Beatty}, {Becker Tjus},
  {Bernardini}, {Besson}, {Binder}, {Bindig}, {Blaufuss}, {Blot}, {Bohm},
  {B{\"o}rner}, {Bos}, {Bose}, {B{\"o}ser}, {Botner}, {Bourbeau}, {Bourbeau},
  {Bradascio}, {Braun}, {Brayeur}, {Brenzke}, {Bretz}, {Bron},
  {Brostean-Kaiser}, {Burgman}, {Carver}, {Casey}, {Casier}, {Cheung},
  {Chirkin}, {Christov}, {Clark}, {Classen}, {Coenders}, {Collin}, {Conrad},
  {Cowen}, {Cross}, {Day}, {de Andr{\'e}}, {De Clercq}, {DeLaunay},
  {Dembinski}, {De Ridder}, {Desiati}, {de Vries}, {de Wasseige}, {de With},
  {DeYoung}, {D{\'\i}az-V{\'e}lez}, {di Lorenzo}, {Dujmovic}, {Dumm},
  {Dunkman}, {Dvorak}, {Eberhardt}, {Ehrhardt}, {Eichmann}, {Eller}, {Evenson},
  {Fahey}, {Fazely}, {Felde}, {Filimonov}, {Finley}, {Flis}, {Franckowiak},
  {Friedman}, {Fuchs}, {Gaisser}, {Gallagher}, {Gerhardt}, {Ghorbani}, {Giang},
  {Glauch}, {Gl{\"u}senkamp}, {Goldschmidt}, {Gonzalez}, {Grant}, {Griffith},
  {Haack}, {Hallgren}, {Halzen}, {Hanson}, {Hebecker}, {Heereman}, {Helbing},
  {Hellauer}, {Hickford}, {Hignight}, {Hill}, {Hoffman}, {Hoffmann},
  {Hokanson-Fasig}, {Hoshina}, {Huang}, {Huber}, {Hultqvist}, {H{\"u}nnefeld},
  {In}, {Ishihara}, {Jacobi}, {Japaridze}, {Jeong}, {Jero}, {Jones},
  {Kalaczynski}, {Kang}, {Kappes}, {Karg}, {Karle}, {Kauer}, {Keivani},
  {Kelley}, {Kheirandish}, {Kim}, {Kim}, {Kintscher}, {Kiryluk}, {Kittler},
  {Klein}, {Kohnen}, {Koirala}, {Kolanoski}, {K{\"o}pke}, {Kopper}, {Kopper},
  {Koschinsky}, {Koskinen}, {Kowalski}, {Krings}, {Kroll}, {Kr{\"u}ckl},
  {Kunnen}, {Kunwar}, {Kurahashi}, {Kuwabara}, {Kyriacou}, {Labare},
  {Lanfranchi}, {Larson}, {Lauber}, {Lesiak-Bzdak}, {Leuermann}, {Liu}, {Lu},
  {L{\"u}nemann}, {Luszczak}, {Madsen}, {Maggi}, {Mahn}, {Mancina}, {Maruyama},
  {Mase}, {Maunu}, {McNally}, {Meagher}, {Medici}, {Meier}, {Menne}, {Merino},
  {Meures}, {Miarecki}, {Micallef}, {Moment{\'e}}, {Montaruli}, {Moore},
  {Moulai}, {Nahnhauer}, {Nakarmi}, {Naumann}, {Neer}, {Niederhausen},
  {Nowicki}, {Nygren}, {Obertacke Pollmann}, {Olivas}, {O'Murchadha},
  {Palczewski}, {Pandya}, {Pankova}, {Peiffer}, {Pepper}, {P{\'e}rez de los
  Heros}, {Pieloth}, {Pinat}, {Price}, {Przybylski}, {Raab}, {R{\"a}del},
  {Rameez}, {Rawlins}, {Rea}, {Reimann}, {Relethford}, {Relich}, {Resconi},
  {Rhode}, {Richman}, {Robertson}, {Rongen}, {Rott}, {Ruhe}, {Ryckbosch},
  {Rysewyk}, {S{\"a}lzer}, {Sanchez Herrera}, {Sandrock}, {Sandroos},
  {Santander}, {Sarkar}, {Sarkar}, {Satalecka}, {Schlunder}, {Schmidt},
  {Schneider}, {Schoenen}, {Sch{\"o}neberg}, {Schumacher}, {Seckel},
  {Seunarine}, {Soedingrekso}, {Soldin}, {Song}, {Spiczak}, {Spiering},
  {Stachurska}, {Stamatikos}, {Stanev}, {Stasik}, {Stettner}, {Steuer},
  {Stezelberger}, {Stokstad}, {St{\"o}ssl}, {Strotjohann}, {Stuttard},
  {Sullivan}, {Sutherland}, {Taboada}, {Tatar}, {Tenholt}, {Ter-Antonyan},
  {Terliuk}, {Te{\v{s}}i{\'c}}, {Tilav}, {Toale}, {Tobin}, {Toscano}, {Tosi},
  {Tselengidou}, {Tung}, {Turcati}, {Turley}, {Ty}, {Unger}, {Usner},
  {Vandenbroucke}, {Van Driessche}, {van Eijndhoven}, {Vanheule}, {van Santen},
  {Vehring}, {Vogel}, {Vraeghe}, {Walck}, {Wallace}, {Wallraff}, {Wandler},
  {Wandkowsky}, {Waza}, {Weaver}, {Weiss}, {Wendt}, {Werthebach}, {Whelan},
  {Wiebe}, {Wiebusch}, {Wille}, {Williams}, {Wills}, {Wolf}, {Wood}, {Woolsey},
  {Woschnagg}, {Xu}, {Xu}, {Xu}, {Yanez}, {Yodh}, {Yoshida}, {Yuan}, {Zoll},
  {IceCube Collaboration}, {Balasubramanian}, {Mate}, {Bhalerao},
  {Bhattacharya}, {Vibhute}, {Dewangan}, {Rao}, {Vadawale}, {AstroSat Cadmium
  Zinc Telluride Imager Team}, {Svinkin}, {Hurley}, {Aptekar}, {Frederiks},
  {Golenetskii}, {Kozlova}, {Lysenko}, {Oleynik}, {Tsvetkova}, {Ulanov},
  {Cline}, {IPN Collaboration}, {Li}, {Xiong}, {Zhang}, {Lu}, {Song}, {Cao},
  {Chang}, {Chen}, {Chen}, {Chen}, {Chen}, {Chen}, {Chen}, {Cui}, {Cui},
  {Deng}, {Dong}, {Du}, {Fu}, {Gao}, {Gao}, {Gao}, {Ge}, {Gu}, {Guan}, {Guo},
  {Han}, {Hu}, {Huang}, {Huo}, {Jia}, {Jiang}, {Jiang}, {Jin}, {Jin}, {Li},
  {Li}, {Li}, {Li}, {Li}, {Li}, {Li}, {Li}, {Li}, {Li}, {Li}, {Liang}, {Liao},
  {Liu}, {Liu}, {Liu}, {Liu}, {Liu}, {Liu}, {Liu}, {Lu}, {Lu}, {Luo}, {Ma},
  {Meng}, {Nang}, {Nie}, {Ou}, {Qu}, {Sai}, {Sun}, {Tan}, {Tao}, {Tao}, {Tuo},
  {Wang}, {Wang}, {Wang}, {Wang}, {Wang}, {Wen}, {Wu}, {Wu}, {Xiao}, {Xu},
  {Xu}, {Yan}, {Yang}, {Yang}, {Yang}, {Zhang}, {Zhang}, {Zhang}, {Zhang},
  {Zhang}, {Zhang}, {Zhang}, {Zhang}, {Zhang}, {Zhang}, {Zhang}, {Zhang},
  {Zhang}, {Zhang}, {Zhang}, {Zhang}, {Zhang}, {Zhang}, {Zhao}, {Zhao}, {Zhao},
  {Zheng}, {Zhu}, {Zhu}, {Zou}, {Insight-HXMT Collaboration}, {Albert},
  {Andr{\'e}}, {Anghinolfi}, {Ardid}, {Aubert}, {Aublin}, {Avgitas}, {Baret},
  {Barrios-Mart{\'\i}}, {Basa}, {Belhorma}, {Bertin}, {Biagi}, {Bormuth},
  {Bourret}, {Bouwhuis}, {Br{\^a}nza{\c{s}}}, {Bruijn}, {Brunner}, {Busto},
  {Capone}, {Caramete}, {Carr}, {Celli}, {Cherkaoui El Moursli}, {Chiarusi},
  {Circella}, {Coelho}, {Coleiro}, {Coniglione}, {Costantini}, {Coyle},
  {Creusot}, {D{\'\i}az}, {Deschamps}, {De Bonis}, {Distefano}, {Di Palma},
  {Domi}, {Donzaud}, {Dornic}, {Drouhin}, {Eberl}, {El Bojaddaini}, {El
  Khayati}, {Els{\"a}sser}, {Enzenh{\"o}fer}, {Ettahiri}, {Fassi}, {Felis},
  {Fusco}, {Gay}, {Giordano}, {Glotin}, {Gr{\'e}goire}, {Ruiz}, {Graf},
  {Hallmann}, {van Haren}, {Heijboer}, {Hello}, {Hern{\'a}ndez-Rey},
  {H{\"o}ssl}, {Hofest{\"a}dt}, {Hugon}, {Illuminati}, {James}, {de Jong},
  {Jongen}, {Kadler}, {Kalekin}, {Katz}, {Kiessling}, {Kouchner}, {Kreter},
  {Kreykenbohm}, {Kulikovskiy}, {Lachaud}, {Lahmann}, {Lef{\`e}vre}, {Leonora},
  {Lotze}, {Loucatos}, {Marcelin}, {Margiotta}, {Marinelli},
  {Mart{\'\i}nez-Mora}, {Mele}, {Melis}, {Michael}, {Migliozzi}, {Moussa},
  {Navas}, {Nezri}, {Organokov}, {P{\u{a}}v{\u{a}}la{\c{s}}}, {Pellegrino},
  {Perrina}, {Piattelli}, {Popa}, {Pradier}, {Quinn}, {Racca}, {Riccobene},
  {S{\'a}nchez-Losa}, {Salda{\~n}a}, {Salvadori}, {Samtleben}, {Sanguineti},
  {Sapienza}, {Sieger}, {Spurio}, {Stolarczyk}, {Taiuti}, {Tayalati},
  {Trovato}, {Turpin}, {T{\"o}nnis}, {Vallage}, {Van Elewyck}, {Versari},
  {Vivolo}, {Vizzoca}, {Wilms}, {Zornoza}, {Z{\'u}{\~n}iga}, {ANTARES
  Collaboration}, {Beardmore}, {Breeveld}, {Burrows}, {Cenko}, {Cusumano},
  {D'A{\`\i}}, {de Pasquale}, {Emery}, {Evans}, {Giommi}, {Gronwall}, {Kennea},
  {Krimm}, {Kuin}, {Lien}, {Marshall}, {Melandri}, {Nousek}, {Oates},
  {Osborne}, {Pagani}, {Page}, {Palmer}, {Perri}, {Siegel}, {Sbarufatti},
  {Tagliaferri}, {Tohuvavohu}, {Swift Collaboration}, {Tavani}, {Verrecchia},
  {Bulgarelli}, {Evangelista}, {Pacciani}, {Feroci}, {Pittori}, {Giuliani},
  {Del Monte}, {Donnarumma}, {Argan}, {Trois}, {Ursi}, {Cardillo}, {Piano},
  {Longo}, {Lucarelli}, {Munar-Adrover}, {Fuschino}, {Labanti}, {Marisaldi},
  {Minervini}, {Fioretti}, {Parmiggiani}, {Gianotti}, {Trifoglio}, {Di Persio},
  {Antonelli}, {Barbiellini}, {Caraveo}, {Cattaneo}, {Costa}, {Colafrancesco},
  {D'Amico}, {Ferrari}, {Morselli}, {Paoletti}, {Picozza}, {Pilia}, {Rappoldi},
  {Soffitta}, {Vercellone}, {AGILE Team}, {Foley}, {Coulter}, {Kilpatrick},
  {Drout}, {Piro}, {Shappee}, {Siebert}, {Simon}, {Ulloa}, {Kasen}, {Madore},
  {Murguia-Berthier}, {Pan}, {Prochaska}, {Ramirez-Ruiz}, {Rest},
  {Rojas-Bravo}, {1M2H Team}, {Berger}, {Soares-Santos}, {Annis}, {Alexander},
  {Allam}, {Balbinot}, {Blanchard}, {Brout}, {Butler}, {Chornock}, {Cook},
  {Cowperthwaite}, {Diehl}, {Drlica-Wagner}, {Drout}, {Durret}, {Eftekhari},
  {Finley}, {Fong}, {Frieman}, {Fryer}, {Garc{\'\i}a-Bellido}, {Gruendl},
  {Hartley}, {Herner}, {Kessler}, {Lin}, {Lopes}, {Louren{\c{c}}o}, {Margutti},
  {Marshall}, {Matheson}, {Medina}, {Metzger}, {Mu{\~n}oz}, {Muir}, {Nicholl},
  {Nugent}, {Palmese}, {Paz-Chinch{\'o}n}, {Quataert}, {Sako}, {Sauseda},
  {Schlegel}, {Scolnic}, {Secco}, {Smith}, {Sobreira}, {Villar}, {Vivas},
  {Wester}, {Williams}, {Yanny}, {Zenteno}, {Zhang}, {Abbott}, {Banerji},
  {Bechtol}, {Benoit-L{\'e}vy}, {Bertin}, {Brooks}, {Buckley-Geer}, {Burke},
  {Capozzi}, {Carnero Rosell}, {Carrasco Kind}, {Castander}, {Crocce}, {Cunha},
  {D'Andrea}, {da Costa}, {Davis}, {DePoy}, {Desai}, {Dietrich}, {Eifler},
  {Fernandez}, {Flaugher}, {Fosalba}, {Gaztanaga}, {Gerdes}, {Giannantonio},
  {Goldstein}, {Gruen}, {Gschwend}, {Gutierrez}, {Honscheid}, {James},
  {Jeltema}, {Johnson}, {Johnson}, {Kent}, {Krause}, {Kron}, {Kuehn}, {Lahav},
  {Lima}, {Maia}, {March}, {Martini}, {McMahon}, {Menanteau}, {Miller},
  {Miquel}, {Mohr}, {Nichol}, {Ogando}, {Plazas}, {Romer}, {Roodman}, {Rykoff},
  {Sanchez}, {Scarpine}, {Schindler}, {Schubnell}, {Sevilla-Noarbe}, {Sheldon},
  {Smith}, {Smith}, {Stebbins}, {Suchyta}, {Swanson}, {Tarle}, {Thomas},
  {Troxel}, {Tucker}, {Vikram}, {Walker}, {Wechsler}, {Weller}, {Carlin},
  {Gill}, {Li}, {Marriner}, {Neilsen}, {Dark Energy Camera GW-EM
  Collaboration}, {DES Collaboration}, {Haislip}, {Kouprianov}, {Reichart},
  {Sand}, {Tartaglia}, {Valenti}, {Yang}, {DLT40 Collaboration}, {Benetti},
  {Brocato}, {Campana}, {Cappellaro}, {Covino}, {D'Avanzo}, {D'Elia}, {Getman},
  {Ghirlanda}, {Ghisellini}, {Limatola}, {Nicastro}, {Palazzi}, {Pian},
  {Piranomonte}, {Possenti}, {Rossi}, {Salafia}, {Tomasella}, {Amati},
  {Antonelli}, {Bernardini}, {Bufano}, {Capaccioli}, {Casella}, {Dadina}, {De
  Cesare}, {Di Paola}, {Giuffrida}, {Giunta}, {Israel}, {Lisi}, {Maiorano},
  {Mapelli}, {Masetti}, {Pescalli}, {Pulone}, {Salvaterra}, {Schipani},
  {Spera}, {Stamerra}, {Stella}, {Testa}, {Turatto}, {Vergani}, {Aresu},
  {Bachetti}, {Buffa}, {Burgay}, {Buttu}, {Caria}, {Carretti}, {Casasola},
  {Castangia}, {Carboni}, {Casu}, {Concu}, {Corongiu}, {Deiana}, {Egron},
  {Fara}, {Gaudiomonte}, {Gusai}, {Ladu}, {Loru}, {Leurini}, {Marongiu},
  {Melis}, {Melis}, {Migoni}, {Milia}, {Navarrini}, {Orlati}, {Ortu}, {Palmas},
  {Pellizzoni}, {Perrodin}, {Pisanu}, {Poppi}, {Righini}, {Saba}, {Serra},
  {Serrau}, {Stagni}, {Surcis}, {Vacca}, {Vargiu}, {Hunt}, {Jin}, {Klose},
  {Kouveliotou}, {Mazzali}, {M{\o}ller}, {Nava}, {Piran}, {Selsing}, {Vergani},
  {Wiersema}, {Toma}, {Higgins}, {Mundell}, {di Serego Alighieri}, {G{\'o}tz},
  {Gao}, {Gomboc}, {Kaper}, {Kobayashi}, {Kopac}, {Mao}, {Starling}, {Steele},
  {van der Horst}, {GRAWITA: GRAvitational Wave Inaf TeAm}, {Acero}, {Atwood},
  {Baldini}, {Barbiellini}, {Bastieri}, {Berenji}, {Bellazzini}, {Bissaldi},
  {Blandford}, {Bloom}, {Bonino}, {Bottacini}, {Bregeon}, {Buehler}, {Buson},
  {Cameron}, {Caputo}, {Caraveo}, {Cavazzuti}, {Chekhtman}, {Cheung}, {Chiang},
  {Ciprini}, {Cohen-Tanugi}, {Cominsky}, {Costantin}, {Cuoco}, {D'Ammando}, {de
  Palma}, {Digel}, {Di Lalla}, {Di Mauro}, {Di Venere}, {Dubois}, {Fegan},
  {Focke}, {Franckowiak}, {Fukazawa}, {Funk}, {Fusco}, {Gargano}, {Gasparrini},
  {Giglietto}, {Giordano}, {Giroletti}, {Glanzman}, {Green}, {Grondin},
  {Guillemot}, {Guiriec}, {Harding}, {Horan}, {J{\'o}hannesson}, {Kamae},
  {Kensei}, {Kuss}, {La Mura}, {Latronico}, {Lemoine-Goumard}, {Longo},
  {Loparco}, {Lovellette}, {Lubrano}, {Magill}, {Maldera}, {Manfreda},
  {Mazziotta}, {McEnery}, {Meyer}, {Michelson}, {Mirabal}, {Monzani},
  {Moretti}, {Morselli}, {Moskalenko}, {Negro}, {Nuss}, {Ojha}, {Omodei},
  {Orienti}, {Orlando}, {Palatiello}, {Paliya}, {Paneque}, {Pesce-Rollins},
  {Piron}, {Porter}, {Principe}, {Rain{\`o}}, {Rando}, {Razzano}, {Razzaque},
  {Reimer}, {Reimer}, {Reposeur}, {Rochester}, {Saz Parkinson}, {Sgr{\`o}},
  {Siskind}, {Spada}, {Spandre}, {Suson}, {Takahashi}, {Tanaka}, {Thayer},
  {Thayer}, {Thompson}, {Tibaldo}, {Torres}, {Torresi}, {Troja}, {Venters},
  {Vianello}, {Zaharijas}, {Fermi Large Area Telescope Collaboration},
  {Allison}, {Bannister}, {Dobie}, {Kaplan}, {Lenc}, {Lynch}, {Murphy},
  {Sadler}, {Australia Telescope Compact Array}, {Hotan}, {James}, {Oslowski},
  {Raja}, {Shannon}, {Whiting}, {Australian SKA Pathfinder}, {Arcavi},
  {Howell}, {McCully}, {Hosseinzadeh}, {Hiramatsu}, {Poznanski}, {Barnes},
  {Zaltzman}, {Vasylyev}, {Maoz}, {Las Cumbres Observatory Group}, {Cooke},
  {Bailes}, {Wolf}, {Deller}, {Lidman}, {Wang}, {Gendre}, {Andreoni}, {Ackley},
  {Pritchard}, {Bessell}, {Chang}, {M{\"o}ller}, {Onken}, {Scalzo},
  {Ridden-Harper}, {Sharp}, {Tucker}, {Farrell}, {Elmer}, {Johnston},
  {Venkatraman Krishnan}, {Keane}, {Green}, {Jameson}, {Hu}, {Ma}, {Sun}, {Wu},
  {Wang}, {Shang}, {Hu}, {Ashley}, {Yuan}, {Li}, {Tao}, {Zhu}, {Zhang},
  {Suntzeff}, {Zhou}, {Yang}, {Orange}, {Morris}, {Cucchiara}, {Giblin},
  {Klotz}, {Staff}, {Thierry}, {Schmidt}, {OzGrav}, {(Deeper}, {Wider},
  {program}, {AST3}, {CAASTRO Collaborations}, {Tanvir}, {Levan}, {Cano}, {de
  Ugarte-Postigo}, {Gonz{\'a}lez-Fern{\'a}ndez}, {Greiner}, {Hjorth}, {Irwin},
  {Kr{\"u}hler}, {Mandel}, {Milvang-Jensen}, {O'Brien}, {Rol}, {Rosetti},
  {Rosswog}, {Rowlinson}, {Steeghs}, {Th{\"o}ne}, {Ulaczyk}, {Watson}, {Bruun},
  {Cutter}, {Figuera Jaimes}, {Fujii}, {Fruchter}, {Gompertz}, {Jakobsson},
  {Hodosan}, {J{\`e}rgensen}, {Kangas}, {Kann}, {Rabus}, {Schr{\o}der},
  {Stanway}, {Wijers}, {VINROUGE Collaboration}, {Lipunov}, {Gorbovskoy},
  {Kornilov}, {Tyurina}, {Balanutsa}, {Kuznetsov}, {Vlasenko}, {Podesta},
  {Lopez}, {Podesta}, {Levato}, {Saffe}, {Mallamaci}, {Budnev}, {Gress},
  {Kuvshinov}, {Gorbunov}, {Vladimirov}, {Zimnukhov}, {Gabovich}, {Yurkov},
  {Sergienko}, {Rebolo}, {Serra-Ricart}, {Tlatov}, {Ishmuhametova}, {MASTER
  Collaboration}, {Abe}, {Aoki}, {Aoki}, {Asakura}, {Baar}, {Barway}, {Bond},
  {Doi}, {Finet}, {Fujiyoshi}, {Furusawa}, {Honda}, {Itoh}, {Kanda},
  {Kawabata}, {Kawabata}, {Kim}, {Koshida}, {Kuroda}, {Lee}, {Liu},
  {Matsubayashi}, {Miyazaki}, {Morihana}, {Morokuma}, {Motohara}, {Murata},
  {Nagai}, {Nagashima}, {Nagayama}, {Nakaoka}, {Nakata}, {Ohsawa}, {Ohshima},
  {Ohta}, {Okita}, {Saito}, {Saito}, {Sako}, {Sekiguchi}, {Sumi}, {Tajitsu},
  {Takahashi}, {Takayama}, {Tamura}, {Tanaka}, {Tanaka}, {Terai}, {Tominaga},
  {Tristram}, {Uemura}, {Utsumi}, {Yamaguchi}, {Yasuda}, {Yoshida}, {Zenko},
  {J-GEM}, {Adams}, {Anupama}, {Bally}, {Barway}, {Bellm}, {Blagorodnova},
  {Cannella}, {Chandra}, {Chatterjee}, {Clarke}, {Cobb}, {Cook}, {Copperwheat},
  {De}, {Emery}, {Feindt}, {Foster}, {Fox}, {Frail}, {Fremling}, {Frohmaier},
  {Garcia}, {Ghosh}, {Giacintucci}, {Goobar}, {Gottlieb}, {Grefenstette},
  {Hallinan}, {Harrison}, {Heida}, {Helou}, {Ho}, {Horesh}, {Hotokezaka}, {Ip},
  {Itoh}, {Jacobs}, {Jencson}, {Kasen}, {Kasliwal}, {Kassim}, {Kim}, {Kiran},
  {Kuin}, {Kulkarni}, {Kupfer}, {Lau}, {Madsen}, {Mazzali}, {Miller},
  {Miyasaka}, {Mooley}, {Myers}, {Nakar}, {Ngeow}, {Nugent}, {Ofek},
  {Palliyaguru}, {Pavana}, {Perley}, {Peters}, {Pike}, {Piran}, {Qi}, {Quimby},
  {Rana}, {Rosswog}, {Rusu}, {Sadler}, {Van Sistine}, {Sollerman}, {Xu}, {Yan},
  {Yatsu}, {Yu}, {Zhang}, {Zhao}, {GROWTH}, {JAGWAR}, {Caltech-NRAO},
  {TTU-NRAO}, {NuSTAR Collaborations}, {Chambers}, {Huber}, {Schultz},
  {Bulger}, {Flewelling}, {Magnier}, {Lowe}, {Wainscoat}, {Waters}, {Willman},
  {Pan-STARRS}, {Ebisawa}, {Hanyu}, {Harita}, {Hashimoto}, {Hidaka}, {Hori},
  {Ishikawa}, {Isobe}, {Iwakiri}, {Kawai}, {Kawai}, {Kawamuro}, {Kawase},
  {Kitaoka}, {Makishima}, {Matsuoka}, {Mihara}, {Morita}, {Morita}, {Nakahira},
  {Nakajima}, {Nakamura}, {Negoro}, {Oda}, {Sakamaki}, {Sasaki}, {Serino},
  {Shidatsu}, {Shimomukai}, {Sugawara}, {Sugita}, {Sugizaki}, {Tachibana},
  {Takao}, {Tanimoto}, {Tomida}, {Tsuboi}, {Tsunemi}, {Ueda}, {Ueno}, {Yamada},
  {Yamaoka}, {Yamauchi}, {Yatabe}, {Yoneyama}, {Yoshii}, {MAXI Team}, {Coward},
  {Crisp}, {Macpherson}, {Andreoni}, {Laugier}, {Noysena}, {Klotz}, {Gendre},
  {Thierry}, {Turpin}, {Consortium}, {Im}, {Choi}, {Kim}, {Yoon}, {Lim}, {Lee},
  {Lee}, {Kim}, {Ko}, {Joe}, {Kwon}, {Kim}, {Lim}, {Choi}, {KU Collaboration},
  {Fynbo}, {Malesani}, {Xu}, {Optical Telescope}, {Smartt}, {Jerkstrand},
  {Kankare}, {Sim}, {Fraser}, {Inserra}, {Maguire}, {Leloudas}, {Magee},
  {Shingles}, {Smith}, {Young}, {Kotak}, {Gal-Yam}, {Lyman}, {Homan},
  {Agliozzo}, {Anderson}, {Angus}, {Ashall}, {Barbarino}, {Bauer}, {Berton},
  {Botticella}, {Bulla}, {Cannizzaro}, {Cartier}, {Cikota}, {Clark}, {De Cia},
  {Della Valle}, {Dennefeld}, {Dessart}, {Dimitriadis}, {Elias-Rosa}, {Firth},
  {Fl{\"o}rs}, {Frohmaier}, {Galbany}, {Gonz{\'a}lez-Gait{\'a}n}, {Gromadzki},
  {Guti{\'e}rrez}, {Hamanowicz}, {Harmanen}, {Heintz}, {Hernandez}, {Hodgkin},
  {Hook}, {Izzo}, {James}, {Jonker}, {Kerzendorf}, {Kostrzewa-Rutkowska},
  {Kromer}, {Kuncarayakti}, {Lawrence}, {Manulis}, {Mattila}, {McBrien},
  {M{\"u}ller}, {Nordin}, {O'Neill}, {Onori}, {Palmerio}, {Pastorello},
  {Patat}, {Pignata}, {Podsiadlowski}, {Razza}, {Reynolds}, {Roy}, {Ruiter},
  {Rybicki}, {Salmon}, {Pumo}, {Prentice}, {Seitenzahl}, {Smith}, {Sollerman},
  {Sullivan}, {Szegedi}, {Taddia}, {Taubenberger}, {Terreran}, {Van Soelen},
  {Vos}, {Walton}, {Wright}, {Wyrzykowski}, {Yaron}, {pre=''(''>ePESSTO},
  {Chen}, {Kr{\"u}hler}, {Schady}, {Wiseman}, {Greiner}, {Rau}, {Schweyer},
  {Klose}, {Nicuesa Guelbenzu}, {GROND}, {Palliyaguru}, {Tech University},
  {Shara}, {Williams}, {Vaisanen}, {Potter}, {Romero Colmenero}, {Crawford},
  {Buckley}, {Mao}, {SALT Group}, {D{\'\i}az}, {Macri}, {Garc{\'\i}a Lambas},
  {Mendes de Oliveira}, {Nilo Castell{\'o}n}, {Ribeiro}, {S{\'a}nchez},
  {Schoenell}, {Abramo}, {Akras}, {Alcaniz}, {Artola}, {Beroiz}, {Bonoli},
  {Cabral}, {Camuccio}, {Chavushyan}, {Coelho}, {Colazo}, {Costa-Duarte},
  {Cuevas Larenas}, {Dom{\'\i}nguez Romero}, {Dultzin}, {Fern{\'a}ndez},
  {Garc{\'\i}a}, {Girardini}, {Gon{\c{c}}alves}, {Gon{\c{c}}alves}, {Gurovich},
  {Jim{\'e}nez-Teja}, {Kanaan}, {Lares}, {Lopes de Oliveira}, {L{\'o}pez-Cruz},
  {Melia}, {Molino}, {Padilla}, {Pe{\~n}uela}, {Placco}, {Qui{\~n}ones},
  {Ram{\'\i}rez Rivera}, {Renzi}, {Riguccini}, {R{\'\i}os-L{\'o}pez},
  {Rodriguez}, {Sampedro}, {Schneiter}, {Sodr{\'e}}, {Starck}, {Torres-Flores},
  {Tornatore}, {Zadro{\.z}ny}, {Castillo}, {TOROS: Transient Robotic
  Observatory of South Collaboration}, {Castro-Tirado}, {Tello}, {Hu}, {Zhang},
  {Cunniffe}, {Castell{\'o}n}, {Hiriart}, {Caballero-Garc{\'\i}a},
  {Jel{\'\i}nek}, {Kub{\'a}nek}, {P{\'e}rez del Pulgar}, {Park}, {Jeong},
  {Castro Cer{\'o}n}, {Pandey}, {Yock}, {Querel}, {Fan}, {Wang}, {BOOTES
  Collaboration}, {Beardsley}, {Brown}, {Crosse}, {Emrich}, {Franzen},
  {Gaensler}, {Horsley}, {Johnston-Hollitt}, {Kenney}, {Morales}, {Pallot},
  {Sokolowski}, {Steele}, {Tingay}, {Trott}, {Walker}, {Wayth}, {Williams},
  {Wu}, {Murchison Widefield Array}, {Yoshida}, {Sakamoto}, {Kawakubo},
  {Yamaoka}, {Takahashi}, {Asaoka}, {Ozawa}, {Torii}, {Shimizu}, {Tamura},
  {Ishizaki}, {Cherry}, {Ricciarini}, {Penacchioni}, {Marrocchesi}, {CALET
  Collaboration}, {Pozanenko}, {Volnova}, {Mazaeva}, {Minaev}, {Krugov},
  {Kusakin}, {Reva}, {Moskvitin}, {Rumyantsev}, {Inasaridze}, {Klunko},
  {Tungalag}, {Schmalz}, {Burhonov}, {IKI-GW Follow-up Collaboration},
  {Abdalla}, {Abramowski}, {Aharonian}, {Ait Benkhali}, {Ang{\"u}ner},
  {Arakawa}, {Arrieta}, {Aubert}, {Backes}, {Balzer}, {Barnard}, {Becherini},
  {Becker Tjus}, {Berge}, {Bernhard}, {Bernl{\"o}hr}, {Blackwell},
  {B{\"o}ttcher}, {Boisson}, {Bolmont}, {Bonnefoy}, {Bordas}, {Bregeon},
  {Brun}, {Brun}, {Bryan}, {B{\"u}chele}, {Bulik}, {Capasso}, {Caroff},
  {Carosi}, {Casanova}, {Cerruti}, {Chakraborty}, {Chaves}, {Chen},
  {Chevalier}, {Colafrancesco}, {Condon}, {Conrad}, {Davids}, {Decock}, {Deil},
  {Devin}, {deWilt}, {Dirson}, {Djannati-Ata{\"\i}}, {Donath}, {O'C. Drury},
  {Dutson}, {Dyks}, {Edwards}, {Egberts}, {Emery}, {Ernenwein}, {Eschbach},
  {Farnier}, {Fegan}, {Fernandes}, {Fiasson}, {Fontaine}, {Funk},
  {F{\"u}ssling}, {Gabici}, {Gallant}, {Garrigoux}, {Gat{\'e}}, {Giavitto},
  {Giebels}, {Glawion}, {Glicenstein}, {Gottschall}, {Grondin}, {Hahn},
  {Haupt}, {Hawkes}, {Heinzelmann}, {Henri}, {Hermann}, {Hinton}, {Hofmann},
  {Hoischen}, {Holch}, {Holler}, {Horns}, {Ivascenko}, {Iwasaki},
  {Jacholkowska}, {Jamrozy}, {Jankowsky}, {Jankowsky}, {Jingo}, {Jouvin},
  {Jung-Richardt}, {Kastendieck}, {Katarzy{\'n}ski}, {Katsuragawa},
  {Kerszberg}, {Khangulyan}, {Kh{\'e}lifi}, {King}, {Klepser}, {Klochkov},
  {Klu{\'z}niak}, {Komin}, {Kosack}, {Krakau}, {Kraus}, {Kr{\"u}ger}, {Laffon},
  {Lamanna}, {Lau}, {Lees}, {Lefaucheur}, {Lemi{\`e}re}, {Lemoine-Goumard},
  {Lenain}, {Leser}, {Lohse}, {Lorentz}, {Liu}, {Lypova}, {Malyshev},
  {Marandon}, {Marcowith}, {Mariaud}, {Marx}, {Maurin}, {Maxted}, {Mayer},
  {Meintjes}, {Meyer}, {Mitchell}, {Moderski}, {Mohamed}, {Mohrmann},
  {Mor{\r{a}}}, {Moulin}, {Murach}, {Nakashima}, {de Naurois}, {Ndiyavala},
  {Niederwanger}, {Niemiec}, {Oakes}, {O'Brien}, {Odaka}, {Ohm}, {Ostrowski},
  {Oya}, {Padovani}, {Panter}, {Parsons}, {Pekeur}, {Pelletier}, {Perennes},
  {Petrucci}, {Peyaud}, {Piel}, {Pita}, {Poireau}, {Poon}, {Prokhorov},
  {Prokoph}, {P{\"u}hlhofer}, {Punch}, {Quirrenbach}, {Raab}, {Rauth},
  {Reimer}, {Reimer}, {Renaud}, {de los Reyes}, {Rieger}, {Rinchiuso},
  {Romoli}, {Rowell}, {Rudak}, {Rulten}, {Sahakian}, {Saito}, {Sanchez},
  {Santangelo}, {Sasaki}, {Schlickeiser}, {Sch{\"u}ssler}, {Schulz},
  {Schwanke}, {Schwemmer}, {Seglar-Arroyo}, {Settimo}, {Seyffert}, {Shafi},
  {Shilon}, {Shiningayamwe}, {Simoni}, {Sol}, {Spanier}, {Spir-Jacob},
  {Stawarz}, {Steenkamp}, {Stegmann}, {Steppa}, {Sushch}, {Takahashi},
  {Tavernet}, {Tavernier}, {Taylor}, {Terrier}, {Tibaldo}, {Tiziani},
  {Tluczykont}, {Trichard}, {Tsirou}, {Tsuji}, {Tuffs}, {Uchiyama}, {van der
  Walt}, {van Eldik}, {van Rensburg}, {van Soelen}, {Vasileiadis}, {Veh},
  {Venter}, {Viana}, {Vincent}, {Vink}, {Voisin}, {V{\"o}lk}, {Vuillaume},
  {Wadiasingh}, {Wagner}, {Wagner}, {Wagner}, {White}, {Wierzcholska},
  {Willmann}, {W{\"o}rnlein}, {Wouters}, {Yang}, {Zaborov}, {Zacharias},
  {Zanin}, {Zdziarski}, {Zech}, {Zefi}, {Ziegler}, {Zorn}, {{\.Z}ywucka},
  {H.~E.~S.~S. Collaboration}, {Fender}, {Broderick}, {Rowlinson}, {Wijers},
  {Stewart}, {ter Veen}, {Shulevski}, {LOFAR Collaboration}, {Kavic},
  {Simonetti}, {League}, {Tsai}, {Obenberger}, {Nathaniel}, {Taylor}, {Dowell},
  {Liebling}, {Estes}, {Lippert}, {Sharma}, {Vincent}, {Farella}, {Wavelength
  Array}, {Abeysekara}, {Albert}, {Alfaro}, {Alvarez}, {Arceo},
  {Arteaga-Vel{\'a}zquez}, {Avila Rojas}, {Ayala Solares}, {Barber}, {Becerra
  Gonzalez}, {Becerril}, {Belmont-Moreno}, {BenZvi}, {Berley}, {Bernal},
  {Braun}, {Brisbois}, {Caballero-Mora}, {Capistr{\'a}n}, {Carrami{\~n}ana},
  {Casanova}, {Castillo}, {Cotti}, {Cotzomi}, {Couti{\~n}o de Le{\'o}n}, {De
  Le{\'o}n}, {De la Fuente}, {Diaz Hernandez}, {Dichiara}, {Dingus},
  {DuVernois}, {D{\'\i}az-V{\'e}lez}, {Ellsworth}, {Engel},
  {Enr{\'\i}quez-Rivera}, {Fiorino}, {Fleischhack}, {Fraija},
  {Garc{\'\i}a-Gonz{\'a}lez}, {Garfias}, {Gerhardt}, {Gonz{\~o}lez Mu{\~n}oz},
  {Gonz{\'a}lez}, {Goodman}, {Hampel-Arias}, {Harding}, {Hernandez},
  {Hernandez-Almada}, {Hona}, {H{\"u}ntemeyer}, {Iriarte}, {Jardin-Blicq},
  {Joshi}, {Kaufmann}, {Kieda}, {Lara}, {Lauer}, {Lennarz}, {Le{\'o}n Vargas},
  {Linnemann}, {Longinotti}, {Raya}, {Luna-Garc{\'\i}a}, {L{\'o}pez-Coto},
  {Malone}, {Marinelli}, {Martinez}, {Martinez-Castellanos},
  {Mart{\'\i}nez-Castro}, {Mart{\'\i}nez-Huerta}, {Matthews},
  {Miranda-Romagnoli}, {Moreno}, {Mostaf{\'a}}, {Nellen}, {Newbold}, {Nisa},
  {Noriega-Papaqui}, {Pelayo}, {Pretz}, {P{\'e}rez-P{\'e}rez}, {Ren}, {Rho},
  {Rivi{\`e}re}, {Rosa-Gonz{\'a}lez}, {Rosenberg}, {Ruiz-Velasco}, {Salazar},
  {Salesa Greus}, {Sandoval}, {Schneider}, {Schoorlemmer}, {Sinnis}, {Smith},
  {Springer}, {Surajbali}, {Tibolla}, {Tollefson}, {Torres}, {Ukwatta},
  {Weisgarber}, {Westerhoff}, {Wisher}, {Wood}, {Yapici}, {Yodh}, {Younk},
  {Zhou}, {{\'A}lvarez}, {HAWC Collaboration}, {Aab}, {Abreu}, {Aglietta},
  {Albuquerque}, {Albury}, {Allekotte}, {Almela}, {Alvarez Castillo},
  {Alvarez-Mu{\~n}iz}, {Anastasi}, {Anchordoqui}, {Andrada}, {Andringa},
  {Aramo}, {Arsene}, {Asorey}, {Assis}, {Avila}, {Badescu}, {Balaceanu},
  {Barbato}, {Barreira Luz}, {Becker}, {Bellido}, {Berat}, {Bertaina},
  {Bertou}, {Biermann}, {Biteau}, {Blaess}, {Blanco}, {Blazek}, {Bleve},
  {Boh{\'a}{\v{c}}ov{\'a}}, {Bonifazi}, {Borodai}, {Botti}, {Brack}, {Brancus},
  {Bretz}, {Bridgeman}, {Briechle}, {Buchholz}, {Bueno}, {Buitink}, {Buscemi},
  {Caballero-Mora}, {Caccianiga}, {Cancio}, {Canfora}, {Caruso}, {Castellina},
  {Catalani}, {Cataldi}, {Cazon}, {Chavez}, {Chinellato}, {Chudoba}, {Clay},
  {Cobos Cerutti}, {Colalillo}, {Coleman}, {Collica}, {Coluccia},
  {Concei{\c{c}}{\~a}o}, {Consolati}, {Contreras}, {Cooper}, {Coutu},
  {Covault}, {Cronin}, {D'Amico}, {Daniel}, {Dasso}, {Daumiller}, {Dawson},
  {Day}, {de Almeida}, {de Jong}, {De Mauro}, {de Mello Neto}, {De Mitri}, {de
  Oliveira}, {de Souza}, {Debatin}, {Deligny}, {D{\'\i}az Castro}, {Diogo},
  {Dobrigkeit}, {D'Olivo}, {Dorosti}, {Dos Anjos}, {Dova}, {Dundovic}, {Ebr},
  {Engel}, {Erdmann}, {Erfani}, {Escobar}, {Espadanal}, {Etchegoyen}, {Falcke},
  {Farmer}, {Farrar}, {Fauth}, {Fazzini}, {Feldbusch}, {Fenu}, {Fick},
  {Figueira}, {Filip{\v{c}}i{\v{c}}}, {Freire}, {Fujii}, {Fuster},
  {Ga{\"\i}or}, {Garc{\'\i}a}, {Gat{\'e}}, {Gemmeke}, {Gherghel-Lascu}, {Ghia},
  {Giaccari}, {Giammarchi}, {Giller}, {G{\l}as}, {Glaser}, {Golup}, {G{\'o}mez
  Berisso}, {G{\'o}mez Vitale}, {Gonz{\'a}lez}, {Gorgi}, {Gottowik}, {Grillo},
  {Grubb}, {Guarino}, {Guedes}, {Halliday}, {Hampel}, {Hansen}, {Harari},
  {Harrison}, {Harvey}, {Haungs}, {Hebbeker}, {Heck}, {Heimann}, {Herve},
  {Hill}, {Hojvat}, {Holt}, {Homola}, {H{\"o}randel}, {Horvath},
  {Hrabovsk{\'y}}, {Huege}, {Hulsman}, {Insolia}, {Isar}, {Jandt}, {Johnsen},
  {Josebachuili}, {Jurysek}, {K{\"a}{\"a}p{\"a}}, {Kampert}, {Keilhauer},
  {Kemmerich}, {Kemp}, {Kieckhafer}, {Klages}, {Kleifges}, {Kleinfeller},
  {Krause}, {Krohm}, {Kuempel}, {Kukec Mezek}, {Kunka}, {Kuotb Awad}, {Lago},
  {LaHurd}, {Lang}, {Lauscher}, {Legumina}, {Leigui de Oliveira},
  {Letessier-Selvon}, {Lhenry-Yvon}, {Link}, {Lo Presti}, {Lopes}, {L{\'o}pez},
  {L{\'o}pez Casado}, {Lorek}, {Luce}, {Lucero}, {Malacari}, {Mallamaci},
  {Mandat}, {Mantsch}, {Mariazzi}, {Maris}, {Marsella}, {Martello}, {Martinez},
  {Mart{\'\i}nez Bravo}, {Mas{\'\i}as Meza}, {Mathes}, {Mathys}, {Matthews},
  {Matthiae}, {Mayotte}, {Mazur}, {Medina}, {Medina-Tanco}, {Melo},
  {Menshikov}, {Merenda}, {Michal}, {Micheletti}, {Middendorf}, {Miramonti},
  {Mitrica}, {Mockler}, {Mollerach}, {Montanet}, {Morello}, {Morlino},
  {M{\"u}ller}, {M{\"u}ller}, {Muller}, {M{\"u}ller}, {Mussa}, {Naranjo},
  {Nguyen}, {Niculescu-Oglinzanu}, {Niechciol}, {Niemietz}, {Niggemann},
  {Nitz}, {Nosek}, {Novotny}, {No{\v{z}}ka}, {N{\'u}{\~n}ez}, {Oikonomou},
  {Olinto}, {Palatka}, {Pallotta}, {Papenbreer}, {Parente}, {Parra}, {Paul},
  {Pech}, {Pedreira}, {P{\c{e}}kala}, {Pe{\~n}a-Rodriguez}, {Pereira},
  {Perlin}, {Perrone}, {Peters}, {Petrera}, {Phuntsok}, {Pierog}, {Pimenta},
  {Pirronello}, {Platino}, {Plum}, {Poh}, {Porowski}, {Prado}, {Privitera},
  {Prouza}, {Quel}, {Querchfeld}, {Quinn}, {Ramos-Pollan}, {Rautenberg},
  {Ravignani}, {Ridky}, {Riehn}, {Risse}, {Ristori}, {Rizi}, {Rodrigues de
  Carvalho}, {Rodriguez Fernandez}, {Rodriguez Rojo}, {Roncoroni}, {Roth},
  {Roulet}, {Rovero}, {Ruehl}, {Saffi}, {Saftoiu}, {Salamida}, {Salazar},
  {Saleh}, {Salina}, {S{\'a}nchez}, {Sanchez-Lucas}, {Santos}, {Santos},
  {Sarazin}, {Sarmento}, {Sarmiento-Cano}, {Sato}, {Schauer}, {Scherini},
  {Schieler}, {Schimp}, {Schmidt}, {Scholten}, {Schov{\'a}nek}, {Schr{\"o}der},
  {Schr{\"o}der}, {Schulz}, {Schumacher}, {Sciutto}, {Segreto}, {Shadkam},
  {Shellard}, {Sigl}, {Silli}, {{\v{S}}m{\'\i}da}, {Snow}, {Sommers},
  {Sonntag}, {Soriano}, {Squartini}, {Stanca}, {Stani{\v{c}}}, {Stasielak},
  {Stassi}, {Stolpovskiy}, {Strafella}, {Streich}, {Suarez},
  {Suarez-Dur{\'a}n}, {Sudholz}, {Suomij{\"a}rvi}, {Supanitsky},
  {{\v{S}}up{\'\i}k}, {Swain}, {Szadkowski}, {Taboada}, {Taborda},
  {Timmermans}, {Todero Peixoto}, {Tomankova}, {Tom{\'e}}, {Torralba Elipe},
  {Travnicek}, {Trini}, {Tueros}, {Ulrich}, {Unger}, {Urban}, {Vald{\'e}s
  Galicia}, {Vali{\~n}o}, {Valore}, {van Aar}, {van Bodegom}, {van den Berg},
  {van Vliet}, {Varela}, {Vargas C{\'a}rdenas}, {V{\'a}zquez}, {Veberi{\v{c}}},
  {Ventura}, {Vergara Quispe}, {Verzi}, {Vicha}, {Villase{\~n}or}, {Vorobiov},
  {Wahlberg}, {Wainberg}, {Walz}, {Watson}, {Weber}, {Weindl}, {Wiede{\'n}ski},
  {Wiencke}, {Wilczy{\'n}ski}, {Wirtz}, {Wittkowski}, {Wundheiler}, {Yang},
  {Yushkov}, {Zas}, {Zavrtanik}, {Zavrtanik}, {Zepeda}, {Zimmermann},
  {Ziolkowski}, {Zong}, {Zuccarello}, {Pierre Auger Collaboration}, {Kim},
  {Schulze}, {Bauer}, {Corral-Santana}, {de Gregorio-Monsalvo},
  {Gonz{\'a}lez-L{\'o}pez}, {Hartmann}, {Ishwara-Chandra}, {Mart{\'\i}n},
  {Mehner}, {Misra}, {Micha{\l}owski}, {Resmi}, {ALMA Collaboration}, {Paragi},
  {Agudo}, {An}, {Beswick}, {Casadio}, {Frey}, {Jonker}, {Kettenis}, {Marcote},
  {Moldon}, {Szomoru}, {van Langevelde}, {Yang}, {Euro VLBI Team}, {Cwiek},
  {Cwiok}, {Czyrkowski}, {Dabrowski}, {Kasprowicz}, {Mankiewicz}, {Nawrocki},
  {Opiela}, {Piotrowski}, {Wrochna}, {Zaremba}, {{\.Z}arnecki}, {Pi of Sky
  Collaboration}, {Haggard}, {Nynka}, {Ruan}, {Chandra Team at McGill
  University}, {Bland}, {Booler}, {Devillepoix}, {de Gois}, {Hancock}, {Howie},
  {Paxman}, {Sansom}, {Towner}, {Desert Fireball Network}, {Tonry}, {Coughlin},
  {Stubbs}, {Denneau}, {Heinze}, {Stalder}, {Weiland}, {ATLAS}, {Eatough},
  {Kramer}, {Kraus}, {Time Resolution Universe Survey}, {Troja}, {Piro},
  {Becerra Gonz{\'a}lez}, {Butler}, {Fox}, {Khandrika}, {Kutyrev}, {Lee},
  {Ricci}, {Ryan}, {S{\'a}nchez-Ram{\'\i}rez}, {Veilleux}, {Watson},
  {Wieringa}, {Burgess}, {van Eerten}, {Fontes}, {Fryer}, {Korobkin},
  {Wollaeger}, {RIMAS}, {RATIR}, {Camilo}, {Foley}, {Goedhart}, {Makhathini},
  {Oozeer}, {Smirnov}, {Fender}, {Woudt}, \& {South
  Africa/MeerKAT}}]{2017ApJ...848L..12A}
---. 2017{\natexlab{b}}, \apjl, 848, L12, \dodoi{10.3847/2041-8213/aa91c9}

\bibitem[{{Aggarwal} {et~al.}(2021){Aggarwal}, {Budav{\'a}ri}, {Deller},
  {Eftekhari}, {James}, {Prochaska}, \& {Tendulkar}}]{2021ApJ...911...95A}
{Aggarwal}, K., {Budav{\'a}ri}, T., {Deller}, A.~T., {et~al.} 2021, \apj, 911,
  95, \dodoi{10.3847/1538-4357/abe8d2}

\bibitem[{{Ahumada} {et~al.}(2020){Ahumada}, {Prieto}, {Almeida}, {Anders},
  {Anderson}, {Andrews}, {Anguiano}, {Arcodia}, {Armengaud}, {Aubert}, {Avila},
  {Avila-Reese}, {Badenes}, {Balland}, {Barger}, {Barrera-Ballesteros}, {Basu},
  {Bautista}, {Beaton}, {Beers}, {Benavides}, {Bender}, {Bernardi}, {Bershady},
  {Beutler}, {Bidin}, {Bird}, {Bizyaev}, {Blanc}, {Blanton}, {Boquien},
  {Borissova}, {Bovy}, {Brandt}, {Brinkmann}, {Brownstein}, {Bundy}, {Bureau},
  {Burgasser}, {Burtin}, {Cano-D{\'\i}az}, {Capasso}, {Cappellari}, {Carrera},
  {Chabanier}, {Chaplin}, {Chapman}, {Cherinka}, {Chiappini}, {Doohyun Choi},
  {Chojnowski}, {Chung}, {Clerc}, {Coffey}, {Comerford}, {Comparat}, {da
  Costa}, {Cousinou}, {Covey}, {Crane}, {Cunha}, {Ilha}, {Dai}, {Damsted},
  {Darling}, {Davidson}, {Davies}, {Dawson}, {De}, {de la Macorra}, {De Lee},
  {Queiroz}, {Deconto Machado}, {de la Torre}, {Dell'Agli}, {du Mas des
  Bourboux}, {Diamond-Stanic}, {Dillon}, {Donor}, {Drory}, {Duckworth},
  {Dwelly}, {Ebelke}, {Eftekharzadeh}, {Davis Eigenbrot}, {Elsworth},
  {Eracleous}, {Erfanianfar}, {Escoffier}, {Fan}, {Farr},
  {Fern{\'a}ndez-Trincado}, {Feuillet}, {Finoguenov}, {Fofie},
  {Fraser-McKelvie}, {Frinchaboy}, {Fromenteau}, {Fu}, {Galbany}, {Garcia},
  {Garc{\'\i}a-Hern{\'a}ndez}, {Oehmichen}, {Ge}, {Maia}, {Geisler}, {Gelfand},
  {Goddy}, {Gonzalez-Perez}, {Grabowski}, {Green}, {Grier}, {Guo}, {Guy},
  {Harding}, {Hasselquist}, {Hawken}, {Hayes}, {Hearty}, {Hekker}, {Hogg},
  {Holtzman}, {Horta}, {Hou}, {Hsieh}, {Huber}, {Hunt}, {Chitham}, {Imig},
  {Jaber}, {Angel}, {Johnson}, {Jones}, {J{\"o}nsson}, {Jullo}, {Kim},
  {Kinemuchi}, {Kirkpatrick}, {Kite}, {Klaene}, {Kneib}, {Kollmeier}, {Kong},
  {Kounkel}, {Krishnarao}, {Lacerna}, {Lan}, {Lane}, {Law}, {Le Goff}, {Leung},
  {Lewis}, {Li}, {Lian}, {Lin}, {Long}, {Longa-Pe{\~n}a}, {Lundgren}, {Lyke},
  {Ted Mackereth}, {MacLeod}, {Majewski}, {Manchado}, {Maraston}, {Martini},
  {Masseron}, {Masters}, {Mathur}, {McDermid}, {Merloni}, {Merrifield},
  {M{\'e}sz{\'a}ros}, {Miglio}, {Minniti}, {Minsley}, {Miyaji}, {Mohammad},
  {Mosser}, {Mueller}, {Muna}, {Mu{\~n}oz-Guti{\'e}rrez}, {Myers}, {Nadathur},
  {Nair}, {Nandra}, {do Nascimento}, {Nevin}, {Newman}, {Nidever}, {Nitschelm},
  {Noterdaeme}, {O'Connell}, {Olmstead}, {Oravetz}, {Oravetz}, {Osorio},
  {Pace}, {Padilla}, {Palanque-Delabrouille}, {Palicio}, {Pan}, {Pan},
  {Parker}, {Paviot}, {Peirani}, {Ram{\'r}ez}, {Penny}, {Percival},
  {Perez-Fournon}, {P{\'e}rez-R{\`a}fols}, {Petitjean}, {Pieri},
  {Pinsonneault}, {Poovelil}, {Povick}, {Prakash}, {Price-Whelan}, {Raddick},
  {Raichoor}, {Ray}, {Rembold}, {Rezaie}, {Riffel}, {Riffel}, {Rix}, {Robin},
  {Roman-Lopes}, {Rom{\'a}n-Z{\'u}{\~n}iga}, {Rose}, {Ross}, {Rossi},
  {Rowlands}, {Rubin}, {Salvato}, {S{\'a}nchez}, {S{\'a}nchez-Menguiano},
  {S{\'a}nchez-Gallego}, {Sayres}, {Schaefer}, {Schiavon}, {Schimoia},
  {Schlafly}, {Schlegel}, {Schneider}, {Schultheis}, {Schwope}, {Seo},
  {Serenelli}, {Shafieloo}, {Shamsi}, {Shao}, {Shen}, {Shetrone}, {Shirley},
  {Aguirre}, {Simon}, {Skrutskie}, {Slosar}, {Smethurst}, {Sobeck}, {Sodi},
  {Souto}, {Stark}, {Stassun}, {Steinmetz}, {Stello}, {Stermer},
  {Storchi-Bergmann}, {Streblyanska}, {Stringfellow}, {Stutz}, {Su{\'a}rez},
  {Sun}, {Taghizadeh-Popp}, {Talbot}, {Tayar}, {Thakar}, {Theriault}, {Thomas},
  {Thomas}, {Tinker}, {Tojeiro}, {Toledo}, {Tremonti}, {Troup}, {Tuttle},
  {Unda-Sanzana}, {Valentini}, {Vargas-Gonz{\'a}lez}, {Vargas-Maga{\~n}a},
  {V{\'a}zquez-Mata}, {Vivek}, {Wake}, {Wang}, {Weaver}, {Weijmans}, {Wild},
  {Wilson}, {Wilson}, {Wolthuis}, {Wood-Vasey}, {Yan}, {Yang}, {Y{\`e}che},
  {Zamora}, {Zarrouk}, {Zasowski}, {Zhang}, {Zhao}, {Zhao}, {Zheng}, {Zheng},
  {Zhu}, \& {Zou}}]{2020ApJS..249....3A}
{Ahumada}, R., {Prieto}, C.~A., {Almeida}, A., {et~al.} 2020, \apjs, 249, 3,
  \dodoi{10.3847/1538-4365/ab929e}

\bibitem[{{Amati} {et~al.}(2002){Amati}, {Frontera}, {Tavani}, {in't Zand},
  {Antonelli}, {Costa}, {Feroci}, {Guidorzi}, {Heise}, {Masetti}, {Montanari},
  {Nicastro}, {Palazzi}, {Pian}, {Piro}, \& {Soffitta}}]{2002A&A...390...81A}
{Amati}, L., {Frontera}, F., {Tavani}, M., {et~al.} 2002, \aap, 390, 81,
  \dodoi{10.1051/0004-6361:20020722}

\bibitem[{{Anderson} {et~al.}(2021){Anderson}, {Hancock}, {Rowlinson},
  {Sokolowski}, {Williams}, {Tian}, {Miller-Jones}, {Hurley-Walker},
  {Bannister}, {Bell}, {James}, {Kaplan}, {Murphy}, {Tingay}, {Meyers},
  {Johnston-Hollitt}, \& {Wayth}}]{2021PASA...38...26A}
{Anderson}, G.~E., {Hancock}, P.~J., {Rowlinson}, A., {et~al.} 2021, \pasa, 38,
  e026, \dodoi{10.1017/pasa.2021.15}

\bibitem[{{Anderson} {et~al.}(2018){Anderson}, {Hallinan}, {Eastwood},
  {Monroe}, {Vedantham}, {Bourke}, {Greenhill}, {Kocz}, {Lazio}, {Price},
  {Schinzel}, {Wang}, \& {Woody}}]{2018ApJ...864...22A}
{Anderson}, M.~M., {Hallinan}, G., {Eastwood}, M.~W., {et~al.} 2018, \apj, 864,
  22, \dodoi{10.3847/1538-4357/aad2d7}

\bibitem[{{Bannister} {et~al.}(2012){Bannister}, {Murphy}, {Gaensler}, \&
  {Reynolds}}]{2012ApJ...757...38B}
{Bannister}, K.~W., {Murphy}, T., {Gaensler}, B.~M., \& {Reynolds}, J.~E. 2012,
  \apj, 757, 38, \dodoi{10.1088/0004-637X/757/1/38}

\bibitem[{{Bannister} {et~al.}(2019){Bannister}, {Deller}, {Phillips},
  {Macquart}, {Prochaska}, {Tejos}, {Ryder}, {Sadler}, {Shannon}, {Simha},
  {Day}, {McQuinn}, {North-Hickey}, {Bhandari}, {Arcus}, {Bennert}, {Burchett},
  {Bouwhuis}, {Dodson}, {Ekers}, {Farah}, {Flynn}, {James}, {Kerr}, {Lenc},
  {Mahony}, {O'Meara}, {Os{\l}owski}, {Qiu}, {Treu}, {U}, {Bateman}, {Bock},
  {Bolton}, {Brown}, {Bunton}, {Chippendale}, {Cooray}, {Cornwell}, {Gupta},
  {Hayman}, {Kesteven}, {Koribalski}, {MacLeod}, {McClure-Griffiths},
  {Neuhold}, {Norris}, {Pilawa}, {Qiao}, {Reynolds}, {Roxby}, {Shimwell},
  {Voronkov}, \& {Wilson}}]{2019Sci...365..565B}
{Bannister}, K.~W., {Deller}, A.~T., {Phillips}, C., {et~al.} 2019, Science,
  365, 565, \dodoi{10.1126/science.aaw5903}

\bibitem[{{Barbier} {et~al.}(2007){Barbier}, {Barthelmy}, {Cummings},
  {Fenimore}, {Gehrels}, {Krimm}, {Markwardt}, {Palmer}, {Parsons}, {Racusin},
  {Sakamoto}, {Sato}, {Stamatikos}, {Tueller}, \&
  {Ukwatta}}]{2007GCN..6623....1B}
{Barbier}, L., {Barthelmy}, S.~D., {Cummings}, J., {et~al.} 2007, GRB
  Coordinates Network, 6623, 1

\bibitem[{{Barthelmy} {et~al.}(2005){Barthelmy}, {Chincarini}, {Burrows},
  {Gehrels}, {Covino}, {Moretti}, {Romano}, {O'Brien}, {Sarazin},
  {Kouveliotou}, {Goad}, {Vaughan}, {Tagliaferri}, {Zhang}, {Antonelli},
  {Campana}, {Cummings}, {D'Avanzo}, {Davies}, {Giommi}, {Grupe}, {Kaneko},
  {Kennea}, {King}, {Kobayashi}, {Melandri}, {Meszaros}, {Nousek}, {Patel},
  {Sakamoto}, \& {Wijers}}]{2005Natur.438..994B}
{Barthelmy}, S.~D., {Chincarini}, G., {Burrows}, D.~N., {et~al.} 2005, \nat,
  438, 994, \dodoi{10.1038/nature04392}

\bibitem[{{Barthelmy} {et~al.}(2009){Barthelmy}, {Baumgartner}, {Cummings},
  {Fenimore}, {Gehrels}, {Krimm}, {Markwardt}, {Marshall}, {Palmer},
  {Sakamoto}, {Sato}, {Stamatikos}, {Tueller}, \&
  {Ukwatta}}]{2009GCN..9494....1B}
{Barthelmy}, S.~D., {Baumgartner}, W.~H., {Cummings}, J.~R., {et~al.} 2009, GRB
  Coordinates Network, 9494, 1

\bibitem[{{Barthelmy} {et~al.}(2017){Barthelmy}, {Cannizzo}, {Cummings},
  {Krimm}, {Lien}, {Markwardt}, {Norris}, {Palmer}, {Sakamoto}, {Stamatikos},
  \& {Ukwatta}}]{2017GCN.21981....1B}
{Barthelmy}, S.~D., {Cannizzo}, J.~K., {Cummings}, J.~R., {et~al.} 2017, GRB
  Coordinates Network, 21981, 1

\bibitem[{{Beardmore} {et~al.}(2015){Beardmore}, {Page}, {Palmer}, \&
  {Ukwatta}}]{2015GCN.17743....1B}
{Beardmore}, A.~P., {Page}, K.~L., {Palmer}, D.~M., \& {Ukwatta}, T.~N. 2015,
  GRB Coordinates Network, 17743, 1

\bibitem[{{Beloborodov}(2017)}]{2017ApJ...843L..26B}
{Beloborodov}, A.~M. 2017, \apjl, 843, L26, \dodoi{10.3847/2041-8213/aa78f3}

\bibitem[{{Beloborodov}(2020)}]{2020ApJ...896..142B}
---. 2020, \apj, 896, 142, \dodoi{10.3847/1538-4357/ab83eb}

\bibitem[{{Beloborodov} \& {Thompson}(2007)}]{2007ApJ...657..967B}
{Beloborodov}, A.~M., \& {Thompson}, C. 2007, \apj, 657, 967,
  \dodoi{10.1086/508917}

\bibitem[{{Berger}(2014)}]{2014ARA&A..52...43B}
{Berger}, E. 2014, \araa, 52, 43, \dodoi{10.1146/annurev-astro-081913-035926}

\bibitem[{{Berger} {et~al.}(2006){Berger}, {Cenko}, \&
  {Rau}}]{2006GCN..5071....1B}
{Berger}, E., {Cenko}, S.~B., \& {Rau}, A. 2006, GRB Coordinates Network, 5071,
  1

\bibitem[{{Berger} {et~al.}(2007){Berger}, {Fox}, {Price}, {Nakar}, {Gal-Yam},
  {Holz}, {Schmidt}, {Cucchiara}, {Cenko}, {Kulkarni}, {Soderberg}, {Frail},
  {Penprase}, {Rau}, {Ofek}, {Burnell}, {Cameron}, {Cowie}, {Dopita}, {Hook},
  {Peterson}, {Podsiadlowski}, {Roth}, {Rutledge}, {Sheppard}, \&
  {Songaila}}]{2007ApJ...664.1000B}
{Berger}, E., {Fox}, D.~B., {Price}, P.~A., {et~al.} 2007, \apj, 664, 1000,
  \dodoi{10.1086/518762}

\bibitem[{{Bhandari} {et~al.}(2020{\natexlab{a}}){Bhandari}, {Bannister},
  {Lenc}, {Cho}, {Ekers}, {Day}, {Deller}, {Flynn}, {James}, {Macquart},
  {Mahony}, {Marnoch}, {Moss}, {Phillips}, {Prochaska}, {Qiu}, {Ryder},
  {Shannon}, {Tejos}, \& {Wong}}]{2020ApJ...901L..20B}
{Bhandari}, S., {Bannister}, K.~W., {Lenc}, E., {et~al.} 2020{\natexlab{a}},
  \apjl, 901, L20, \dodoi{10.3847/2041-8213/abb462}

\bibitem[{{Bhandari} {et~al.}(2020{\natexlab{b}}){Bhandari}, {Sadler},
  {Prochaska}, {Simha}, {Ryder}, {Marnoch}, {Bannister}, {Macquart}, {Flynn},
  {Shannon}, {Tejos}, {Corro-Guerra}, {Day}, {Deller}, {Ekers}, {Lopez},
  {Mahony}, {Nu{\~n}ez}, \& {Phillips}}]{2020ApJ...895L..37B}
{Bhandari}, S., {Sadler}, E.~M., {Prochaska}, J.~X., {et~al.}
  2020{\natexlab{b}}, \apjl, 895, L37, \dodoi{10.3847/2041-8213/ab672e}

\bibitem[{{Bhandari} {et~al.}(2022){Bhandari}, {Heintz}, {Aggarwal}, {Marnoch},
  {Day}, {Sydnor}, {Burke-Spolaor}, {Law}, {Xavier Prochaska}, {Tejos},
  {Bannister}, {Butler}, {Deller}, {Ekers}, {Flynn}, {Fong}, {James}, {Lazio},
  {Luo}, {Mahony}, {Ryder}, {Sadler}, {Shannon}, {Han}, {Lee}, \&
  {Zhang}}]{2022AJ....163...69B}
{Bhandari}, S., {Heintz}, K.~E., {Aggarwal}, K., {et~al.} 2022, \aj, 163, 69,
  \dodoi{10.3847/1538-3881/ac3aec}

\bibitem[{{Bhardwaj} {et~al.}(2021{\natexlab{a}}){Bhardwaj}, {Gaensler},
  {Kaspi}, {Landecker}, {Mckinven}, {Michilli}, {Pleunis}, {Tendulkar},
  {Andersen}, {Boyle}, {Cassanelli}, {Chawla}, {Cook}, {Dobbs}, {Fonseca},
  {Kaczmarek}, {Leung}, {Masui}, {Mnchmeyer}, {Ng}, {Rafiei-Ravandi}, {Scholz},
  {Shin}, {Smith}, {Stairs}, \& {Zwaniga}}]{2021ApJ...910L..18B}
{Bhardwaj}, M., {Gaensler}, B.~M., {Kaspi}, V.~M., {et~al.} 2021{\natexlab{a}},
  \apjl, 910, L18, \dodoi{10.3847/2041-8213/abeaa6}

\bibitem[{{Bhardwaj} {et~al.}(2021{\natexlab{b}}){Bhardwaj}, {Kirichenko},
  {Michilli}, {Mayya}, {Kaspi}, {Gaensler}, {Rahman}, {Tendulkar}, {Fonseca},
  {Josephy}, {Leung}, {Merryfield}, {Petroff}, {Pleunis}, {Sanghavi}, {Scholz},
  {Shin}, {Smith}, \& {Stairs}}]{2021ApJ...919L..24B}
{Bhardwaj}, M., {Kirichenko}, A.~Y., {Michilli}, D., {et~al.}
  2021{\natexlab{b}}, \apjl, 919, L24, \dodoi{10.3847/2041-8213/ac223b}

\bibitem[{{Bloom} {et~al.}(2002){Bloom}, {Kulkarni}, \&
  {Djorgovski}}]{2002AJ....123.1111B}
{Bloom}, J.~S., {Kulkarni}, S.~R., \& {Djorgovski}, S.~G. 2002, \aj, 123, 1111,
  \dodoi{10.1086/338893}

\bibitem[{{Bloom} {et~al.}(2006{\natexlab{a}}){Bloom}, {Perley}, {Kocevski},
  {Butler}, {Prochaska}, \& {Chen}}]{2006GCN..5238....1B}
{Bloom}, J.~S., {Perley}, D., {Kocevski}, D., {et~al.} 2006{\natexlab{a}}, GRB
  Coordinates Network, 5238, 1

\bibitem[{{Bloom} {et~al.}(2006{\natexlab{b}}){Bloom}, {Prochaska}, {Pooley},
  {Blake}, {Foley}, {Jha}, {Ramirez-Ruiz}, {Granot}, {Filippenko},
  {Sigurdsson}, {Barth}, {Chen}, {Cooper}, {Falco}, {Gal}, {Gerke}, {Gladders},
  {Greene}, {Hennanwi}, {Ho}, {Hurley}, {Koester}, {Li}, {Lubin}, {Newman},
  {Perley}, {Squires}, \& {Wood-Vasey}}]{2006ApJ...638..354B}
{Bloom}, J.~S., {Prochaska}, J.~X., {Pooley}, D., {et~al.} 2006{\natexlab{b}},
  \apj, 638, 354, \dodoi{10.1086/498107}

\bibitem[{{Bloom} {et~al.}(2007){Bloom}, {Perley}, {Chen}, {Butler},
  {Prochaska}, {Kocevski}, {Blake}, {Szentgyorgyi}, {Falco}, \&
  {Starr}}]{2007ApJ...654..878B}
{Bloom}, J.~S., {Perley}, D.~A., {Chen}, H.~W., {et~al.} 2007, \apj, 654, 878,
  \dodoi{10.1086/509114}

\bibitem[{{Bochenek} {et~al.}(2020){Bochenek}, {Ravi}, {Belov}, {Hallinan},
  {Kocz}, {Kulkarni}, \& {McKenna}}]{2020Natur.587...59B}
{Bochenek}, C.~D., {Ravi}, V., {Belov}, K.~V., {et~al.} 2020, \nat, 587, 59,
  \dodoi{10.1038/s41586-020-2872-x}

\bibitem[{{Bochenek} {et~al.}(2021){Bochenek}, {Ravi}, \&
  {Dong}}]{2021ApJ...907L..31B}
{Bochenek}, C.~D., {Ravi}, V., \& {Dong}, D. 2021, \apjl, 907, L31,
  \dodoi{10.3847/2041-8213/abd634}

\bibitem[{{Briggs} {et~al.}(1996){Briggs}, {Paciesas}, {Pendleton}, {Meegan},
  {Fishman}, {Horack}, {Brock}, {Kouveliotou}, {Hartmann}, \&
  {Hakkila}}]{1996ApJ...459...40B}
{Briggs}, M.~S., {Paciesas}, W.~S., {Pendleton}, G.~N., {et~al.} 1996, \apj,
  459, 40, \dodoi{10.1086/176867}

\bibitem[{{Butler} {et~al.}(2005){Butler}, {Ricker}, {Atteia}, {Kawai}, {Lamb},
  {Woosley}, {Arimoto}, {Donaghy}, {Fenimore}, {Galassi}, {Graziani}, {Kotoku},
  {Maetou}, {Matsuoka}, {Nakagawa}, {Sakamoto}, {Sato}, {Shirasaki}, {Suzuki},
  {Tamagawa}, {Tanaka}, {Yamamoto}, {Yoshida}, {Crew}, {Doty}, {Prigozhin},
  {Vanderspek}, {Villasenor}, {Jernigan}, {Levine}, {Azzibrouck}, {Braga},
  {Manchanda}, {Pizzichini}, {Boer}, {Olive}, {Dezalay}, \&
  {Hurley}}]{2005GCN..3570....1B}
{Butler}, N., {Ricker}, G., {Atteia}, J.~L., {et~al.} 2005, GRB Coordinates
  Network, 3570, 1

\bibitem[{{Caleb} {et~al.}(2016){Caleb}, {Flynn}, {Bailes}, {Barr}, {Hunstead},
  {Keane}, {Ravi}, \& {van Straten}}]{2016MNRAS.458..708C}
{Caleb}, M., {Flynn}, C., {Bailes}, M., {et~al.} 2016, \mnras, 458, 708,
  \dodoi{10.1093/mnras/stw175}

\bibitem[{{Cannizzo} {et~al.}(2006){Cannizzo}, {Barbier}, {Barthelmy},
  {Gehrels}, {Markwardt}, {Osborne}, {Page}, {Palmer}, \&
  {Stamatikos}}]{2006GCN..5904....1C}
{Cannizzo}, K.~C., {Barbier}, L.~M., {Barthelmy}, S.~D., {et~al.} 2006, GRB
  Coordinates Network, 5904, 1

\bibitem[{{Casentini} {et~al.}(2020){Casentini}, {Verrecchia}, {Tavani},
  {Ursi}, {Antonelli}, {Argan}, {Barbiellini}, {Bulgarelli}, {Caraveo},
  {Cardillo}, {Cattaneo}, {Chen}, {Costa}, {Donnarumma}, {Feroci}, {Ferrari},
  {Fuschino}, {Galli}, {Giuliani}, {Labanti}, {Lazzarotto}, {Lipari}, {Longo},
  {Lucarelli}, {Marisaldi}, {Morselli}, {Paoletti}, {Parmiggiani},
  {Pellizzoni}, {Piano}, {Pilia}, {Pittori}, \&
  {Vercellone}}]{2020ApJ...890L..32C}
{Casentini}, C., {Verrecchia}, F., {Tavani}, M., {et~al.} 2020, \apjl, 890,
  L32, \dodoi{10.3847/2041-8213/ab720a}

\bibitem[{{Champion} {et~al.}(2016){Champion}, {Petroff}, {Kramer}, {Keith},
  {Bailes}, {Barr}, {Bates}, {Bhat}, {Burgay}, {Burke-Spolaor}, {Flynn},
  {Jameson}, {Johnston}, {Ng}, {Levin}, {Possenti}, {Stappers}, {van Straten},
  {Thornton}, {Tiburzi}, \& {Lyne}}]{2016MNRAS.460L..30C}
{Champion}, D.~J., {Petroff}, E., {Kramer}, M., {et~al.} 2016, \mnras, 460,
  L30, \dodoi{10.1093/mnrasl/slw069}

\bibitem[{{Chatterjee} {et~al.}(2017){Chatterjee}, {Law}, {Wharton},
  {Burke-Spolaor}, {Hessels}, {Bower}, {Cordes}, {Tendulkar}, {Bassa},
  {Demorest}, {Butler}, {Seymour}, {Scholz}, {Abruzzo}, {Bogdanov}, {Kaspi},
  {Keimpema}, {Lazio}, {Marcote}, {McLaughlin}, {Paragi}, {Ransom}, {Rupen},
  {Spitler}, \& {van Langevelde}}]{2017Natur.541...58C}
{Chatterjee}, S., {Law}, C.~J., {Wharton}, R.~S., {et~al.} 2017, \nat, 541, 58,
  \dodoi{10.1038/nature20797}

\bibitem[{{Chen} {et~al.}(2021){Chen}, {Wang}, \& {Tong}}]{2021JHEAp..31....1C}
{Chen}, X., {Wang}, W., \& {Tong}, H. 2021, Journal of High Energy
  Astrophysics, 31, 1, \dodoi{10.1016/j.jheap.2021.04.002}

\bibitem[{{Cheng} {et~al.}(2020){Cheng}, {Zhang}, \&
  {Wang}}]{2020MNRAS.491.1498C}
{Cheng}, Y., {Zhang}, G.~Q., \& {Wang}, F.~Y. 2020, \mnras, 491, 1498,
  \dodoi{10.1093/mnras/stz3085}

\bibitem[{{CHIME/FRB Collaboration} {et~al.}(2019){CHIME/FRB Collaboration},
  {Andersen}, {Bandura}, {Bhardwaj}, {Boubel}, {Boyce}, {Boyle}, {Brar},
  {Cassanelli}, {Chawla}, {Cubranic}, {Deng}, {Dobbs}, {Fandino}, {Fonseca},
  {Gaensler}, {Gilbert}, {Giri}, {Good}, {Halpern}, {Hill}, {Hinshaw},
  {H{\"o}fer}, {Josephy}, {Kaspi}, {Kothes}, {Landecker}, {Lang}, {Li}, {Lin},
  {Masui}, {Mena-Parra}, {Merryfield}, {Mckinven}, {Michilli}, {Milutinovic},
  {Naidu}, {Newburgh}, {Ng}, {Patel}, {Pen}, {Pinsonneault-Marotte}, {Pleunis},
  {Rafiei-Ravandi}, {Rahman}, {Ransom}, {Renard}, {Scholz}, {Siegel}, {Singh},
  {Smith}, {Stairs}, {Tendulkar}, {Tretyakov}, {Vanderlinde}, {Yadav}, \&
  {Zwaniga}}]{2019ApJ...885L..24C}
{CHIME/FRB Collaboration}, {Andersen}, B.~C., {Bandura}, K., {et~al.} 2019,
  \apjl, 885, L24, \dodoi{10.3847/2041-8213/ab4a80}

\bibitem[{{CHIME/FRB Collaboration} {et~al.}(2020){CHIME/FRB Collaboration},
  {Andersen}, {Bandura}, {Bhardwaj}, {Bij}, {Boyce}, {Boyle}, {Brar},
  {Cassanelli}, {Chawla}, {Chen}, {Cliche}, {Cook}, {Cubranic}, {Curtin},
  {Denman}, {Dobbs}, {Dong}, {Fandino}, {Fonseca}, {Gaensler}, {Giri}, {Good},
  {Halpern}, {Hill}, {Hinshaw}, {H{\"o}fer}, {Josephy}, {Kania}, {Kaspi},
  {Landecker}, {Leung}, {Li}, {Lin}, {Masui}, {McKinven}, {Mena-Parra},
  {Merryfield}, {Meyers}, {Michilli}, {Milutinovic}, {Mirhosseini},
  {M{\"u}nchmeyer}, {Naidu}, {Newburgh}, {Ng}, {Patel}, {Pen},
  {Pinsonneault-Marotte}, {Pleunis}, {Quine}, {Rafiei-Ravandi}, {Rahman},
  {Ransom}, {Renard}, {Sanghavi}, {Scholz}, {Shaw}, {Shin}, {Siegel}, {Singh},
  {Smegal}, {Smith}, {Stairs}, {Tan}, {Tendulkar}, {Tretyakov}, {Vanderlinde},
  {Wang}, {Wulf}, \& {Zwaniga}}]{2020Natur.587...54C}
{CHIME/FRB Collaboration}, {Andersen}, B.~C., {Bandura}, K.~M., {et~al.} 2020,
  \nat, 587, 54, \dodoi{10.1038/s41586-020-2863-y}

\bibitem[{{Chittidi} {et~al.}(2021){Chittidi}, {Simha}, {Mannings},
  {Prochaska}, {Ryder}, {Rafelski}, {Neeleman}, {Macquart}, {Tejos},
  {Jorgenson}, {Day}, {Marnoch}, {Bhandari}, {Deller}, {Qiu}, {Bannister},
  {Shannon}, \& {Heintz}}]{2021ApJ...922..173C}
{Chittidi}, J.~S., {Simha}, S., {Mannings}, A., {et~al.} 2021, \apj, 922, 173,
  \dodoi{10.3847/1538-4357/ac2818}

\bibitem[{{Church} {et~al.}(2011){Church}, {Levan}, {Davies}, \&
  {Tanvir}}]{2011MNRAS.413.2004C}
{Church}, R.~P., {Levan}, A.~J., {Davies}, M.~B., \& {Tanvir}, N. 2011, \mnras,
  413, 2004, \dodoi{10.1111/j.1365-2966.2011.18277.x}

\bibitem[{{Church} {et~al.}(2012){Church}, {Levan}, {Davies}, \&
  {Tanvir}}]{2012MSAIS..21..104C}
---. 2012, Memorie della Societa Astronomica Italiana Supplementi, 21, 104.
\newblock \doarXiv{1110.4209}

\bibitem[{{Colpi} {et~al.}(2000){Colpi}, {Geppert}, \&
  {Page}}]{2000ApJ...529L..29C}
{Colpi}, M., {Geppert}, U., \& {Page}, D. 2000, \apjl, 529, L29,
  \dodoi{10.1086/312448}

\bibitem[{{Cook} {et~al.}(2023){Cook}, {Bhardwaj}, {Gaensler}, {Scholz},
  {Eadie}, {Hill}, {Kaspi}, {Masui}, {Curtin}, {Dong}, {Fonseca},
  {Herrera-Martin}, {Kaczmarek}, {Lanman}, {Lazda}, {Leung}, {Meyers},
  {Michilli}, {Pandhi}, {Pearlman}, {Pleunis}, {Ransom}, {Rahman}, {Sand},
  {Shin}, {Smith}, {Stairs}, \& {Stenning}}]{2023ApJ...946...58C}
{Cook}, A.~M., {Bhardwaj}, M., {Gaensler}, B.~M., {et~al.} 2023, \apj, 946, 58,
  \dodoi{10.3847/1538-4357/acbbd0}

\bibitem[{{Cook} {et~al.}(1994){Cook}, {Shapiro}, \&
  {Teukolsky}}]{1994ApJ...424..823C}
{Cook}, G.~B., {Shapiro}, S.~L., \& {Teukolsky}, S.~A. 1994, \apj, 424, 823,
  \dodoi{10.1086/173934}

\bibitem[{{Cordes} \& {Chatterjee}(2019)}]{2019ARA&A..57..417C}
{Cordes}, J.~M., \& {Chatterjee}, S. 2019, \araa, 57, 417,
  \dodoi{10.1146/annurev-astro-091918-104501}

\bibitem[{{Cordes} \& {Lazio}(2002)}]{2002astro.ph..7156C}
{Cordes}, J.~M., \& {Lazio}, T.~J.~W. 2002, arXiv e-prints, astro.
\newblock \doarXiv{astro-ph/0207156}

\bibitem[{{Cordes} {et~al.}(2021){Cordes}, {Ocker}, \&
  {Chatterjee}}]{2021arXiv210801172C}
{Cordes}, J.~M., {Ocker}, S.~K., \& {Chatterjee}, S. 2021, arXiv e-prints,
  arXiv:2108.01172.
\newblock \doarXiv{2108.01172}

\bibitem[{{Cunningham} {et~al.}(2020){Cunningham}, {Cenko}, {Burns},
  {Goldstein}, {Lien}, \& {Kocevski}}]{2020AAS...23543903C}
{Cunningham}, V., {Cenko}, S., {Burns}, E., {et~al.} 2020, in American
  Astronomical Society Meeting Abstracts, Vol. 235, American Astronomical
  Society Meeting Abstracts \#235, 439.03

\bibitem[{{Cunningham} {et~al.}(2019){Cunningham}, {Cenko}, {Burns},
  {Goldstein}, {Lien}, {Kocevski}, {Briggs}, {Connaughton}, {Miller},
  {Racusin}, \& {Stanbro}}]{2019ApJ...879...40C}
{Cunningham}, V., {Cenko}, S.~B., {Burns}, E., {et~al.} 2019, \apj, 879, 40,
  \dodoi{10.3847/1538-4357/ab2235}

\bibitem[{{Cunningham} \& {Cenko}(2018)}]{2018AAS...23124308C}
{Cunningham}, V.~A., \& {Cenko}, B. 2018, in American Astronomical Society
  Meeting Abstracts, Vol. 231, American Astronomical Society Meeting Abstracts
  \#231, 243.08

\bibitem[{{Curtin} {et~al.}(2022){Curtin}, {Tendulkar}, {Josephy}, {Chawla},
  {Andersen}, {Kaspi}, {Bhardwaj}, {Cassanelli}, {Cook}, {Dong}, {Fonseca},
  {Gaensler}, {Kaczmarek}, {Lanmnan}, {Leung}, {Pearlman}, {Petroff},
  {Pleunis}, {Rafiei-Ravandi}, {Ransom}, {Shin}, {Scholz}, {Smith}, \&
  {Stairs}}]{2022arXiv220800803C}
{Curtin}, A.~P., {Tendulkar}, S.~P., {Josephy}, A., {et~al.} 2022, arXiv
  e-prints, arXiv:2208.00803.
\newblock \doarXiv{2208.00803}

\bibitem[{{Dai} {et~al.}(2016){Dai}, {Wang}, {Wu}, \&
  {Huang}}]{2016ApJ...829...27D}
{Dai}, Z.~G., {Wang}, J.~S., {Wu}, X.~F., \& {Huang}, Y.~F. 2016, \apj, 829,
  27, \dodoi{10.3847/0004-637X/829/1/27}

\bibitem[{{Day} {et~al.}(2021){Day}, {Deller}, {James}, {Lenc}, {Bhandari},
  {Shannon}, \& {Bannister}}]{2021PASA...38...50D}
{Day}, C.~K., {Deller}, A.~T., {James}, C.~W., {et~al.} 2021, \pasa, 38, e050,
  \dodoi{10.1017/pasa.2021.40}

\bibitem[{{Dehman} {et~al.}(2020){Dehman}, {Vigan{\`o}}, {Rea}, {Pons},
  {Perna}, \& {Garcia-Garcia}}]{2020ApJ...902L..32D}
{Dehman}, C., {Vigan{\`o}}, D., {Rea}, N., {et~al.} 2020, \apjl, 902, L32,
  \dodoi{10.3847/2041-8213/abbda9}

\bibitem[{{DeLaunay} {et~al.}(2016){DeLaunay}, {Fox}, {Murase},
  {M{\'e}sz{\'a}ros}, {Keivani}, {Messick}, {Mostaf{\'a}}, {Oikonomou},
  {Te{\v{s}}i{\'c}}, \& {Turley}}]{2016ApJ...832L...1D}
{DeLaunay}, J.~J., {Fox}, D.~B., {Murase}, K., {et~al.} 2016, \apjl, 832, L1,
  \dodoi{10.3847/2041-8205/832/1/L1}

\bibitem[{{D'Elia} {et~al.}(2011){D'Elia}, {Barthelmy}, {Beardmore}, {Burrows},
  {Campana}, {Chester}, {Gehrels}, {Gendre}, {Guidorzi}, {Holland}, {Kennea},
  {Krimm}, {Kuin}, {Mangano}, {Markwardt}, {Marshall}, {O'Brien}, {Osborne},
  {Page}, {Palmer}, {Romano}, {Sakamoto}, {Sbarufatti}, {Siegel}, {Stamatikos},
  {Starling}, {Stratta}, {Swenson}, {Troja}, {Ukwatta}, \&
  {Zhang}}]{2011GCN.12578....1D}
{D'Elia}, V., {Barthelmy}, S.~D., {Beardmore}, A.~P., {et~al.} 2011, GRB
  Coordinates Network, 12578, 1

\bibitem[{{Deng} {et~al.}(2019){Deng}, {Wei}, \& {Wu}}]{2019JHEAp..23....1D}
{Deng}, C.-M., {Wei}, J.-J., \& {Wu}, X.-F. 2019, Journal of High Energy
  Astrophysics, 23, 1, \dodoi{10.1016/j.jheap.2019.05.001}

\bibitem[{{Deng} {et~al.}(2021){Deng}, {Zhong}, \& {Dai}}]{2021ApJ...922...98D}
{Deng}, C.-M., {Zhong}, S.-Q., \& {Dai}, Z.-G. 2021, \apj, 922, 98,
  \dodoi{10.3847/1538-4357/ac30db}

\bibitem[{{Deng} \& {Zhang}(2014)}]{2014ApJ...783L..35D}
{Deng}, W., \& {Zhang}, B. 2014, \apjl, 783, L35,
  \dodoi{10.1088/2041-8205/783/2/L35}

\bibitem[{{Dessenne} {et~al.}(1996){Dessenne}, {Green}, {Warner},
  {Titterington}, {Waldram}, {Barthelmy}, {Butterworth}, {Cline}, {Gehrels},
  {Palmer}, {Fishman}, {Kouveliotou}, \& {Meegan}}]{1996MNRAS.281..977D}
{Dessenne}, C.~A.~C., {Green}, D.~A., {Warner}, P.~J., {et~al.} 1996, \mnras,
  281, 977, \dodoi{10.1093/mnras/281.3.977}

\bibitem[{{Du}(2020)}]{2020ApJ...901...75D}
{Du}, S. 2020, \apj, 901, 75, \dodoi{10.3847/1538-4357/abaf4d}

\bibitem[{{Falcke} \& {Rezzolla}(2014)}]{2014A&A...562A.137F}
{Falcke}, H., \& {Rezzolla}, L. 2014, \aap, 562, A137,
  \dodoi{10.1051/0004-6361/201321996}

\bibitem[{{Faucher-Gigu{\`e}re} {et~al.}(2011){Faucher-Gigu{\`e}re},
  {Kere{\v{s}}}, \& {Ma}}]{2011MNRAS.417.2982F}
{Faucher-Gigu{\`e}re}, C.-A., {Kere{\v{s}}}, D., \& {Ma}, C.-P. 2011, \mnras,
  417, 2982, \dodoi{10.1111/j.1365-2966.2011.19457.x}

\bibitem[{{Fermi Gbm Team} \& {Likely Short Grb}(2021)}]{2021GCN.30248....1F}
{Fermi Gbm Team}, \& {Likely Short Grb}, T. D. O.~A. 2021, GRB Coordinates
  Network, 30248, 1

\bibitem[{{Fletcher} \& {Fermi-GBM Team}(2021)}]{2021GCN.29536....1F}
{Fletcher}, C., \& {Fermi-GBM Team}. 2021, GRB Coordinates Network, 29536, 1

\bibitem[{{Fong} \& {Berger}(2013)}]{2013ApJ...776...18F}
{Fong}, W., \& {Berger}, E. 2013, \apj, 776, 18,
  \dodoi{10.1088/0004-637X/776/1/18}

\bibitem[{{Fong} {et~al.}(2010){Fong}, {Berger}, \&
  {Fox}}]{2010ApJ...708....9F}
{Fong}, W., {Berger}, E., \& {Fox}, D.~B. 2010, \apj, 708, 9,
  \dodoi{10.1088/0004-637X/708/1/9}

\bibitem[{{Fong} {et~al.}(2013){Fong}, {Berger}, {Chornock}, {Margutti},
  {Levan}, {Tanvir}, {Tunnicliffe}, {Czekala}, {Fox}, {Perley}, {Cenko},
  {Zauderer}, {Laskar}, {Persson}, {Monson}, {Kelson}, {Birk}, {Murphy},
  {Servillat}, \& {Anglada}}]{2013ApJ...769...56F}
{Fong}, W., {Berger}, E., {Chornock}, R., {et~al.} 2013, \apj, 769, 56,
  \dodoi{10.1088/0004-637X/769/1/56}

\bibitem[{{Fox} {et~al.}(2005){Fox}, {Frail}, {Price}, {Kulkarni}, {Berger},
  {Piran}, {Soderberg}, {Cenko}, {Cameron}, {Gal-Yam}, {Kasliwal}, {Moon},
  {Harrison}, {Nakar}, {Schmidt}, {Penprase}, {Chevalier}, {Kumar}, {Roth},
  {Watson}, {Lee}, {Shectman}, {Phillips}, {Roth}, {McCarthy}, {Rauch},
  {Cowie}, {Peterson}, {Rich}, {Kawai}, {Aoki}, {Kosugi}, {Totani}, {Park},
  {MacFadyen}, \& {Hurley}}]{2005Natur.437..845F}
{Fox}, D.~B., {Frail}, D.~A., {Price}, P.~A., {et~al.} 2005, \nat, 437, 845,
  \dodoi{10.1038/nature04189}

\bibitem[{{Fukugita} {et~al.}(1998){Fukugita}, {Hogan}, \&
  {Peebles}}]{1998ApJ...503..518F}
{Fukugita}, M., {Hogan}, C.~J., \& {Peebles}, P.~J.~E. 1998, \apj, 503, 518,
  \dodoi{10.1086/306025}

\bibitem[{{Gao} {et~al.}(2016){Gao}, {Zhang}, \&
  {L{\"u}}}]{2016PhRvD..93d4065G}
{Gao}, H., {Zhang}, B., \& {L{\"u}}, H.-J. 2016, \prd, 93, 044065,
  \dodoi{10.1103/PhysRevD.93.044065}

\bibitem[{{Gehrels} {et~al.}(2005){Gehrels}, {Sarazin}, {O'Brien}, {Zhang},
  {Barbier}, {Barthelmy}, {Blustin}, {Burrows}, {Cannizzo}, {Cummings}, {Goad},
  {Holland}, {Hurkett}, {Kennea}, {Levan}, {Markwardt}, {Mason}, {Meszaros},
  {Page}, {Palmer}, {Rol}, {Sakamoto}, {Willingale}, {Angelini}, {Beardmore},
  {Boyd}, {Breeveld}, {Campana}, {Chester}, {Chincarini}, {Cominsky},
  {Cusumano}, {de Pasquale}, {Fenimore}, {Giommi}, {Gronwall}, {Grupe}, {Hill},
  {Hinshaw}, {Hjorth}, {Hullinger}, {Hurley}, {Klose}, {Kobayashi},
  {Kouveliotou}, {Krimm}, {Mangano}, {Marshall}, {McGowan}, {Moretti},
  {Mushotzky}, {Nakazawa}, {Norris}, {Nousek}, {Osborne}, {Page}, {Parsons},
  {Patel}, {Perri}, {Poole}, {Romano}, {Roming}, {Rosen}, {Sato}, {Schady},
  {Smale}, {Sollerman}, {Starling}, {Still}, {Suzuki}, {Tagliaferri},
  {Takahashi}, {Tashiro}, {Tueller}, {Wells}, {White}, \&
  {Wijers}}]{2005Natur.437..851G}
{Gehrels}, N., {Sarazin}, C.~L., {O'Brien}, P.~T., {et~al.} 2005, \nat, 437,
  851, \dodoi{10.1038/nature04142}

\bibitem[{{Geng} {et~al.}(2021){Geng}, {Li}, \& {Huang}}]{2021Innov...200152G}
{Geng}, J., {Li}, B., \& {Huang}, Y. 2021, The Innovation, 2, 100152,
  \dodoi{10.1016/j.xinn.2021.100152}

\bibitem[{{Gohar} \& {Flynn}(2022)}]{2022MNRAS.509.5265G}
{Gohar}, N., \& {Flynn}, C. 2022, \mnras, 509, 5265,
  \dodoi{10.1093/mnras/stab3349}

\bibitem[{{Goldstein} {et~al.}(2017){Goldstein}, {Veres}, {Burns}, {Briggs},
  {Hamburg}, {Kocevski}, {Wilson-Hodge}, {Preece}, {Poolakkil}, {Roberts},
  {Hui}, {Connaughton}, {Racusin}, {von Kienlin}, {Dal Canton}, {Christensen},
  {Littenberg}, {Siellez}, {Blackburn}, {Broida}, {Bissaldi}, {Cleveland},
  {Gibby}, {Giles}, {Kippen}, {McBreen}, {McEnery}, {Meegan}, {Paciesas}, \&
  {Stanbro}}]{2017ApJ...848L..14G}
{Goldstein}, A., {Veres}, P., {Burns}, E., {et~al.} 2017, \apjl, 848, L14,
  \dodoi{10.3847/2041-8213/aa8f41}

\bibitem[{{Golenetskii} {et~al.}(2008){Golenetskii}, {Aptekar}, {Mazets},
  {Pal'Shin}, {Frederiks}, {Cline}, {Cummings}, {Barthelmy}, {Gehrels}, \&
  {Krimm}}]{2008GCN..8676....1G}
{Golenetskii}, S., {Aptekar}, R., {Mazets}, E., {et~al.} 2008, GRB Coordinates
  Network, 8676, 1

\bibitem[{{Gourdji} {et~al.}(2020){Gourdji}, {Rowlinson}, {Wijers}, \&
  {Goldstein}}]{2020MNRAS.497.3131G}
{Gourdji}, K., {Rowlinson}, A., {Wijers}, R.~A.~M.~J., \& {Goldstein}, A. 2020,
  \mnras, 497, 3131, \dodoi{10.1093/mnras/staa2128}

\bibitem[{{Grupe} {et~al.}(2009){Grupe}, {Cummings}, {Gronwall}, {Guidorzi},
  {Markwardt}, {O'Brien}, {Page}, {Palmer}, {Romano}, {Siegel}, {Ukwatta}, \&
  {Vetere}}]{2009GCN..9945....1G}
{Grupe}, D., {Cummings}, J.~R., {Gronwall}, C., {et~al.} 2009, GRB Coordinates
  Network, 9945, 1

\bibitem[{{Hafen} {et~al.}(2019){Hafen}, {Faucher-Gigu{\`e}re},
  {Angl{\'e}s-Alc{\'a}zar}, {Stern}, {Kere{\v{s}}}, {Hummels}, {Esmerian},
  {Garrison-Kimmel}, {El-Badry}, {Wetzel}, {Chan}, {Hopkins}, \&
  {Murray}}]{2019MNRAS.488.1248H}
{Hafen}, Z., {Faucher-Gigu{\`e}re}, C.-A., {Angl{\'e}s-Alc{\'a}zar}, D.,
  {et~al.} 2019, \mnras, 488, 1248, \dodoi{10.1093/mnras/stz1773}

\bibitem[{{Hallinan} {et~al.}(2019){Hallinan}, {Dong}, \&
  {Ravi}}]{2019ATel13018....1H}
{Hallinan}, G., {Dong}, D., \& {Ravi}, V. 2019, The Astronomer's Telegram,
  13018, 1

\bibitem[{{Halpern} \& {Mirabal}(2006{\natexlab{a}})}]{2006GCN..5072....1H}
{Halpern}, J.~P., \& {Mirabal}, N. 2006{\natexlab{a}}, GRB Coordinates Network,
  5072, 1

\bibitem[{{Halpern} \& {Mirabal}(2006{\natexlab{b}})}]{2006GCN..5066....1H}
---. 2006{\natexlab{b}}, GRB Coordinates Network, 5066, 1

\bibitem[{{Hardy} {et~al.}(2017){Hardy}, {Dhillon}, {Spitler}, {Littlefair},
  {Ashley}, {De Cia}, {Green}, {Jaroenjittichai}, {Keane}, {Kerry}, {Kramer},
  {Malesani}, {Marsh}, {Parsons}, {Possenti}, {Rattanasoon}, \&
  {Sahman}}]{2017MNRAS.472.2800H}
{Hardy}, L.~K., {Dhillon}, V.~S., {Spitler}, L.~G., {et~al.} 2017, \mnras, 472,
  2800, \dodoi{10.1093/mnras/stx2153}

\bibitem[{{Heintz} {et~al.}(2020){Heintz}, {Prochaska}, {Simha}, {Platts},
  {Fong}, {Tejos}, {Ryder}, {Aggerwal}, {Bhandari}, {Day}, {Deller},
  {Kilpatrick}, {Law}, {Macquart}, {Mannings}, {Marnoch}, {Sadler}, \&
  {Shannon}}]{2020ApJ...903..152H}
{Heintz}, K.~E., {Prochaska}, J.~X., {Simha}, S., {et~al.} 2020, \apj, 903,
  152, \dodoi{10.3847/1538-4357/abb6fb}

\bibitem[{{Hut} {et~al.}(1992){Hut}, {McMillan}, {Goodman}, {Mateo}, {Phinney},
  {Pryor}, {Richer}, {Verbunt}, \& {Weinberg}}]{1992PASP..104..981H}
{Hut}, P., {McMillan}, S., {Goodman}, J., {et~al.} 1992, \pasp, 104, 981,
  \dodoi{10.1086/133085}

\bibitem[{{Ioka} \& {Zhang}(2020)}]{2020ApJ...893L..26I}
{Ioka}, K., \& {Zhang}, B. 2020, \apjl, 893, L26,
  \dodoi{10.3847/2041-8213/ab83fb}

\bibitem[{{Israel} {et~al.}(2016){Israel}, {Esposito}, {Rea}, {Coti Zelati},
  {Tiengo}, {Campana}, {Mereghetti}, {Rodriguez Castillo}, {G{\"o}tz},
  {Burgay}, {Possenti}, {Zane}, {Turolla}, {Perna}, {Cannizzaro}, \&
  {Pons}}]{2016MNRAS.457.3448I}
{Israel}, G.~L., {Esposito}, P., {Rea}, N., {et~al.} 2016, \mnras, 457, 3448,
  \dodoi{10.1093/mnras/stw008}

\bibitem[{{James} {et~al.}(2019){James}, {Anderson}, {Wen}, {Bosveld}, {Chu},
  {Kovalam}, {Slaven-Blair}, \& {Williams}}]{2019MNRAS.489L..75J}
{James}, C.~W., {Anderson}, G.~E., {Wen}, L., {et~al.} 2019, \mnras, 489, L75,
  \dodoi{10.1093/mnrasl/slz129}

\bibitem[{{Kann} {et~al.}(2006){Kann}, {Klose}, \&
  {Ferrero}}]{2006GCN..5062....1K}
{Kann}, D.~A., {Klose}, S., \& {Ferrero}, P. 2006, GRB Coordinates Network,
  5062, 1

\bibitem[{{Kaplan} {et~al.}(2015){Kaplan}, {Rowlinson}, {Bannister}, {Bell},
  {Croft}, {Murphy}, {Tingay}, {Wayth}, \& {Williams}}]{2015ApJ...814L..25K}
{Kaplan}, D.~L., {Rowlinson}, A., {Bannister}, K.~W., {et~al.} 2015, \apjl,
  814, L25, \dodoi{10.1088/2041-8205/814/2/L25}

\bibitem[{{Kashiyama} {et~al.}(2013){Kashiyama}, {Ioka}, \&
  {M{\'e}sz{\'a}ros}}]{2013ApJ...776L..39K}
{Kashiyama}, K., {Ioka}, K., \& {M{\'e}sz{\'a}ros}, P. 2013, \apjl, 776, L39,
  \dodoi{10.1088/2041-8205/776/2/L39}

\bibitem[{{Katz}(2016)}]{2016ApJ...826..226K}
{Katz}, J.~I. 2016, \apj, 826, 226, \dodoi{10.3847/0004-637X/826/2/226}

\bibitem[{{Keating} \& {Pen}(2020)}]{2020MNRAS.496L.106K}
{Keating}, L.~C., \& {Pen}, U.-L. 2020, \mnras, 496, L106,
  \dodoi{10.1093/mnrasl/slaa095}

\bibitem[{{Kirsten} {et~al.}(2022){Kirsten}, {Marcote}, {Nimmo}, {Hessels},
  {Bhardwaj}, {Tendulkar}, {Keimpema}, {Yang}, {Snelders}, {Scholz},
  {Pearlman}, {Law}, {Peters}, {Giroletti}, {Paragi}, {Bassa}, {Hewitt},
  {Bach}, {Bezrukovs}, {Burgay}, {Buttaccio}, {Conway}, {Corongiu}, {Feiler},
  {Forss{\'e}n}, {Gawro{\'n}ski}, {Karuppusamy}, {Kharinov}, {Lindqvist},
  {Maccaferri}, {Melnikov}, {Ould-Boukattine}, {Possenti}, {Surcis}, {Wang},
  {Yuan}, {Aggarwal}, {Anna-Thomas}, {Bower}, {Blaauw}, {Burke-Spolaor},
  {Cassanelli}, {Clarke}, {Fonseca}, {Gaensler}, {Gopinath}, {Kaspi}, {Kassim},
  {Lazio}, {Leung}, {Li}, {Lin}, {Masui}, {Mckinven}, {Michilli}, {Mikhailov},
  {Ng}, {Orbidans}, {Pen}, {Petroff}, {Rahman}, {Ransom}, {Shin}, {Smith},
  {Stairs}, \& {Vlemmings}}]{2022Natur.602..585K}
{Kirsten}, F., {Marcote}, B., {Nimmo}, K., {et~al.} 2022, \nat, 602, 585,
  \dodoi{10.1038/s41586-021-04354-w}

\bibitem[{{Kouveliotou} {et~al.}(1993){Kouveliotou}, {Meegan}, {Fishman},
  {Bhat}, {Briggs}, {Koshut}, {Paciesas}, \& {Pendleton}}]{1993ApJ...413L.101K}
{Kouveliotou}, C., {Meegan}, C.~A., {Fishman}, G.~J., {et~al.} 1993, \apjl,
  413, L101, \dodoi{10.1086/186969}

\bibitem[{{Krimm} {et~al.}(2005){Krimm}, {Barbier}, {Barthelmy}, {Cummings},
  {Fenimore}, {Gehrels}, {Hinshaw}, {Hullinger}, {Markwardt}, {McLean},
  {Palmer}, {Parsons}, {Sakamoto}, \& {Tueller}}]{2005GCN..3667....1K}
{Krimm}, H., {Barbier}, L., {Barthelmy}, S., {et~al.} 2005, GRB Coordinates
  Network, 3667, 1

\bibitem[{{Krimm} {et~al.}(2006){Krimm}, {Barbier}, {Barthelmy}, {Cummings},
  {Fenimore}, {Gehrels}, {Hullinger}, {Koss}, {Markwardt}, {Palmer}, {Parsons},
  {Sakamoto}, {Sato}, {Schady}, {Stamatikos}, \&
  {Tueller}}]{2006GCN..5704....1K}
---. 2006, GRB Coordinates Network, 5704, 1

\bibitem[{{Kulkarni} {et~al.}(2014){Kulkarni}, {Ofek}, {Neill}, {Zheng}, \&
  {Juric}}]{2014ApJ...797...70K}
{Kulkarni}, S.~R., {Ofek}, E.~O., {Neill}, J.~D., {Zheng}, Z., \& {Juric}, M.
  2014, \apj, 797, 70, \dodoi{10.1088/0004-637X/797/1/70}

\bibitem[{{Kumar} \& {Bo{\v{s}}njak}(2020)}]{2020MNRAS.494.2385K}
{Kumar}, P., \& {Bo{\v{s}}njak}, {\v{Z}}. 2020, \mnras, 494, 2385,
  \dodoi{10.1093/mnras/staa774}

\bibitem[{{Kumar} \& {Zhang}(2015)}]{2015PhR...561....1K}
{Kumar}, P., \& {Zhang}, B. 2015, \physrep, 561, 1,
  \dodoi{10.1016/j.physrep.2014.09.008}

\bibitem[{{Law} {et~al.}(2020){Law}, {Butler}, {Prochaska}, {Zackay},
  {Burke-Spolaor}, {Mannings}, {Tejos}, {Josephy}, {Andersen}, {Chawla},
  {Heintz}, {Aggarwal}, {Bower}, {Demorest}, {Kilpatrick}, {Lazio}, {Linford},
  {Mckinven}, {Tendulkar}, \& {Simha}}]{2020ApJ...899..161L}
{Law}, C.~J., {Butler}, B.~J., {Prochaska}, J.~X., {et~al.} 2020, \apj, 899,
  161, \dodoi{10.3847/1538-4357/aba4ac}

\bibitem[{{Law} {et~al.}(2023){Law}, {Sharma}, {Ravi}, {Chen}, {Catha},
  {Connor}, {Faber}, {Hallinan}, {Harnach}, {Hellbourg}, {Hobbs}, {Hodge},
  {Hodges}, {Lamb}, {Rasmussen}, {Sherman}, {Shi}, {Simard}, {Squillace},
  {Weinreb}, {Woody}, \& {Yadlapalli}}]{2023arXiv230703344L}
{Law}, C.~J., {Sharma}, K., {Ravi}, V., {et~al.} 2023, arXiv e-prints,
  arXiv:2307.03344, \dodoi{10.48550/arXiv.2307.03344}

\bibitem[{{Li} {et~al.}(2021){Li}, {Lin}, {Xiong}, {Ge}, {Li}, {Li}, {Lu},
  {Zhang}, {Tuo}, {Nang}, {Zhang}, {Xiao}, {Chen}, {Song}, {Xu}, {Liu}, {Jia},
  {Cao}, {Qu}, {Zhang}, {Gu}, {Liao}, {Zhao}, {Tan}, {Nie}, {Zhao}, {Zheng},
  {Zheng}, {Luo}, {Cai}, {Li}, {Xue}, {Bu}, {Chang}, {Chen}, {Chen}, {Chen},
  {Chen}, {Chen}, {Cui}, {Cui}, {Deng}, {Dong}, {Du}, {Fu}, {Gao}, {Gao},
  {Gao}, {Gu}, {Guan}, {Guo}, {Han}, {Huang}, {Huo}, {Jiang}, {Jiang}, {Jin},
  {Jin}, {Kong}, {Li}, {Li}, {Li}, {Li}, {Li}, {Li}, {Li}, {Liang}, {Liu},
  {Liu}, {Liu}, {Liu}, {Liu}, {Lu}, {Lu}, {Luo}, {Ma}, {Meng}, {Ou}, {Sai},
  {Shang}, {Song}, {Sun}, {Tao}, {Wang}, {Wang}, {Wang}, {Wang}, {Wang}, {Wen},
  {Wu}, {Wu}, {Wu}, {Xiao}, {Xu}, {Yang}, {Yang}, {Yang}, {Yang}, {Yi}, {Yin},
  {You}, {Zhang}, {Zhang}, {Zhang}, {Zhang}, {Zhang}, {Zhang}, {Zhang},
  {Zhang}, {Zhang}, {Zhang}, {Zhang}, {Zhang}, {Zhang}, {Zhang}, {Zhang},
  {Zhang}, {Zhou}, {Zhou}, {Zhu}, {Zhu}, \& {Zhuang}}]{2021NatAs...5..378L}
{Li}, C.~K., {Lin}, L., {Xiong}, S.~L., {et~al.} 2021, Nature Astronomy, 5,
  378, \dodoi{10.1038/s41550-021-01302-6}

\bibitem[{{Li} {et~al.}(2022){Li}, {Li}, {Zhong}, {Xia}, {Xie}, {Wang}, \&
  {Dai}}]{2022arXiv220306994L}
{Li}, L., {Li}, Q.-C., {Zhong}, S.-Q., {et~al.} 2022, arXiv e-prints,
  arXiv:2203.06994.
\newblock \doarXiv{2203.06994}

\bibitem[{{Li} \& {Zhang}(2020)}]{2020ApJ...899L...6L}
{Li}, Y., \& {Zhang}, B. 2020, \apjl, 899, L6, \dodoi{10.3847/2041-8213/aba907}

\bibitem[{{Li} {et~al.}(2019){Li}, {Zhang}, {Nagamine}, \&
  {Shi}}]{2019ApJ...884L..26L}
{Li}, Y., {Zhang}, B., {Nagamine}, K., \& {Shi}, J. 2019, \apjl, 884, L26,
  \dodoi{10.3847/2041-8213/ab3e41}

\bibitem[{{Liang} \& {Zhang}(2005)}]{2005ApJ...633..611L}
{Liang}, E., \& {Zhang}, B. 2005, \apj, 633, 611, \dodoi{10.1086/491594}

\bibitem[{{Liang} {et~al.}(2017){Liang}, {Xia}, {Duan}, {Shen}, {Li}, \&
  {Fan}}]{2017PhRvD..95f3531L}
{Liang}, Y.-F., {Xia}, Z.-Q., {Duan}, K.-K., {et~al.} 2017, \prd, 95, 063531,
  \dodoi{10.1103/PhysRevD.95.063531}

\bibitem[{{Lin} \& {Totani}(2020)}]{2020MNRAS.498.2384L}
{Lin}, H., \& {Totani}, T. 2020, \mnras, 498, 2384,
  \dodoi{10.1093/mnras/staa2418}

\bibitem[{{Lipunov} {et~al.}(2006){Lipunov}, {Kornilov}, {Tyurina}, {Belinski},
  {Kuvshinov}, {Gorbovskoy}, {Krylov}, {Borisov}, {Sankovich}, {Vladimirov},
  {Gritsyk}, \& {Korobkin}}]{2006GCN..5056....1L}
{Lipunov}, V., {Kornilov}, V., {Tyurina}, N., {et~al.} 2006, GRB Coordinates
  Network, 5056, 1

\bibitem[{{Liu} {et~al.}(2022){Liu}, {Wang}, {Xin}, {Zhou}, {Chen}, {Li},
  {Yang}, {Luo}, {Li}, {Xiong}, {Wang}, {Han}, {Song}, {Wei}, {Liang}, \&
  {Zhang}}]{2022RAA....22f5002L}
{Liu}, H.-Y., {Wang}, X.-G., {Xin}, L.-P., {et~al.} 2022, Research in Astronomy
  and Astrophysics, 22, 065002, \dodoi{10.1088/1674-4527/ac65e6}

\bibitem[{{Liu} {et~al.}(2016){Liu}, {Romero}, {Liu}, \&
  {Li}}]{2016ApJ...826...82L}
{Liu}, T., {Romero}, G.~E., {Liu}, M.-L., \& {Li}, A. 2016, \apj, 826, 82,
  \dodoi{10.3847/0004-637X/826/1/82}

\bibitem[{{Lorimer} {et~al.}(2007){Lorimer}, {Bailes}, {McLaughlin},
  {Narkevic}, \& {Crawford}}]{2007Sci...318..777L}
{Lorimer}, D.~R., {Bailes}, M., {McLaughlin}, M.~A., {Narkevic}, D.~J., \&
  {Crawford}, F. 2007, Science, 318, 777, \dodoi{10.1126/science.1147532}

\bibitem[{{L{\"u}} {et~al.}(2019){L{\"u}}, {Lan}, \&
  {Liang}}]{2019ApJ...871...54L}
{L{\"u}}, H.-J., {Lan}, L., \& {Liang}, E.-W. 2019, \apj, 871, 54,
  \dodoi{10.3847/1538-4357/aaf71d}

\bibitem[{{L{\"u}} {et~al.}(2015){L{\"u}}, {Zhang}, {Lei}, {Li}, \&
  {Lasky}}]{2015ApJ...805...89L}
{L{\"u}}, H.-J., {Zhang}, B., {Lei}, W.-H., {Li}, Y., \& {Lasky}, P.~D. 2015,
  \apj, 805, 89, \dodoi{10.1088/0004-637X/805/2/89}

\bibitem[{{Lu} \& {Kumar}(2016)}]{2016MNRAS.461L.122L}
{Lu}, W., \& {Kumar}, P. 2016, \mnras, 461, L122, \dodoi{10.1093/mnrasl/slw113}

\bibitem[{{Lu} \& {Kumar}(2018)}]{2018MNRAS.477.2470L}
---. 2018, \mnras, 477, 2470, \dodoi{10.1093/mnras/sty716}

\bibitem[{{Lu} {et~al.}(2020){Lu}, {Kumar}, \& {Zhang}}]{2020MNRAS.498.1397L}
{Lu}, W., {Kumar}, P., \& {Zhang}, B. 2020, \mnras, 498, 1397,
  \dodoi{10.1093/mnras/staa2450}

\bibitem[{{Luo} {et~al.}(2020){Luo}, {Wang}, {Men}, {Zhang}, {Jiang}, {Xu},
  {Wang}, {Lee}, {Han}, {Zhang}, {Caballero}, {Chen}, {Chen}, {Gan}, {Guo},
  {Hao}, {Huang}, {Jiang}, {Li}, {Li}, {Li}, {Luo}, {Pan}, {Pei}, {Qian},
  {Sun}, {Wang}, {Wang}, {Wen}, {Xu}, {Xu}, {Yan}, {Yan}, {Yu}, {Yuan},
  {Zhang}, \& {Zhu}}]{2020Natur.586..693L}
{Luo}, R., {Wang}, B.~J., {Men}, Y.~P., {et~al.} 2020, \nat, 586, 693,
  \dodoi{10.1038/s41586-020-2827-2}

\bibitem[{{Lyubarsky}(2014)}]{2014MNRAS.442L...9L}
{Lyubarsky}, Y. 2014, \mnras, 442, L9, \dodoi{10.1093/mnrasl/slu046}

\bibitem[{{Lyubarsky}(2020)}]{2020ApJ...897....1L}
---. 2020, \apj, 897, 1, \dodoi{10.3847/1538-4357/ab97b5}

\bibitem[{{Lyutikov}(2021)}]{2021arXiv211008435L}
{Lyutikov}, M. 2021, arXiv e-prints, arXiv:2110.08435.
\newblock \doarXiv{2110.08435}

\bibitem[{{Macquart} {et~al.}(2020){Macquart}, {Prochaska}, {McQuinn},
  {Bannister}, {Bhandari}, {Day}, {Deller}, {Ekers}, {James}, {Marnoch},
  {Os{\l}owski}, {Phillips}, {Ryder}, {Scott}, {Shannon}, \&
  {Tejos}}]{2020Natur.581..391M}
{Macquart}, J.~P., {Prochaska}, J.~X., {McQuinn}, M., {et~al.} 2020, \nat, 581,
  391, \dodoi{10.1038/s41586-020-2300-2}

\bibitem[{{Mahony} {et~al.}(2018){Mahony}, {Ekers}, {Macquart}, {Sadler},
  {Bannister}, {Bhandari}, {Flynn}, {Koribalski}, {Prochaska}, {Ryder},
  {Shannon}, {Tejos}, {Whiting}, \& {Wong}}]{2018ApJ...867L..10M}
{Mahony}, E.~K., {Ekers}, R.~D., {Macquart}, J.-P., {et~al.} 2018, \apjl, 867,
  L10, \dodoi{10.3847/2041-8213/aae7cb}

\bibitem[{{Mannings} {et~al.}(2021){Mannings}, {Fong}, {Simha}, {Prochaska},
  {Rafelski}, {Kilpatrick}, {Tejos}, {Heintz}, {Bannister}, {Bhandari}, {Day},
  {Deller}, {Ryder}, {Shannon}, \& {Tendulkar}}]{2021ApJ...917...75M}
{Mannings}, A.~G., {Fong}, W.-f., {Simha}, S., {et~al.} 2021, \apj, 917, 75,
  \dodoi{10.3847/1538-4357/abff56}

\bibitem[{{Mao} {et~al.}(2008){Mao}, {Baumgartner}, {Burrows}, {Chester},
  {Gehrels}, {Guidorzi}, {Holland}, {Kennea}, {Markwardt}, {Marshall},
  {O'Brien}, {Palmer}, {Romano}, {Ukwatta}, \& {vanden
  Berk}}]{2008GCN..7665....1M}
{Mao}, J., {Baumgartner}, W.~H., {Burrows}, D.~N., {et~al.} 2008, GRB
  Coordinates Network, 7665, 1

\bibitem[{{Marcote} {et~al.}(2017){Marcote}, {Paragi}, {Hessels}, {Keimpema},
  {van Langevelde}, {Huang}, {Bassa}, {Bogdanov}, {Bower}, {Burke-Spolaor},
  {Butler}, {Campbell}, {Chatterjee}, {Cordes}, {Demorest}, {Garrett}, {Ghosh},
  {Kaspi}, {Law}, {Lazio}, {McLaughlin}, {Ransom}, {Salter}, {Scholz},
  {Seymour}, {Siemion}, {Spitler}, {Tendulkar}, \&
  {Wharton}}]{2017ApJ...834L...8M}
{Marcote}, B., {Paragi}, Z., {Hessels}, J.~W.~T., {et~al.} 2017, \apjl, 834,
  L8, \dodoi{10.3847/2041-8213/834/2/L8}

\bibitem[{{Marcote} {et~al.}(2020){Marcote}, {Nimmo}, {Hessels}, {Tendulkar},
  {Bassa}, {Paragi}, {Keimpema}, {Bhardwaj}, {Karuppusamy}, {Kaspi}, {Law},
  {Michilli}, {Aggarwal}, {Andersen}, {Archibald}, {Bandura}, {Bower}, {Boyle},
  {Brar}, {Burke-Spolaor}, {Butler}, {Cassanelli}, {Chawla}, {Demorest},
  {Dobbs}, {Fonseca}, {Giri}, {Good}, {Gourdji}, {Josephy}, {Kirichenko},
  {Kirsten}, {Landecker}, {Lang}, {Lazio}, {Li}, {Lin}, {Linford}, {Masui},
  {Mena-Parra}, {Naidu}, {Ng}, {Patel}, {Pen}, {Pleunis}, {Rafiei-Ravandi},
  {Rahman}, {Renard}, {Scholz}, {Siegel}, {Smith}, {Stairs}, {Vanderlinde}, \&
  {Zwaniga}}]{2020Natur.577..190M}
{Marcote}, B., {Nimmo}, K., {Hessels}, J.~W.~T., {et~al.} 2020, \nat, 577, 190,
  \dodoi{10.1038/s41586-019-1866-z}

\bibitem[{{Margalit} {et~al.}(2019){Margalit}, {Berger}, \&
  {Metzger}}]{2019ApJ...886..110M}
{Margalit}, B., {Berger}, E., \& {Metzger}, B.~D. 2019, \apj, 886, 110,
  \dodoi{10.3847/1538-4357/ab4c31}

\bibitem[{{Margalit} \& {Metzger}(2018)}]{2018ApJ...868L...4M}
{Margalit}, B., \& {Metzger}, B.~D. 2018, \apjl, 868, L4,
  \dodoi{10.3847/2041-8213/aaedad}

\bibitem[{{Margalit} \& {Metzger}(2019)}]{2019ApJ...880L..15M}
---. 2019, \apjl, 880, L15, \dodoi{10.3847/2041-8213/ab2ae2}

\bibitem[{{McQuinn}(2014)}]{2014ApJ...780L..33M}
{McQuinn}, M. 2014, \apjl, 780, L33, \dodoi{10.1088/2041-8205/780/2/L33}

\bibitem[{{Mereghetti} {et~al.}(2020){Mereghetti}, {Savchenko}, {Ferrigno},
  {G{\"o}tz}, {Rigoselli}, {Tiengo}, {Bazzano}, {Bozzo}, {Coleiro},
  {Courvoisier}, {Doyle}, {Goldwurm}, {Hanlon}, {Jourdain}, {von Kienlin},
  {Lutovinov}, {Martin-Carrillo}, {Molkov}, {Natalucci}, {Onori}, {Panessa},
  {Rodi}, {Rodriguez}, {S{\'a}nchez-Fern{\'a}ndez}, {Sunyaev}, \&
  {Ubertini}}]{2020ApJ...898L..29M}
{Mereghetti}, S., {Savchenko}, V., {Ferrigno}, C., {et~al.} 2020, \apjl, 898,
  L29, \dodoi{10.3847/2041-8213/aba2cf}

\bibitem[{{Metzger} {et~al.}(2017){Metzger}, {Berger}, \&
  {Margalit}}]{2017ApJ...841...14M}
{Metzger}, B.~D., {Berger}, E., \& {Margalit}, B. 2017, \apj, 841, 14,
  \dodoi{10.3847/1538-4357/aa633d}

\bibitem[{{Metzger} {et~al.}(2019){Metzger}, {Margalit}, \&
  {Sironi}}]{2019MNRAS.485.4091M}
{Metzger}, B.~D., {Margalit}, B., \& {Sironi}, L. 2019, \mnras, 485, 4091,
  \dodoi{10.1093/mnras/stz700}

\bibitem[{{Meurs} {et~al.}(2006){Meurs}, {Vergani}, {O'Maoileidigh},
  {Malesani}, \& {Gualandi}}]{2006GCN..5074....1M}
{Meurs}, E.~J.~A., {Vergani}, S.~D., {O'Maoileidigh}, C., {Malesani}, D., \&
  {Gualandi}, R. 2006, GRB Coordinates Network, 5074, 1

\bibitem[{{Micha{\l}owski}(2021)}]{2021ApJ...920L..21M}
{Micha{\l}owski}, M.~J. 2021, \apjl, 920, L21, \dodoi{10.3847/2041-8213/ac2b35}

\bibitem[{{Mu{\~n}oz} \& {Loeb}(2018)}]{2018PhRvD..98j3518M}
{Mu{\~n}oz}, J.~B., \& {Loeb}, A. 2018, \prd, 98, 103518,
  \dodoi{10.1103/PhysRevD.98.103518}

\bibitem[{{Murase} {et~al.}(2016){Murase}, {Kashiyama}, \&
  {M{\'e}sz{\'a}ros}}]{2016MNRAS.461.1498M}
{Murase}, K., {Kashiyama}, K., \& {M{\'e}sz{\'a}ros}, P. 2016, \mnras, 461,
  1498, \dodoi{10.1093/mnras/stw1328}

\bibitem[{{NASA/IPAC Extragalactic Database
  (NED)}(2019)}]{https://doi.org/10.26132/ned1}
{NASA/IPAC Extragalactic Database (NED)}. 2019, NASA/IPAC Extragalactic
  Database (NED),  IPAC, \dodoi{10.26132/NED1}

\bibitem[{{Navarro} {et~al.}(1997){Navarro}, {Frenk}, \&
  {White}}]{1997ApJ...490..493N}
{Navarro}, J.~F., {Frenk}, C.~S., \& {White}, S. D.~M. 1997, \apj, 490, 493,
  \dodoi{10.1086/304888}

\bibitem[{{Nicholl} {et~al.}(2017){Nicholl}, {Williams}, {Berger}, {Villar},
  {Alexander}, {Eftekhari}, \& {Metzger}}]{2017ApJ...843...84N}
{Nicholl}, M., {Williams}, P.~K.~G., {Berger}, E., {et~al.} 2017, \apj, 843,
  84, \dodoi{10.3847/1538-4357/aa794d}

\bibitem[{{Nimmo} {et~al.}(2021){Nimmo}, {Hewitt}, {Hessels}, {Kirsten},
  {Marcote}, {Bach}, {Blaauw}, {Burgay}, {Corongiu}, {Feiler}, {Gawro{\'n}ski},
  {Giroletti}, {Karuppusamy}, {Keimpema}, {Kharinov}, {Lindqvist},
  {Maccaferri}, {Melnikov}, {Mikhailov}, {Ould-Boukattine}, {Paragi}, {Pilia},
  {Possenti}, {Snelders}, {Surcis}, {Trudu}, {Venturi}, {Vlemmings}, {Wang},
  {Yang}, \& {Yuan}}]{2021arXiv211101600N}
{Nimmo}, K., {Hewitt}, D.~M., {Hessels}, J.~W.~T., {et~al.} 2021, arXiv
  e-prints, arXiv:2111.01600.
\newblock \doarXiv{2111.01600}

\bibitem[{{Norris} {et~al.}(2010{\natexlab{a}}){Norris}, {Ukwatta},
  {Barthelmy}, {Gehrels}, {Stamatikos}, \& {Sakamoto}}]{2010GCN.11113....1N}
{Norris}, J., {Ukwatta}, T.~N., {Barthelmy}, S.~D., {et~al.}
  2010{\natexlab{a}}, GRB Coordinates Network, 11113, 1

\bibitem[{{Norris} {et~al.}(2010{\natexlab{b}}){Norris}, {Barthelmy},
  {Gehrels}, \& {Grupe}}]{2010GCN.10427....1N}
{Norris}, J.~P., {Barthelmy}, S.~D., {Gehrels}, N., \& {Grupe}, D.
  2010{\natexlab{b}}, GRB Coordinates Network, 10427, 1

\bibitem[{{Obenberger} {et~al.}(2014){Obenberger}, {Hartman}, {Taylor},
  {Craig}, {Dowell}, {Helmboldt}, {Henning}, {Schinzel}, \&
  {Wilson}}]{2014ApJ...785...27O}
{Obenberger}, K.~S., {Hartman}, J.~M., {Taylor}, G.~B., {et~al.} 2014, \apj,
  785, 27, \dodoi{10.1088/0004-637X/785/1/27}

\bibitem[{{Palaniswamy} {et~al.}(2014){Palaniswamy}, {Wayth}, {Trott},
  {McCallum}, {Tingay}, \& {Reynolds}}]{2014ApJ...790...63P}
{Palaniswamy}, D., {Wayth}, R.~B., {Trott}, C.~M., {et~al.} 2014, \apj, 790,
  63, \dodoi{10.1088/0004-637X/790/1/63}

\bibitem[{{Pal'Shin} {et~al.}(2008){Pal'Shin}, {Golenetskii}, {Aptekar},
  {Mazets}, {Frederiks}, {Cline}, {Sakamoto}, {Barthelmy}, {Baumgartner},
  {Beardmore}, {Cummings}, {Fenimore}, {Gehrels}, {Krimm}, {Markwardt},
  {McLean}, {Palmer}, {Sato}, {Stamatikos}, {Tueller}, \&
  {Ukwatta}}]{2008GCN..8256....1P}
{Pal'Shin}, V., {Golenetskii}, S., {Aptekar}, R., {et~al.} 2008, GRB
  Coordinates Network, 8256, 1

\bibitem[{{Pan} {et~al.}(2022){Pan}, {Yang}, \& {Yagi}}]{2022arXiv220808808P}
{Pan}, Z., {Yang}, H., \& {Yagi}, K. 2022, arXiv e-prints, arXiv:2208.08808.
\newblock \doarXiv{2208.08808}

\bibitem[{{Petroff} {et~al.}(2019){Petroff}, {Hessels}, \&
  {Lorimer}}]{2019A&ARv..27....4P}
{Petroff}, E., {Hessels}, J.~W.~T., \& {Lorimer}, D.~R. 2019, \aapr, 27, 4,
  \dodoi{10.1007/s00159-019-0116-6}

\bibitem[{{Petroff} \& {Yaron}(2020)}]{2020TNSAN.160....1P}
{Petroff}, E., \& {Yaron}, O. 2020, Transient Name Server AstroNote, 160, 1

\bibitem[{{Petroff} {et~al.}(2015){Petroff}, {Bailes}, {Barr}, {Barsdell},
  {Bhat}, {Bian}, {Burke-Spolaor}, {Caleb}, {Champion}, {Chandra}, {Da Costa},
  {Delvaux}, {Flynn}, {Gehrels}, {Greiner}, {Jameson}, {Johnston}, {Kasliwal},
  {Keane}, {Keller}, {Kocz}, {Kramer}, {Leloudas}, {Malesani}, {Mulchaey},
  {Ng}, {Ofek}, {Perley}, {Possenti}, {Schmidt}, {Shen}, {Stappers},
  {Tisserand}, {van Straten}, \& {Wolf}}]{2015MNRAS.447..246P}
{Petroff}, E., {Bailes}, M., {Barr}, E.~D., {et~al.} 2015, \mnras, 447, 246,
  \dodoi{10.1093/mnras/stu2419}

\bibitem[{{Piro} {et~al.}(2021){Piro}, {Bruni}, {Troja}, {O'Connor}, {Panessa},
  {Ricci}, {Zhang}, {Burgay}, {Dichiara}, {Lee}, {Lotti}, {Niu}, {Pilia},
  {Possenti}, {Trudu}, {Xu}, {Zhu}, {Kutyrev}, \&
  {Veilleux}}]{2021A&A...656L..15P}
{Piro}, L., {Bruni}, G., {Troja}, E., {et~al.} 2021, \aap, 656, L15,
  \dodoi{10.1051/0004-6361/202141903}

\bibitem[{{Planck Collaboration} {et~al.}(2016){Planck Collaboration}, {Ade},
  {Aghanim}, {Arnaud}, {Ashdown}, {Aumont}, {Baccigalupi}, {Banday},
  {Barreiro}, {Bartlett}, {Bartolo}, {Battaner}, {Battye}, {Benabed},
  {Beno{\^\i}t}, {Benoit-L{\'e}vy}, {Bernard}, {Bersanelli}, {Bielewicz},
  {Bock}, {Bonaldi}, {Bonavera}, {Bond}, {Borrill}, {Bouchet}, {Boulanger},
  {Bucher}, {Burigana}, {Butler}, {Calabrese}, {Cardoso}, {Catalano},
  {Challinor}, {Chamballu}, {Chary}, {Chiang}, {Chluba}, {Christensen},
  {Church}, {Clements}, {Colombi}, {Colombo}, {Combet}, {Coulais}, {Crill},
  {Curto}, {Cuttaia}, {Danese}, {Davies}, {Davis}, {de Bernardis}, {de Rosa},
  {de Zotti}, {Delabrouille}, {D{\'e}sert}, {Di Valentino}, {Dickinson},
  {Diego}, {Dolag}, {Dole}, {Donzelli}, {Dor{\'e}}, {Douspis}, {Ducout},
  {Dunkley}, {Dupac}, {Efstathiou}, {Elsner}, {En{\ss}lin}, {Eriksen},
  {Farhang}, {Fergusson}, {Finelli}, {Forni}, {Frailis}, {Fraisse},
  {Franceschi}, {Frejsel}, {Galeotta}, {Galli}, {Ganga}, {Gauthier}, {Gerbino},
  {Ghosh}, {Giard}, {Giraud-H{\'e}raud}, {Giusarma}, {Gjerl{\o}w},
  {Gonz{\'a}lez-Nuevo}, {G{\'o}rski}, {Gratton}, {Gregorio}, {Gruppuso},
  {Gudmundsson}, {Hamann}, {Hansen}, {Hanson}, {Harrison}, {Helou},
  {Henrot-Versill{\'e}}, {Hern{\'a}ndez-Monteagudo}, {Herranz}, {Hildebrandt},
  {Hivon}, {Hobson}, {Holmes}, {Hornstrup}, {Hovest}, {Huang}, {Huffenberger},
  {Hurier}, {Jaffe}, {Jaffe}, {Jones}, {Juvela}, {Keih{\"a}nen}, {Keskitalo},
  {Kisner}, {Kneissl}, {Knoche}, {Knox}, {Kunz}, {Kurki-Suonio}, {Lagache},
  {L{\"a}hteenm{\"a}ki}, {Lamarre}, {Lasenby}, {Lattanzi}, {Lawrence}, {Leahy},
  {Leonardi}, {Lesgourgues}, {Levrier}, {Lewis}, {Liguori}, {Lilje},
  {Linden-V{\o}rnle}, {L{\'o}pez-Caniego}, {Lubin}, {Mac{\'\i}as-P{\'e}rez},
  {Maggio}, {Maino}, {Mandolesi}, {Mangilli}, {Marchini}, {Maris}, {Martin},
  {Martinelli}, {Mart{\'\i}nez-Gonz{\'a}lez}, {Masi}, {Matarrese}, {McGehee},
  {Meinhold}, {Melchiorri}, {Melin}, {Mendes}, {Mennella}, {Migliaccio},
  {Millea}, {Mitra}, {Miville-Desch{\^e}nes}, {Moneti}, {Montier}, {Morgante},
  {Mortlock}, {Moss}, {Munshi}, {Murphy}, {Naselsky}, {Nati}, {Natoli},
  {Netterfield}, {N{\o}rgaard-Nielsen}, {Noviello}, {Novikov}, {Novikov},
  {Oxborrow}, {Paci}, {Pagano}, {Pajot}, {Paladini}, {Paoletti}, {Partridge},
  {Pasian}, {Patanchon}, {Pearson}, {Perdereau}, {Perotto}, {Perrotta},
  {Pettorino}, {Piacentini}, {Piat}, {Pierpaoli}, {Pietrobon}, {Plaszczynski},
  {Pointecouteau}, {Polenta}, {Popa}, {Pratt}, {Pr{\'e}zeau}, {Prunet},
  {Puget}, {Rachen}, {Reach}, {Rebolo}, {Reinecke}, {Remazeilles}, {Renault},
  {Renzi}, {Ristorcelli}, {Rocha}, {Rosset}, {Rossetti}, {Roudier},
  {Rouill{\'e} d'Orfeuil}, {Rowan-Robinson}, {Rubi{\~n}o-Mart{\'\i}n},
  {Rusholme}, {Said}, {Salvatelli}, {Salvati}, {Sandri}, {Santos},
  {Savelainen}, {Savini}, {Scott}, {Seiffert}, {Serra}, {Shellard}, {Spencer},
  {Spinelli}, {Stolyarov}, {Stompor}, {Sudiwala}, {Sunyaev}, {Sutton},
  {Suur-Uski}, {Sygnet}, {Tauber}, {Terenzi}, {Toffolatti}, {Tomasi},
  {Tristram}, {Trombetti}, {Tucci}, {Tuovinen}, {T{\"u}rler}, {Umana},
  {Valenziano}, {Valiviita}, {Van Tent}, {Vielva}, {Villa}, {Wade}, {Wandelt},
  {Wehus}, {White}, {White}, {Wilkinson}, {Yvon}, {Zacchei}, \&
  {Zonca}}]{2016A&A...594A..13P}
{Planck Collaboration}, {Ade}, P.~A.~R., {Aghanim}, N., {et~al.} 2016, \aap,
  594, A13, \dodoi{10.1051/0004-6361/201525830}

\bibitem[{{Planck Collaboration} {et~al.}(2020){Planck Collaboration},
  {Aghanim}, {Akrami}, {Ashdown}, {Aumont}, {Baccigalupi}, {Ballardini},
  {Banday}, {Barreiro}, {Bartolo}, {Basak}, {Battye}, {Benabed}, {Bernard},
  {Bersanelli}, {Bielewicz}, {Bock}, {Bond}, {Borrill}, {Bouchet}, {Boulanger},
  {Bucher}, {Burigana}, {Butler}, {Calabrese}, {Cardoso}, {Carron},
  {Challinor}, {Chiang}, {Chluba}, {Colombo}, {Combet}, {Contreras}, {Crill},
  {Cuttaia}, {de Bernardis}, {de Zotti}, {Delabrouille}, {Delouis}, {Di
  Valentino}, {Diego}, {Dor{\'e}}, {Douspis}, {Ducout}, {Dupac}, {Dusini},
  {Efstathiou}, {Elsner}, {En{\ss}lin}, {Eriksen}, {Fantaye}, {Farhang},
  {Fergusson}, {Fernandez-Cobos}, {Finelli}, {Forastieri}, {Frailis},
  {Fraisse}, {Franceschi}, {Frolov}, {Galeotta}, {Galli}, {Ganga},
  {G{\'e}nova-Santos}, {Gerbino}, {Ghosh}, {Gonz{\'a}lez-Nuevo}, {G{\'o}rski},
  {Gratton}, {Gruppuso}, {Gudmundsson}, {Hamann}, {Handley}, {Hansen},
  {Herranz}, {Hildebrandt}, {Hivon}, {Huang}, {Jaffe}, {Jones}, {Karakci},
  {Keih{\"a}nen}, {Keskitalo}, {Kiiveri}, {Kim}, {Kisner}, {Knox},
  {Krachmalnicoff}, {Kunz}, {Kurki-Suonio}, {Lagache}, {Lamarre}, {Lasenby},
  {Lattanzi}, {Lawrence}, {Le Jeune}, {Lemos}, {Lesgourgues}, {Levrier},
  {Lewis}, {Liguori}, {Lilje}, {Lilley}, {Lindholm}, {L{\'o}pez-Caniego},
  {Lubin}, {Ma}, {Mac{\'\i}as-P{\'e}rez}, {Maggio}, {Maino}, {Mandolesi},
  {Mangilli}, {Marcos-Caballero}, {Maris}, {Martin}, {Martinelli},
  {Mart{\'\i}nez-Gonz{\'a}lez}, {Matarrese}, {Mauri}, {McEwen}, {Meinhold},
  {Melchiorri}, {Mennella}, {Migliaccio}, {Millea}, {Mitra},
  {Miville-Desch{\^e}nes}, {Molinari}, {Montier}, {Morgante}, {Moss}, {Natoli},
  {N{\o}rgaard-Nielsen}, {Pagano}, {Paoletti}, {Partridge}, {Patanchon},
  {Peiris}, {Perrotta}, {Pettorino}, {Piacentini}, {Polastri}, {Polenta},
  {Puget}, {Rachen}, {Reinecke}, {Remazeilles}, {Renzi}, {Rocha}, {Rosset},
  {Roudier}, {Rubi{\~n}o-Mart{\'\i}n}, {Ruiz-Granados}, {Salvati}, {Sandri},
  {Savelainen}, {Scott}, {Shellard}, {Sirignano}, {Sirri}, {Spencer},
  {Sunyaev}, {Suur-Uski}, {Tauber}, {Tavagnacco}, {Tenti}, {Toffolatti},
  {Tomasi}, {Trombetti}, {Valenziano}, {Valiviita}, {Van Tent}, {Vibert},
  {Vielva}, {Villa}, {Vittorio}, {Wandelt}, {Wehus}, {White}, {White},
  {Zacchei}, \& {Zonca}}]{2020A&A...641A...6P}
{Planck Collaboration}, {Aghanim}, N., {Akrami}, Y., {et~al.} 2020, \aap, 641,
  A6, \dodoi{10.1051/0004-6361/201833910}

\bibitem[{{Platts} {et~al.}(2019){Platts}, {Weltman}, {Walters}, {Tendulkar},
  {Gordin}, \& {Kandhai}}]{2019PhR...821....1P}
{Platts}, E., {Weltman}, A., {Walters}, A., {et~al.} 2019, \physrep, 821, 1,
  \dodoi{10.1016/j.physrep.2019.06.003}

\bibitem[{{Pleunis} {et~al.}(2021){Pleunis}, {Good}, {Kaspi}, {Mckinven},
  {Ransom}, {Scholz}, {Bandura}, {Bhardwaj}, {Boyle}, {Brar}, {Cassanelli},
  {Chawla}, {Fengqiu}, {Dong}, {Fonseca}, {Gaensler}, {Josephy}, {Kaczmarek},
  {Leung}, {Lin}, {Masui}, {Mena-Parra}, {Michilli}, {Ng}, {Patel},
  {Rafiei-Ravandi}, {Rahman}, {Sanghavi}, {Shin}, {Smith}, {Stairs}, \&
  {Tendulkar}}]{2021arXiv210604356P}
{Pleunis}, Z., {Good}, D.~C., {Kaspi}, V.~M., {et~al.} 2021, arXiv e-prints,
  arXiv:2106.04356.
\newblock \doarXiv{2106.04356}

\bibitem[{{Poole} \& {Troja}(2006)}]{2006GCN..5069....1P}
{Poole}, T.~S., \& {Troja}, E. 2006, GRB Coordinates Network, 5069, 1

\bibitem[{{Pooley} {et~al.}(2003){Pooley}, {Lewin}, {Anderson}, {Baumgardt},
  {Filippenko}, {Gaensler}, {Homer}, {Hut}, {Kaspi}, {Makino}, {Margon},
  {McMillan}, {Portegies Zwart}, {van der Klis}, \&
  {Verbunt}}]{2003ApJ...591L.131P}
{Pooley}, D., {Lewin}, W. H.~G., {Anderson}, S.~F., {et~al.} 2003, \apjl, 591,
  L131, \dodoi{10.1086/377074}

\bibitem[{{Popov} \& {Postnov}(2013)}]{2013arXiv1307.4924P}
{Popov}, S.~B., \& {Postnov}, K.~A. 2013, arXiv e-prints, arXiv:1307.4924.
\newblock \doarXiv{1307.4924}

\bibitem[{{Price} {et~al.}(2006){Price}, {Berger}, {Fox}, {Cenko}, \&
  {Rau}}]{2006GCN..5077....1P}
{Price}, P.~A., {Berger}, E., {Fox}, D.~B., {Cenko}, S.~B., \& {Rau}, A. 2006,
  GRB Coordinates Network, 5077, 1

\bibitem[{{Prieto} {et~al.}(2008){Prieto}, {Stanek}, \&
  {Beacom}}]{2008ApJ...673..999P}
{Prieto}, J.~L., {Stanek}, K.~Z., \& {Beacom}, J.~F. 2008, \apj, 673, 999,
  \dodoi{10.1086/524654}

\bibitem[{{Prochaska} \& {Zheng}(2019)}]{2019MNRAS.485..648P}
{Prochaska}, J.~X., \& {Zheng}, Y. 2019, \mnras, 485, 648,
  \dodoi{10.1093/mnras/stz261}

\bibitem[{{Prochaska} {et~al.}(2019){Prochaska}, {Macquart}, {McQuinn},
  {Simha}, {Shannon}, {Day}, {Marnoch}, {Ryder}, {Deller}, {Bannister},
  {Bhandari}, {Bordoloi}, {Bunton}, {Cho}, {Flynn}, {Mahony}, {Phillips},
  {Qiu}, \& {Tejos}}]{2019Sci...366..231P}
{Prochaska}, J.~X., {Macquart}, J.-P., {McQuinn}, M., {et~al.} 2019, Science,
  366, 231, \dodoi{10.1126/science.aay0073}

\bibitem[{{Ravi} {et~al.}(2019){Ravi}, {Catha}, {D'Addario}, {Djorgovski},
  {Hallinan}, {Hobbs}, {Kocz}, {Kulkarni}, {Shi}, {Vedantham}, {Weinreb}, \&
  {Woody}}]{2019Natur.572..352R}
{Ravi}, V., {Catha}, M., {D'Addario}, L., {et~al.} 2019, \nat, 572, 352,
  \dodoi{10.1038/s41586-019-1389-7}

\bibitem[{{Rowlinson} {et~al.}(2019){Rowlinson}, {Gourdji}, {van der Meulen},
  {Meyers}, {Shimwell}, {ter Veen}, {Wijers}, {Kuiack}, {Shulevski},
  {Broderick}, {van der Horst}, {Tasse}, {Hardcastle}, {Mechev}, \&
  {Williams}}]{2019MNRAS.490.3483R}
{Rowlinson}, A., {Gourdji}, K., {van der Meulen}, K., {et~al.} 2019, \mnras,
  490, 3483, \dodoi{10.1093/mnras/stz2866}

\bibitem[{{Rowlinson} {et~al.}(2021){Rowlinson}, {Starling}, {Gourdji},
  {Anderson}, {ter Veen}, {Mandhai}, {Wijers}, {Shimwell}, \& {van der
  Horst}}]{2021MNRAS.506.5268R}
{Rowlinson}, A., {Starling}, R.~L.~C., {Gourdji}, K., {et~al.} 2021, \mnras,
  506, 5268, \dodoi{10.1093/mnras/stab2060}

\bibitem[{{Rumyantsev} {et~al.}(2006){Rumyantsev}, {Karimov}, {Salyamov},
  {Pavlenko}, {Efimov}, {Pozanenko}, \& {Ibrahimov}}]{2006GCN..5184....1R}
{Rumyantsev}, V., {Karimov}, R., {Salyamov}, R., {et~al.} 2006, GRB Coordinates
  Network, 5184, 1

\bibitem[{{Safarzadeh} {et~al.}(2020){Safarzadeh}, {Prochaska}, {Heintz}, \&
  {Fong}}]{2020ApJ...905L..30S}
{Safarzadeh}, M., {Prochaska}, J.~X., {Heintz}, K.~E., \& {Fong}, W.-f. 2020,
  \apjl, 905, L30, \dodoi{10.3847/2041-8213/abd03e}

\bibitem[{{Sakamoto} {et~al.}(2016){Sakamoto}, {Barthelmy}, {Cummings},
  {Gehrels}, {Gibson}, {Krimm}, {Lien}, {Markwardt}, {Norris}, {Palmer},
  {Stamatikos}, \& {Ukwatta}}]{2016GCN.19276....1S}
{Sakamoto}, T., {Barthelmy}, S.~D., {Cummings}, J.~R., {et~al.} 2016, GRB
  Coordinates Network, 19276, 1

\bibitem[{{Sato} {et~al.}(2006){Sato}, {Barbier}, {Barthelmy}, {Cummings},
  {Fenimore}, {Gehrels}, {Hullinger}, {Krimm}, {Koss}, {Markwardt}, {Norris},
  {Palmer}, {Parsons}, {Sakamoto}, {Stamatikos}, {Troja}, \&
  {Tueller}}]{2006GCN..5064....1S}
{Sato}, G., {Barbier}, L., {Barthelmy}, S., {et~al.} 2006, GRB Coordinates
  Network, 5064, 1

\bibitem[{{Shannon} {et~al.}(2018){Shannon}, {Macquart}, {Bannister}, {Ekers},
  {James}, {Os{\l}owski}, {Qiu}, {Sammons}, {Hotan}, {Voronkov}, {Beresford},
  {Brothers}, {Brown}, {Bunton}, {Chippendale}, {Haskins}, {Leach},
  {Marquarding}, {McConnell}, {Pilawa}, {Sadler}, {Troup}, {Tuthill},
  {Whiting}, {Allison}, {Anderson}, {Bell}, {Collier}, {G{\"u}rkan}, {Heald},
  \& {Riseley}}]{2018Natur.562..386S}
{Shannon}, R.~M., {Macquart}, J.~P., {Bannister}, K.~W., {et~al.} 2018, \nat,
  562, 386, \dodoi{10.1038/s41586-018-0588-y}

\bibitem[{{Takahashi} {et~al.}(2006{\natexlab{a}}){Takahashi}, {Uehara},
  {Yoshida}, {Kobayashi}, {Koshiishi}, {Tanaka}, {Nakagawa}, {Sugita},
  {Yamaoka}, \& {Yoshida}}]{2006GCN..5073....1T}
{Takahashi}, I., {Uehara}, T., {Yoshida}, K., {et~al.} 2006{\natexlab{a}}, GRB
  Coordinates Network, 5073, 1

\bibitem[{{Takahashi} {et~al.}(2006{\natexlab{b}}){Takahashi}, {Uehara},
  {Yoshida}, {Kobayashi}, {Tanaka}, {Nakagawa}, {Sugita}, {Yamaoka}, \&
  {Yoshida}}]{2006GCN..5065....1T}
---. 2006{\natexlab{b}}, GRB Coordinates Network, 5065, 1

\bibitem[{{Tavani} {et~al.}(2020){Tavani}, {Verrecchia}, {Casentini}, {Perri},
  {Ursi}, {Pittori}, {Lucarelli}, {Pilia}, {Corongiu}, {Bernardi}, {Naldi}, \&
  {Pupillo}}]{2020ATel13446....1T}
{Tavani}, M., {Verrecchia}, F., {Casentini}, C., {et~al.} 2020, The
  Astronomer's Telegram, 13446, 1

\bibitem[{{Tendulkar} {et~al.}(2017){Tendulkar}, {Bassa}, {Cordes}, {Bower},
  {Law}, {Chatterjee}, {Adams}, {Bogdanov}, {Burke-Spolaor}, {Butler},
  {Demorest}, {Hessels}, {Kaspi}, {Lazio}, {Maddox}, {Marcote}, {McLaughlin},
  {Paragi}, {Ransom}, {Scholz}, {Seymour}, {Spitler}, {van Langevelde}, \&
  {Wharton}}]{2017ApJ...834L...7T}
{Tendulkar}, S.~P., {Bassa}, C.~G., {Cordes}, J.~M., {et~al.} 2017, \apjl, 834,
  L7, \dodoi{10.3847/2041-8213/834/2/L7}

\bibitem[{{Tendulkar} {et~al.}(2021){Tendulkar}, {Gil de Paz}, {Kirichenko},
  {Hessels}, {Bhardwaj}, {{\'A}vila}, {Bassa}, {Chawla}, {Fonseca}, {Kaspi},
  {Keimpema}, {Kirsten}, {Lazio}, {Marcote}, {Masui}, {Nimmo}, {Paragi},
  {Rahman}, {Pay{\'a}}, {Scholz}, \& {Stairs}}]{2021ApJ...908L..12T}
{Tendulkar}, S.~P., {Gil de Paz}, A., {Kirichenko}, A.~Y., {et~al.} 2021,
  \apjl, 908, L12, \dodoi{10.3847/2041-8213/abdb38}

\bibitem[{{The CHIME/FRB Collaboration} {et~al.}(2021){The CHIME/FRB
  Collaboration}, {:}, {Amiri}, {Andersen}, {Bandura}, {Berger}, {Bhardwaj},
  {Boyce}, {Boyle}, {Brar}, {Breitman}, {Cassanelli}, {Chawla}, {Chen},
  {Cliche}, {Cook}, {Cubranic}, {Curtin}, {Deng}, {Dobbs}, {Fengqiu}, {Dong},
  {Eadie}, {Fandino}, {Fonseca}, {Gaensler}, {Giri}, {Good}, {Halpern}, {Hill},
  {Hinshaw}, {Josephy}, {Kaczmarek}, {Kader}, {Kania}, {Kaspi}, {Landecker},
  {Lang}, {Leung}, {Li}, {Lin}, {Masui}, {Mckinven}, {Mena-Parra},
  {Merryfield}, {Meyers}, {Michilli}, {Milutinovic}, {Mirhosseini},
  {M{\"u}nchmeyer}, {Naidu}, {Newburgh}, {Ng}, {Patel}, {Pen}, {Petroff},
  {Pinsonneault-Marotte}, {Pleunis}, {Rafiei-Ravandi}, {Rahman}, {Ransom},
  {Renard}, {Sanghavi}, {Scholz}, {Shaw}, {Shin}, {Siegel}, {Sikora}, {Singh},
  {Smith}, {Stairs}, {Tan}, {Tendulkar}, {Vanderlinde}, {Wang}, {Wulf}, \&
  {Zwaniga}}]{2021arXiv210604352T}
{The CHIME/FRB Collaboration}, {:}, {Amiri}, M., {et~al.} 2021, arXiv e-prints,
  arXiv:2106.04352.
\newblock \doarXiv{2106.04352}

\bibitem[{{The LIGO Scientific Collaboration} {et~al.}(2022){The LIGO
  Scientific Collaboration}, {the Virgo Collaboration}, {the KAGRA
  Collaboration}, {the CHIME/FRB Collaboration}, {:}, {Abbott}, {Abbott},
  {Acernese}, {Ackley}, {Adams}, {Adhikari}, {Adhikari}, {Adya}, {Affeldt},
  {Agarwal}, {Agathos}, {Agatsuma}, {Aggarwal}, {Aguiar}, {Aiello}, {Ain},
  {Ajith}, {Akutsu}, {Albanesi}, {Allocca}, {Altin}, {Amato}, {Anand}, {Anand},
  {Ananyeva}, {Anderson}, {Anderson}, {Ando}, {Andrade}, {Andres},
  {Andri{\'c}}, {Angelova}, {Ansoldi}, {Antelis}, {Antier}, {Appert}, {Arai},
  {Arai}, {Arai}, {Araki}, {Araya}, {Araya}, {Areeda}, {Ar{\`e}ne}, {Aritomi},
  {Arnaud}, {Aronson}, {Arun}, {Asada}, {Asali}, {Ashton}, {Aso}, {Assiduo},
  {Aston}, {Astone}, {Aubin}, {Austin}, {Babak}, {Badaracco}, {Bader},
  {Badger}, {Bae}, {Bae}, {Baer}, {Bagnasco}, {Bai}, {Baiotti}, {Baird},
  {Bajpai}, {Ball}, {Ballardin}, {Ballmer}, {Balsamo}, {Baltus}, {Banagiri},
  {Bankar}, {Barayoga}, {Barbieri}, {Barish}, {Barker}, {Barneo}, {Barone},
  {Barr}, {Barsotti}, {Barsuglia}, {Barta}, {Bartlett}, {Barton}, {Bartos},
  {Bassiri}, {Basti}, {Bawaj}, {Bayley}, {Baylor}, {Bazzan}, {B{\'e}csy},
  {Bedakihale}, {Bejger}, {Belahcene}, {Benedetto}, {Beniwal}, {Bennett},
  {Bentley}, {BenYaala}, {Bergamin}, {Berger}, {Bernuzzi}, {Berry},
  {Bersanetti}, {Bertolini}, {Betzwieser}, {Beveridge}, {Bhandare}, {Bhardwaj},
  {Bhattacharjee}, {Bhaumik}, {Bilenko}, {Billingsley}, {Bini}, {Birney},
  {Birnholtz}, {Biscans}, {Bischi}, {Biscoveanu}, {Bisht}, {Biswas}, {Bitossi},
  {Bizouard}, {Blackburn}, {Blair}, {Blair}, {Blair}, {Bobba}, {Bode}, {Boer},
  {Bogaert}, {Boldrini}, {Bonavena}, {Bondu}, {Bonilla}, {Bonnand}, {Booker},
  {Boom}, {Bork}, {Boschi}, {Bose}, {Bose}, {Bossilkov}, {Boudart},
  {Bouffanais}, {Bozzi}, {Bradaschia}, {Brady}, {Bramley}, {Branch},
  {Branchesi}, {Brau}, {Breschi}, {Briant}, {Briggs}, {Brillet}, {Brinkmann},
  {Brockill}, {Brooks}, {Brooks}, {Brown}, {Brunett}, {Bruno}, {Bruntz},
  {Bryant}, {Buchanan}, {Bulik}, {Bulten}, {Buonanno}, {Buscicchio},
  {Buskulic}, {Buy}, {Byer}, {Cadonati}, {Cagnoli}, {Cahillane}, {Calder{\'o}n
  Bustillo}, {Callaghan}, {Callister}, {Calloni}, {Cameron}, {Camp}, {Canepa},
  {Canevarolo}, {Cannavacciuolo}, {Cannon}, {Cao}, {Cao}, {Capocasa}, {Capote},
  {Carapella}, {Carbognani}, {Carlin}, {Carney}, {Carpinelli}, {Carrillo},
  {Carullo}, {Carver}, {Casanueva Diaz}, {Casentini}, {Castaldi}, {Caudill},
  {Cavagli{\`a}}, {Cavalier}, {Cavalieri}, {Ceasar}, {Cella},
  {Cerd{\'a}-Dur{\'a}n}, {Cesarini}, {Chaibi}, {Chakravarti}, {Chalathadka
  Subrahmanya}, {Champion}, {Chan}, {Chan}, {Chan}, {Chan}, {Chan}, {Chandra},
  {Chanial}, {Chao}, {Charlton}, {Chase}, {Chassande-Mottin}, {Chatterjee},
  {Chatterjee}, {Chatterjee}, {Chaturvedi}, {Chaty}, {Chen}, {Chen}, {Chen},
  {Chen}, {Chen}, {Chen}, {Chen}, {Chen}, {Cheng}, {Cheong}, {Cheung}, {Chia},
  {Chiadini}, {Chiang}, {Chiarini}, {Chierici}, {Chincarini}, {Chiofalo},
  {Chiummo}, {Cho}, {Cho}, {Choudhary}, {Choudhary}, {Christensen}, {Chu},
  {Chu}, {Chu}, {Chua}, {Chung}, {Ciani}, {Ciecielag}, {Cie{\'s}lar},
  {Cifaldi}, {Ciobanu}, {Ciolfi}, {Cipriano}, {Cirone}, {Clara}, {Clark},
  {Clark}, {Clarke}, {Clearwater}, {Clesse}, {Cleva}, {Coccia}, {Codazzo},
  {Cohadon}, {Cohen}, {Cohen}, {Colleoni}, {Collette}, {Colombo}, {Colpi},
  {Compton}, {Constancio}, {Conti}, {Cooper}, {Corban}, {Corbitt},
  {Cordero-Carri{\'o}n}, {Corezzi}, {Corley}, {Cornish}, {Corre}, {Corsi},
  {Cortese}, {Costa}, {Cotesta}, {Coughlin}, {Coulon}, {Countryman}, {Cousins},
  {Couvares}, {Coward}, {Cowart}, {Coyne}, {Coyne}, {Creighton}, {Creighton},
  {Criswell}, {Croquette}, {Crowder}, {Cudell}, {Cullen}, {Cumming},
  {Cummings}, {Cunningham}, {Cuoco}, {Cury{\l}o}, {Dabadie}, {Dal Canton},
  {Dall'Osso}, {D{\'a}lya}, {Dana}, {DaneshgaranBajastani}, {D'Angelo},
  {Danilishin}, {D'Antonio}, {Danzmann}, {Darsow-Fromm}, {Dasgupta}, {Datrier},
  {Datta}, {Dattilo}, {Dave}, {Davier}, {Davies}, {Davis}, {Davis}, {Daw},
  {Dean}, {DeBra}, {Deenadayalan}, {Degallaix}, {De Laurentis},
  {Del{\'e}glise}, {Del Favero}, {De Lillo}, {De Lillo}, {Del Pozzo},
  {DeMarchi}, {De Matteis}, {D'Emilio}, {Demos}, {Dent}, {Depasse}, {De
  Pietri}, {De Rosa}, {De Rossi}, {DeSalvo}, {De Simone}, {Dhurandhar},
  {D{\'\i}az}, {Diaz-Ortiz}, {Didio}, {Dietrich}, {Di Fiore}, {Di Fronzo}, {Di
  Giorgio}, {Di Giovanni}, {Di Giovanni}, {Di Girolamo}, {Di Lieto}, {Ding},
  {Di Pace}, {Di Palma}, {Di Renzo}, {Divakarla}, {Dmitriev}, {Doctor},
  {D'Onofrio}, {Donovan}, {Dooley}, {Doravari}, {Dorrington}, {Drago},
  {Driggers}, {Drori}, {Ducoin}, {Dupej}, {Durante}, {D'Urso}, {Duverne},
  {Dwyer}, {Eassa}, {Easter}, {Ebersold}, {Eckhardt}, {Eddolls}, {Edelman},
  {Edo}, {Edy}, {Effler}, {Eguchi}, {Eichholz}, {Eikenberry}, {Eisenmann},
  {Eisenstein}, {Ejlli}, {Engelby}, {Enomoto}, {Errico}, {Essick},
  {Estell{\'e}s}, {Estevez}, {Etienne}, {Etzel}, {Evans}, {Evans}, {Ewing},
  {Fafone}, {Fair}, {Fairhurst}, {Farah}, {Farinon}, {Farr}, {Farr}, {Farrow},
  {Fauchon-Jones}, {Favaro}, {Favata}, {Fays}, {Fazio}, {Feicht}, {Fejer},
  {Fenyvesi}, {Ferguson}, {Fernandez-Galiana}, {Ferrante}, {Ferreira},
  {Fidecaro}, {Figura}, {Fiori}, {Fishbach}, {Fisher}, {Fittipaldi}, {Fiumara},
  {Flaminio}, {Floden}, {Fong}, {Font}, {Fornal}, {Forsyth}, {Franke},
  {Frasca}, {Frasconi}, {Frederick}, {Freed}, {Frei}, {Freise}, {Frey},
  {Fritschel}, {Frolov}, {Fronz{\'e}}, {Fujii}, {Fujikawa}, {Fukunaga},
  {Fukushima}, {Fulda}, {Fyffe}, {Gabbard}, {Gadre}, {Gair}, {Gais},
  {Galaudage}, {Gamba}, {Ganapathy}, {Ganguly}, {Gao}, {Gaonkar}, {Garaventa},
  {Garc{\'\i}a-N{\'u}{\~n}ez}, {Garc{\'\i}a-Quir{\'o}s}, {Garufi}, {Gateley},
  {Gaudio}, {Gayathri}, {Ge}, {Gemme}, {Gennai}, {George}, {Gerberding},
  {Gergely}, {Gewecke}, {Ghonge}, {Ghosh}, {Ghosh}, {Ghosh}, {Ghosh},
  {Giacomazzo}, {Giacoppo}, {Giaime}, {Giardina}, {Gibson}, {Gier}, {Giesler},
  {Giri}, {Gissi}, {Glanzer}, {Gleckl}, {Godwin}, {Goetz}, {Goetz}, {Gohlke},
  {Goncharov}, {Gonz{\'a}lez}, {Gopakumar}, {Gosselin}, {Gouaty}, {Gould},
  {Grace}, {Grado}, {Granata}, {Granata}, {Grant}, {Gras}, {Grassia}, {Gray},
  {Gray}, {Greco}, {Green}, {Green}, {Gretarsson}, {Gretarsson}, {Griffith},
  {Griffiths}, {Griggs}, {Grignani}, {Grimaldi}, {Grimm}, {Grote}, {Grunewald},
  {Gruning}, {Guerra}, {Guidi}, {Guimaraes}, {Guix{\'e}}, {Gulati}, {Guo},
  {Guo}, {Gupta}, {Gupta}, {Gupta}, {Gustafson}, {Gustafson}, {Guzman}, {Ha},
  {Haegel}, {Hagiwara}, {Haino}, {Halim}, {Hall}, {Hamilton}, {Hammond}, {Han},
  {Haney}, {Hanks}, {Hanna}, {Hannam}, {Hannuksela}, {Hansen}, {Hansen},
  {Hanson}, {Harder}, {Hardwick}, {Haris}, {Harms}, {Harry}, {Harry},
  {Hartwig}, {Hasegawa}, {Haskell}, {Hasskew}, {Haster}, {Hattori}, {Haughian},
  {Hayakawa}, {Hayama}, {Hayes}, {Healy}, {Heidmann}, {Heidt}, {Heintze},
  {Heinze}, {Heinzel}, {Heitmann}, {Hellman}, {Hello}, {Helmling-Cornell},
  {Hemming}, {Hendry}, {Heng}, {Hennes}, {Hennig}, {Hennig}, {Hernandez},
  {Hernandez Vivanco}, {Heurs}, {Hild}, {Hill}, {Himemoto}, {Hines},
  {Hiranuma}, {Hirata}, {Hirose}, {Hochheim}, {Hofman}, {Hohmann}, {Holcomb},
  {Holland}, {Hollows}, {Holmes}, {Holt}, {Holz}, {Hong}, {Hopkins}, {Hough},
  {Hourihane}, {Howell}, {Hoy}, {Hoyland}, {Hreibi}, {Hsieh}, {Hsu}, {Huang},
  {Huang}, {Huang}, {Huang}, {Huang}, {Huang}, {H{\"u}bner}, {Huddart},
  {Hughey}, {Hui}, {Hui}, {Husa}, {Huttner}, {Huxford}, {Huynh-Dinh}, {Ide},
  {Idzkowski}, {Iess}, {Ikenoue}, {Imam}, {Inayoshi}, {Ingram}, {Inoue},
  {Ioka}, {Isi}, {Isleif}, {Ito}, {Itoh}, {Iyer}, {Izumi}, {JaberianHamedan},
  {Jacqmin}, {Jadhav}, {Jadhav}, {James}, {Jan}, {Jani}, {Janquart},
  {Janssens}, {Janthalur}, {Jaranowski}, {Jariwala}, {Jaume}, {Jenkins},
  {Jenner}, {Jeon}, {Jeunon}, {Jia}, {Jin}, {Johns}, {Jones}, {Jones}, {Jones},
  {Jones}, {Jones}, {Jonker}, {Ju}, {Jung}, {Jung}, {Junker}, {Juste},
  {Kaihotsu}, {Kajita}, {Kakizaki}, {Kalaghatgi}, {Kalogera}, {Kamai},
  {Kamiizumi}, {Kanda}, {Kandhasamy}, {Kang}, {Kanner}, {Kao}, {Kapadia},
  {Kapasi}, {Karat}, {Karathanasis}, {Karki}, {Kashyap}, {Kasprzack},
  {Kastaun}, {Katsanevas}, {Katsavounidis}, {Katzman}, {Kaur}, {Kawabe},
  {Kawaguchi}, {Kawai}, {Kawasaki}, {K{\'e}f{\'e}lian}, {Keitel}, {Key},
  {Khadka}, {Khalili}, {Khan}, {Khazanov}, {Khetan}, {Khursheed}, {Kijbunchoo},
  {Kim}, {Kim}, {Kim}, {Kim}, {Kim}, {Kim}, {Kimball}, {Kimura},
  {Kinley-Hanlon}, {Kirchhoff}, {Kissel}, {Kita}, {Kitazawa}, {Kleybolte},
  {Klimenko}, {Knee}, {Knowles}, {Knyazev}, {Koch}, {Koekoek}, {Kojima},
  {Kokeyama}, {Koley}, {Kolitsidou}, {Kolstein}, {Komori}, {Kondrashov},
  {Kong}, {Kontos}, {Koper}, {Korobko}, {Kotake}, {Kovalam}, {Kozak},
  {Kozakai}, {Kozu}, {Kringel}, {Krishnendu}, {Kr{\'o}lak}, {Kuehn}, {Kuei},
  {Kuijer}, {Kumar}, {Kumar}, {Kumar}, {Kumar}, {Kume}, {Kuns}, {Kuo}, {Kuo},
  {Kuromiya}, {Kuroyanagi}, {Kusayanagi}, {Kuwahara}, {Kwak}, {Lagabbe},
  {Laghi}, {Lalande}, {Lam}, {Lamberts}, {Landry}, {Lane}, {Lang}, {Lange},
  {Lantz}, {La Rosa}, {Lartaux-Vollard}, {Lasky}, {Laxen}, {Lazzarini},
  {Lazzaro}, {Leaci}, {Leavey}, {Lecoeuche}, {Lee}, {Lee}, {Lee}, {Lee}, {Lee},
  {Lee}, {Lehmann}, {Lema{\^\i}tre}, {Leonardi}, {Leroy}, {Letendre},
  {Levesque}, {Levin}, {Leviton}, {Leyde}, {Li}, {Li}, {Li}, {Li}, {Li}, {Li},
  {Lin}, {Lin}, {Lin}, {Lin}, {Lin}, {Linde}, {Linker}, {Linley}, {Littenberg},
  {Liu}, {Liu}, {Liu}, {Liu}, {Llamas}, {Llorens-Monteagudo}, {Lo}, {Lockwood},
  {London}, {Longo}, {Lopez}, {Lopez Portilla}, {Lorenzini}, {Loriette},
  {Lormand}, {Losurdo}, {Lott}, {Lough}, {Lousto}, {Lovelace}, {Lucaccioni},
  {L{\"u}ck}, {Lumaca}, {Lundgren}, {Luo}, {Lynam}, {Macas}, {MacInnis},
  {Macleod}, {MacMillan}, {Macquet}, {Maga{\~n}a Hernandez}, {Magazz{\`u}},
  {Magee}, {Maggiore}, {Magnozzi}, {Mahesh}, {Majorana}, {Makarem},
  {Maksimovic}, {Maliakal}, {Malik}, {Man}, {Mandic}, {Mangano}, {Mango},
  {Mansell}, {Manske}, {Mantovani}, {Mapelli}, {Marchesoni}, {Marchio},
  {Marion}, {Mark}, {M{\'a}rka}, {M{\'a}rka}, {Markakis}, {Markosyan},
  {Markowitz}, {Maros}, {Marquina}, {Marsat}, {Martelli}, {Martin}, {Martin},
  {Martinez}, {Martinez}, {Martinez}, {Martinovic}, {Martynov}, {Marx},
  {Masalehdan}, {Mason}, {Massera}, {Masserot}, {Massinger}, {Masso-Reid},
  {Mastrogiovanni}, {Matas}, {Mateu-Lucena}, {Matichard}, {Matiushechkina},
  {Mavalvala}, {McCann}, {McCarthy}, {McClelland}, {McClincy}, {McCormick},
  {McCuller}, {McGhee}, {McGuire}, {McIsaac}, {McIver}, {McRae}, {McWilliams},
  {Meacher}, {Mehmet}, {Mehta}, {Meijer}, {Melatos}, {Melchor}, {Mendell},
  {Menendez-Vazquez}, {Menoni}, {Mercer}, {Mereni}, {Merfeld}, {Merilh},
  {Merritt}, {Merzougui}, {Meshkov}, {Messenger}, {Messick}, {Meyers},
  {Meylahn}, {Mhaske}, {Miani}, {Miao}, {Michaloliakos}, {Michel}, {Michimura},
  {Middleton}, {Milano}, {Miller}, {Miller}, {Miller}, {Millhouse}, {Mills},
  {Milotti}, {Minazzoli}, {Minenkov}, {Mio}, {Mir}, {Miravet-Ten{\'e}s},
  {Mishra}, {Mishra}, {Mistry}, {Mitra}, {Mitrofanov}, {Mitselmakher},
  {Mittleman}, {Miyakawa}, {Miyamoto}, {Miyazaki}, {Miyo}, {Miyoki}, {Mo},
  {Moguel}, {Mogushi}, {Mohapatra}, {Mohite}, {Molina}, {Molina-Ruiz},
  {Mondin}, {Montani}, {Moore}, {Moraru}, {Morawski}, {More}, {Moreno},
  {Moreno}, {Mori}, {Morisaki}, {Moriwaki}, {Mours}, {Mow-Lowry}, {Mozzon},
  {Muciaccia}, {Mukherjee}, {Mukherjee}, {Mukherjee}, {Mukherjee}, {Mukherjee},
  {Mukund}, {Mullavey}, {Munch}, {Mu{\~n}iz}, {Murray}, {Musenich}, {Muusse},
  {Nadji}, {Nagano}, {Nagano}, {Nagar}, {Nakamura}, {Nakano}, {Nakano},
  {Nakashima}, {Nakayama}, {Napolano}, {Nardecchia}, {Narikawa}, {Naticchioni},
  {Nayak}, {Nayak}, {Negishi}, {Neil}, {Neilson}, {Nelemans}, {Nelson}, {Nery},
  {Neubauer}, {Neunzert}, {Ng}, {Ng}, {Nguyen}, {Nguyen}, {Nguyen}, {Nguyen
  Quynh}, {Ni}, {Nichols}, {Nishizawa}, {Nissanke}, {Nitoglia}, {Nocera},
  {Norman}, {North}, {Nozaki}, {Nuttall}, {Oberling}, {O'Brien}, {Obuchi},
  {O'Dell}, {Oelker}, {Ogaki}, {Oganesyan}, {Oh}, {Oh}, {Oh}, {Ohashi},
  {Ohishi}, {Ohkawa}, {Ohme}, {Ohta}, {Okada}, {Okutani}, {Okutomi},
  {Olivetto}, {Oohara}, {Ooi}, {Oram}, {O'Reilly}, {Ormiston}, {Ormsby},
  {Ortega}, {O'Shaughnessy}, {O'Shea}, {Oshino}, {Ossokine}, {Osthelder},
  {Otabe}, {Ottaway}, {Overmier}, {Pace}, {Pagano}, {Page}, {Pagliaroli},
  {Pai}, {Pai}, {Palamos}, {Palashov}, {Palomba}, {Pan}, {Pan}, {Panda},
  {Pang}, {Pang}, {Pankow}, {Pannarale}, {Pant}, {Panther}, {Paoletti},
  {Paoli}, {Paolone}, {Parisi}, {Park}, {Park}, {Parker}, {Pascucci},
  {Pasqualetti}, {Passaquieti}, {Passuello}, {Patel}, {Pathak}, {Patricelli},
  {Patron}, {Patrone}, {Paul}, {Payne}, {Pedraza}, {Pegoraro}, {Pele},
  {Pe{\~n}a Arellano}, {Penn}, {Perego}, {Pereira}, {Pereira}, {Perez},
  {P{\'e}rigois}, {Perkins}, {Perreca}, {Perri{\`e}s}, {Petermann},
  {Petterson}, {Pfeiffer}, {Pham}, {Phukon}, {Piccinni}, {Pichot},
  {Piendibene}, {Piergiovanni}, {Pierini}, {Pierro}, {Pillant}, {Pillas},
  {Pilo}, {Pinard}, {Pinto}, {Pinto}, {Piotrzkowski}, {Pirello}, {Pitkin},
  {Placidi}, {Planas}, {Plastino}, {Pluchar}, {Poggiani}, {Polini}, {Pong},
  {Ponrathnam}, {Popolizio}, {Porter}, {Poulton}, {Powell}, {Pracchia},
  {Pradier}, {Prajapati}, {Prasai}, {Prasanna}, {Pratten}, {Principe}, {Prodi},
  {Prokhorov}, {Prosposito}, {Prudenzi}, {Puecher}, {Punturo}, {Puosi},
  {Puppo}, {P{\"u}rrer}, {Qi}, {Quetschke}, {Quitzow-James}, {Raab},
  {Raaijmakers}, {Radkins}, {Radulesco}, {Raffai}, {Rail}, {Raja}, {Rajan},
  {Ramirez}, {Ramirez}, {Ramos-Buades}, {Rana}, {Rapagnani}, {Rapol}, {Ray},
  {Raymond}, {Raza}, {Razzano}, {Read}, {Rees}, {Regimbau}, {Rei}, {Reid},
  {Reid}, {Reitze}, {Relton}, {Renzini}, {Rettegno}, {Rezac}, {Ricci},
  {Richards}, {Richardson}, {Richardson}, {Riemenschneider}, {Riles},
  {Rinaldi}, {Rink}, {Rizzo}, {Robertson}, {Robie}, {Robinet}, {Rocchi},
  {Rodriguez}, {Rolland}, {Rollins}, {Romanelli}, {Romano}, {Romel},
  {Romero-Rodr{\'\i}guez}, {Romero-Shaw}, {Romie}, {Ronchini}, {Rosa}, {Rose},
  {Rosi{\'n}ska}, {Ross}, {Rowan}, {Rowlinson}, {Roy}, {Roy}, {Roy}, {Rozza},
  {Ruggi}, {Ryan}, {Sachdev}, {Sadecki}, {Sadiq}, {Sago}, {Saito}, {Saito},
  {Sakai}, {Sakai}, {Sakellariadou}, {Sakuno}, {Salafia}, {Salconi}, {Saleem},
  {Salemi}, {Samajdar}, {Sanchez}, {Sanchez}, {Sanchez}, {Sanchis-Gual},
  {Sanders}, {Sanuy}, {Saravanan}, {Sarin}, {Sassolas}, {Satari}, {Sato},
  {Sato}, {Sauter}, {Savage}, {Sawada}, {Sawant}, {Sawant}, {Sayah},
  {Schaetzl}, {Scheel}, {Scheuer}, {Schiworski}, {Schmidt}, {Schmidt},
  {Schnabel}, {Schneewind}, {Schofield}, {Sch{\"o}nbeck}, {Schulte}, {Schutz},
  {Schwartz}, {Scott}, {Scott}, {Seglar-Arroyo}, {Sekiguchi}, {Sekiguchi},
  {Sellers}, {Sengupta}, {Sentenac}, {Seo}, {Sequino}, {Sergeev}, {Setyawati},
  {Shaffer}, {Shahriar}, {Shams}, {Shao}, {Sharma}, {Sharma}, {Shawhan},
  {Shcheblanov}, {Shibagaki}, {Shikauchi}, {Shimizu}, {Shimoda}, {Shimode},
  {Shinkai}, {Shishido}, {Shoda}, {Shoemaker}, {Shoemaker}, {ShyamSundar},
  {Sieniawska}, {Sigg}, {Singer}, {Singh}, {Singh}, {Singha}, {Sintes},
  {Sipala}, {Skliris}, {Slagmolen}, {Slaven-Blair}, {Smetana}, {Smith},
  {Smith}, {Soldateschi}, {Somala}, {Somiya}, {Son}, {Soni}, {Soni}, {Sordini},
  {Sorrentino}, {Sorrentino}, {Sotani}, {Soulard}, {Souradeep}, {Sowell},
  {Spagnuolo}, {Spencer}, {Spera}, {Srinivasan}, {Srivastava}, {Srivastava},
  {Staats}, {Stachie}, {Steer}, {Steinlechner}, {Steinlechner}, {Stops},
  {Stover}, {Strain}, {Strang}, {Stratta}, {Strunk}, {Sturani}, {Stuver},
  {Sudhagar}, {Sudhir}, {Sugimoto}, {Suh}, {Summerscales}, {Sun}, {Sun},
  {Sunil}, {Sur}, {Suresh}, {Sutton}, {Suzuki}, {Suzuki}, {Swinkels},
  {Szczepa{\'n}czyk}, {Szewczyk}, {Tacca}, {Tagoshi}, {Tait}, {Takahashi},
  {Takahashi}, {Takamori}, {Takano}, {Takeda}, {Takeda}, {Talbot}, {Talbot},
  {Tanaka}, {Tanaka}, {Tanaka}, {Tanaka}, {Tanaka}, {Tanasijczuk}, {Tanioka},
  {Tanner}, {Tao}, {Tao}, {Tapia San Mart{\'\i}n}, {Taranto}, {Tasson},
  {Telada}, {Tenorio}, {Terhune}, {Terkowski}, {Thirugnanasambandam}, {Thomas},
  {Thomas}, {Thompson}, {Thondapu}, {Thorne}, {Thrane}, {Tiwari}, {Tiwari},
  {Tiwari}, {Toivonen}, {Toland}, {Tolley}, {Tomaru}, {Tomigami}, {Tomura},
  {Tonelli}, {Torres-Forn{\'e}}, {Torrie}, {Tosta e Melo}, {T{\"o}yr{\"a}},
  {Trapananti}, {Travasso}, {Traylor}, {Trevor}, {Tringali}, {Tripathee},
  {Troiano}, {Trovato}, {Trozzo}, {Trudeau}, {Tsai}, {Tsai}, {Tsang}, {Tsang},
  {Tsao}, {Tse}, {Tso}, {Tsubono}, {Tsuchida}, {Tsukada}, {Tsuna}, {Tsutsui},
  {Tsuzuki}, {Turbang}, {Turconi}, {Tuyenbayev}, {Ubhi}, {Uchikata},
  {Uchiyama}, {Udall}, {Ueda}, {Uehara}, {Ueno}, {Ueshima}, {Unnikrishnan},
  {Uraguchi}, {Urban}, {Ushiba}, {Utina}, {Vahlbruch}, {Vajente}, {Vajpeyi},
  {Valdes}, {Valentini}, {Valsan}, {van Bakel}, {van Beuzekom}, {van den
  Brand}, {Van Den Broeck}, {Vander-Hyde}, {van der Schaaf}, {van Heijningen},
  {Vanosky}, {van Putten}, {van Remortel}, {Vardaro}, {Vargas}, {Varma},
  {Vas{\'u}th}, {Vecchio}, {Vedovato}, {Veitch}, {Veitch}, {Venneberg},
  {Venugopalan}, {Verkindt}, {Verma}, {Verma}, {Veske}, {Vetrano},
  {Vicer{\'e}}, {Vidyant}, {Viets}, {Vijaykumar}, {Villa-Ortega}, {Vinet},
  {Virtuoso}, {Vitale}, {Vo}, {Vocca}, {von Reis}, {von Wrangel}, {Vorvick},
  {Vyatchanin}, {Wade}, {Wade}, {Wagner}, {Walet}, {Walker}, {Wallace},
  {Wallace}, {Walsh}, {Wang}, {Wang}, {Wang}, {Ward}, {Warner}, {Was},
  {Washimi}, {Washington}, {Watada}, {Watchi}, {Weaver}, {Webster}, {Weinert},
  {Weinstein}, {Weiss}, {Weller}, {Wellmann}, {Wen}, {We{\ss}els}, {Wette},
  {Whelan}, {White}, {Whiting}, {Whittle}, {Wilken}, {Williams}, {Williams},
  {Williamson}, {Willis}, {Willke}, {Wilson}, {Winkler}, {Wipf}, {Wlodarczyk},
  {Woan}, {Woehler}, {Wofford}, {Wong}, {Wu}, {Wu}, {Wu}, {Wu}, {Wysocki},
  {Xiao}, {Xu}, {Yamada}, {Yamamoto}, {Yamamoto}, {Yamamoto}, {Yamamoto},
  {Yamashita}, {Yamazaki}, {Yang}, {Yang}, {Yang}, {Yang}, {Yang}, {Yap},
  {Yeeles}, {Yelikar}, {Ying}, {Yokogawa}, {Yokoyama}, {Yokozawa}, {Yoo},
  {Yoshioka}, {Yu}, {Yu}, {Yuzurihara}, {Zadro{\.z}ny}, {Zanolin}, {Zeidler},
  {Zelenova}, {Zendri}, {Zevin}, {Zhan}, {Zhang}, {Zhang}, {Zhang}, {Zhang},
  {Zhang}, {Zhao}, {Zhao}, {Zhao}, {Zhao}, {Zhou}, {Zhou}, {Zhu}, {Zhu},
  {Zimmerman}, {Zucker}, {Zweizig}, {Bhardwaj}, {Boyle}, {Cassanelli}, {Dong},
  {Fonseca}, {Kaspi}, {Leung}, {Masui}, {Meyers}, {Michilli}, {Ng}, {Pearlman},
  {Petroff}, {Pleunis}, {Rafiei-Ravandi}, {Rahman}, {Ransom}, {Scholz}, {Shin},
  {Smith}, {Stairs}, {Tendulkar}, \& {Zwaniga}}]{2022arXiv220312038T}
{The LIGO Scientific Collaboration}, {the Virgo Collaboration}, {the KAGRA
  Collaboration}, {et~al.} 2022, arXiv e-prints, arXiv:2203.12038.
\newblock \doarXiv{2203.12038}

\bibitem[{{Thomas} {et~al.}(2009){Thomas}, {Saglia}, {Bender}, {Thomas},
  {Gebhardt}, {Magorrian}, {Corsini}, \& {Wegner}}]{2009ApJ...691..770T}
{Thomas}, J., {Saglia}, R.~P., {Bender}, R., {et~al.} 2009, \apj, 691, 770,
  \dodoi{10.1088/0004-637X/691/1/770}

\bibitem[{{Thornton} {et~al.}(2013){Thornton}, {Stappers}, {Bailes},
  {Barsdell}, {Bates}, {Bhat}, {Burgay}, {Burke-Spolaor}, {Champion}, {Coster},
  {D'Amico}, {Jameson}, {Johnston}, {Keith}, {Kramer}, {Levin}, {Milia}, {Ng},
  {Possenti}, \& {van Straten}}]{2013Sci...341...53T}
{Thornton}, D., {Stappers}, B., {Bailes}, M., {et~al.} 2013, Science, 341, 53,
  \dodoi{10.1126/science.1236789}

\bibitem[{{Tian} {et~al.}(2022){Tian}, {Anderson}, {Hancock}, {Miller-Jones},
  {Sokolowski}, {Rowlinson}, {Williams}, {Morgan}, {Hurley-Walker}, {Kaplan},
  {Murphy}, {Tingay}, {Johnston-Hollitt}, {Bannister}, {Bell}, \&
  {Meyers}}]{2022PASA...39....3T}
{Tian}, J., {Anderson}, G.~E., {Hancock}, P.~J., {et~al.} 2022, \pasa, 39,
  e003, \dodoi{10.1017/pasa.2021.58}

\bibitem[{{Totani}(2013)}]{2013PASJ...65L..12T}
{Totani}, T. 2013, \pasj, 65, L12, \dodoi{10.1093/pasj/65.5.L12}

\bibitem[{{Troja} {et~al.}(2006{\natexlab{a}}){Troja}, {Burrows}, \&
  {Gehrels}}]{2006GCN..5093....1T}
{Troja}, E., {Burrows}, D.~N., \& {Gehrels}, N. 2006{\natexlab{a}}, GRB
  Coordinates Network, 5093, 1

\bibitem[{{Troja} {et~al.}(2008){Troja}, {King}, {O'Brien}, {Lyons}, \&
  {Cusumano}}]{2008MNRAS.385L..10T}
{Troja}, E., {King}, A.~R., {O'Brien}, P.~T., {Lyons}, N., \& {Cusumano}, G.
  2008, \mnras, 385, L10, \dodoi{10.1111/j.1745-3933.2007.00421.x}

\bibitem[{{Troja} {et~al.}(2006{\natexlab{b}}){Troja}, {Barthelmy}, {Boyd},
  {Burrows}, {Cummings}, {Gehrels}, {Holland}, {Hunsberger}, {Kennea}, {Krimm},
  {La Parola}, {Mangano}, {Marshall}, {O'Brien}, {Page}, {Palmer}, {Sakamoto},
  {Tagliaferri}, \& {vanden Berk}}]{2006GCN..5055....1T}
{Troja}, E., {Barthelmy}, S.~D., {Boyd}, P.~T., {et~al.} 2006{\natexlab{b}},
  GRB Coordinates Network, 5055, 1

\bibitem[{{Ukwatta} {et~al.}(2008){Ukwatta}, {Baumgartner}, {Chester},
  {Gehrels}, {Holland}, {Hoversten}, {Immler}, {Kennea}, {Mangano}, {Marshall},
  {Palmer}, {Sakamoto}, {Sbarufatti}, \& {vanden Berk}}]{2008GCN..7203....1U}
{Ukwatta}, T.~N., {Baumgartner}, W.~H., {Chester}, M.~M., {et~al.} 2008, GRB
  Coordinates Network, 7203, 1

\bibitem[{{Verbunt}(2003)}]{2003ASPC..296..245V}
{Verbunt}, F. 2003, in Astronomical Society of the Pacific Conference Series,
  Vol. 296, New Horizons in Globular Cluster Astronomy, ed. G.~{Piotto},
  G.~{Meylan}, S.~G. {Djorgovski}, \& M.~{Riello}, 245.
\newblock \doarXiv{astro-ph/0210057}

\bibitem[{{Vink} \& {Kuiper}(2006)}]{2006MNRAS.370L..14V}
{Vink}, J., \& {Kuiper}, L. 2006, \mnras, 370, L14,
  \dodoi{10.1111/j.1745-3933.2006.00178.x}

\bibitem[{{Wadiasingh} \& {Timokhin}(2019)}]{2019ApJ...879....4W}
{Wadiasingh}, Z., \& {Timokhin}, A. 2019, \apj, 879, 4,
  \dodoi{10.3847/1538-4357/ab2240}

\bibitem[{{Wang} {et~al.}(2020{\natexlab{a}}){Wang}, {Wang}, {Yang}, {Yu},
  {Zuo}, \& {Dai}}]{2020ApJ...891...72W}
{Wang}, F.~Y., {Wang}, Y.~Y., {Yang}, Y.-P., {et~al.} 2020{\natexlab{a}}, \apj,
  891, 72, \dodoi{10.3847/1538-4357/ab74d0}

\bibitem[{{Wang} {et~al.}(2016){Wang}, {Yang}, {Wu}, {Dai}, \&
  {Wang}}]{2016ApJ...822L...7W}
{Wang}, J.-S., {Yang}, Y.-P., {Wu}, X.-F., {Dai}, Z.-G., \& {Wang}, F.-Y. 2016,
  \apjl, 822, L7, \dodoi{10.3847/2041-8205/822/1/L7}

\bibitem[{{Wang} {et~al.}(2021){Wang}, {Xu}, {Wang}, {Du}, {Cheng}, {Zheng}, \&
  {Xu}}]{2021MNRAS.507.2208W}
{Wang}, W.-H., {Xu}, H., {Wang}, W.-Y., {et~al.} 2021, \mnras, 507, 2208,
  \dodoi{10.1093/mnras/stab2213}

\bibitem[{{Wang} {et~al.}(2020{\natexlab{b}}){Wang}, {Li}, {Yang}, {Luo},
  {Zhang}, {Lin}, {Liang}, \& {Qin}}]{2020ApJ...894L..22W}
{Wang}, X.-G., {Li}, L., {Yang}, Y.-P., {et~al.} 2020{\natexlab{b}}, \apjl,
  894, L22, \dodoi{10.3847/2041-8213/ab8d1d}

\bibitem[{{Wang} {et~al.}(2018){Wang}, {Zhang}, {Liang}, {Lu}, {Lin}, {Li}, \&
  {Li}}]{2018ApJ...859..160W}
{Wang}, X.-G., {Zhang}, B., {Liang}, E.-W., {et~al.} 2018, \apj, 859, 160,
  \dodoi{10.3847/1538-4357/aabc13}

\bibitem[{{Wang} \& {Nitz}(2022)}]{2022arXiv220317222W}
{Wang}, Y.-F., \& {Nitz}, A.~H. 2022, arXiv e-prints, arXiv:2203.17222.
\newblock \doarXiv{2203.17222}

\bibitem[{{Xi} {et~al.}(2017){Xi}, {Tam}, {Peng}, \&
  {Wang}}]{2017ApJ...842L...8X}
{Xi}, S.-Q., {Tam}, P.-H.~T., {Peng}, F.-K., \& {Wang}, X.-Y. 2017, \apjl, 842,
  L8, \dodoi{10.3847/2041-8213/aa74cf}

\bibitem[{{Xin} {et~al.}(2021){Xin}, {Li}, {Wang}, {Han}, {Qiu}, {Cai}, {Niu},
  {Lu}, {Liang}, {Dai}, {Wang}, {Wang}, {Huang}, {Wu}, {Li}, {Feng}, {Deng},
  {Sun}, {Yang}, \& {Wei}}]{2021ApJ...922...78X}
{Xin}, L.~P., {Li}, H.~L., {Wang}, J., {et~al.} 2021, \apj, 922, 78,
  \dodoi{10.3847/1538-4357/ac1daf}

\bibitem[{{Xu} {et~al.}(2021){Xu}, {Niu}, {Chen}, {Lee}, {Zhu}, {Dong},
  {Zhang}, {Jiang}, {Wang}, {Xu}, {Zhang}, {Fu}, {Filippenko}, {Peng}, {Zhou},
  {Zhang}, {Wang}, {Feng}, {Li}, {Brink}, {Li}, {Lu}, {Yang}, {Caballero},
  {Cai}, {Chen}, {Dai}, {Djorgovski}, {Esamdin}, {Gan}, {Guhathakurta}, {Han},
  {Hao}, {Huang}, {Jiang}, {Li}, {Li}, {Li}, {Li}, {Li}, {Liu}, {Luo}, {Men},
  {Niu}, {Peng}, {Qian}, {Song}, {Stern}, {Stockton}, {Sun}, {Wang}, {Wang},
  {Wang}, {Wang}, {Wu}, {Xiao}, {Xiong}, {Xu}, {Xu}, {Yang}, {Yang}, {Yao},
  {Yi}, {Yue}, {Yu}, {Yu}, {Yuan}, {Zhang}, {Zhang}, {Zhang}, {Zhao}, {Zheng},
  {Zhu}, \& {Zou}}]{2021arXiv211111764X}
{Xu}, H., {Niu}, J.~R., {Chen}, P., {et~al.} 2021, arXiv e-prints,
  arXiv:2111.11764.
\newblock \doarXiv{2111.11764}

\bibitem[{{Yamasaki} {et~al.}(2016){Yamasaki}, {Totani}, \&
  {Kawanaka}}]{2016MNRAS.460.2875Y}
{Yamasaki}, S., {Totani}, T., \& {Kawanaka}, N. 2016, \mnras, 460, 2875,
  \dodoi{10.1093/mnras/stw1206}

\bibitem[{{Yang} \& {Zhang}(2018)}]{2018ApJ...868...31Y}
{Yang}, Y.-P., \& {Zhang}, B. 2018, \apj, 868, 31,
  \dodoi{10.3847/1538-4357/aae685}

\bibitem[{{Yang} \& {Zhang}(2021)}]{2021ApJ...919...89Y}
---. 2021, \apj, 919, 89, \dodoi{10.3847/1538-4357/ac14b5}

\bibitem[{{Yao} {et~al.}(2017){Yao}, {Manchester}, \&
  {Wang}}]{2017ApJ...835...29Y}
{Yao}, J.~M., {Manchester}, R.~N., \& {Wang}, N. 2017, \apj, 835, 29,
  \dodoi{10.3847/1538-4357/835/1/29}

\bibitem[{{Yi} {et~al.}(2022){Yi}, {Du}, \& {Liu}}]{2022ApJ...924...69Y}
{Yi}, S.-X., {Du}, M., \& {Liu}, T. 2022, \apj, 924, 69,
  \dodoi{10.3847/1538-4357/ac35e7}

\bibitem[{{Zhai} {et~al.}(2006){Zhai}, {Qiu}, {Wei}, {Hu}, {Deng}, {Wang},
  {Huang}, \& {Urata}}]{2006GCN..5057....1Z}
{Zhai}, M., {Qiu}, Y.~L., {Wei}, J.~Y., {et~al.} 2006, GRB Coordinates Network,
  5057, 1

\bibitem[{{Zhang}(2014)}]{2014ApJ...780L..21Z}
{Zhang}, B. 2014, \apjl, 780, L21, \dodoi{10.1088/2041-8205/780/2/L21}

\bibitem[{{Zhang}(2016)}]{2016ApJ...827L..31Z}
---. 2016, \apjl, 827, L31, \dodoi{10.3847/2041-8205/827/2/L31}

\bibitem[{{Zhang}(2017)}]{2017ApJ...836L..32Z}
---. 2017, \apjl, 836, L32, \dodoi{10.3847/2041-8213/aa5ded}

\bibitem[{{Zhang}(2018{\natexlab{a}})}]{2018ApJ...867L..21Z}
---. 2018{\natexlab{a}}, \apjl, 867, L21, \dodoi{10.3847/2041-8213/aae8e3}

\bibitem[{{Zhang}(2018{\natexlab{b}})}]{2018pgrb.book.....Z}
---. 2018{\natexlab{b}}, {The Physics of Gamma-Ray Bursts},
  \dodoi{10.1017/9781139226530}

\bibitem[{{Zhang}(2020)}]{2020ApJ...890L..24Z}
---. 2020, \apjl, 890, L24, \dodoi{10.3847/2041-8213/ab7244}

\bibitem[{{Zhang}(2022)}]{2022ApJ...925...53Z}
---. 2022, \apj, 925, 53, \dodoi{10.3847/1538-4357/ac3979}

\bibitem[{{Zhang} {et~al.}(2009){Zhang}, {Zhang}, {Virgili}, {Liang}, {Kann},
  {Wu}, {Proga}, {Lv}, {Toma}, {M{\'e}sz{\'a}ros}, {Burrows}, {Roming}, \&
  {Gehrels}}]{2009ApJ...703.1696Z}
{Zhang}, B., {Zhang}, B.-B., {Virgili}, F.~J., {et~al.} 2009, \apj, 703, 1696,
  \dodoi{10.1088/0004-637X/703/2/1696}

\bibitem[{{Zhang} \& {Zhang}(2017)}]{2017ApJ...843L..13Z}
{Zhang}, B.-B., \& {Zhang}, B. 2017, \apjl, 843, L13,
  \dodoi{10.3847/2041-8213/aa7633}

\bibitem[{{Zhang} \& {Wang}(2019)}]{2019MNRAS.487.3672Z}
{Zhang}, G.~Q., \& {Wang}, F.~Y. 2019, \mnras, 487, 3672,
  \dodoi{10.1093/mnras/stz1566}

\bibitem[{{Zhang} {et~al.}(2020){Zhang}, {Yu}, {He}, \&
  {Wang}}]{2020ApJ...900..170Z}
{Zhang}, G.~Q., {Yu}, H., {He}, J.~H., \& {Wang}, F.~Y. 2020, \apj, 900, 170,
  \dodoi{10.3847/1538-4357/abaa4a}

\bibitem[{{Zhang} \& {Zhang}(2022)}]{2022ApJ...924L..14Z}
{Zhang}, R.~C., \& {Zhang}, B. 2022, \apjl, 924, L14,
  \dodoi{10.3847/2041-8213/ac46ad}

\bibitem[{{Zhang} {et~al.}(2021){Zhang}, {Zhang}, {Li}, \&
  {Lorimer}}]{2021MNRAS.501..157Z}
{Zhang}, R.~C., {Zhang}, B., {Li}, Y., \& {Lorimer}, D.~R. 2021, \mnras, 501,
  157, \dodoi{10.1093/mnras/staa3537}

\end{thebibliography}
\bibliographystyle{aasjournal}

\clearpage
\begin{appendix}

    \section{SGRB SAMPLE}
    \renewcommand\arraystretch{1.5}
    \setlength{\tabcolsep}{1pt}
	\begin{longtable}{c|c|c|c|c|c|c|c|c|c|c|c|c|c}
    \caption{SGRB sample.}
    \label{table2}\\
        
        \hline
        Name&T$_{90}$&RA&Dec&Err$^{a}$&Redshift$^{b}$&Ref&Name&T$_{90}$&RA&DEC&Err$^{a}$&redshift$^{b}$&Ref\\
	    \hline
	    \endfirsthead
	    
        \hline
        Name&T$_{90}$&RA&Dec&Err$^{a}$&Redshift$^{b}$&Ref&Name&T$_{90}$&RA&DEC&Err$^{a}$&redshift$^{b}$&Ref\\
	    \hline
	    \endhead
	    
	    \hline
	    \endfoot
        GRB150101B&0.018&188.02&-10.93&0.03&0.134&(12)(13)&GRB140402A&0.031&207.592&5.971&2.8&n/a&(12)(13)\\
        GRB100628A&0.036&225.943&-31.653&2.1&n/a&(12)(13)&GRB090515&0.036&164.15&14.44&0.048&n/a&(12)(13)\\
        GRB130822A&0.04&27.92&-3.21&0.043&n/a&(12)(13)&GRB210119A&0.05&282.8&-61.8&0.05&n/a&(12)(13)\\
        GRB070923&0.05&184.623&-38.294&2.1&n/a&(12)(13)&GRB210119A&0.06&282.822&-61.767&1.5&n/a&(12)(13)\\
        GRB170112A&0.06&15.232&-17.233&2.5&n/a&(12)(13)&GRB150101A&0.06&312.6&36.73&0.045&n/a&(12)(13)\\
        GRB050925&0.07&303.49&34.329&1.4&n/a&(12)(13)&GRB090417A&0.072&34.993&-7.141&2.8&n/a&(12)(13)\\
        GRB050509B&0.073&189.06&28.98&0.063&0.226&(12)(13)&GRB190326A&0.08&341.652&39.914&1.6&n/a&(12)(13)\\
        GRB180718A&0.08&336.019&2.79&3.0&n/a&(12)(13)&GRB070810B&0.08&8.952&8.822&2.5&n/a&(12)(13)\\
        GRB110420B&0.084&320.045&-41.277&2.2&n/a&(12)(13)&GRB070209&0.09&46.213&-47.376&2.8&n/a&(12)(13)\\
        GRB051105A&0.093&265.279&34.916&2.6&n/a&(12)(13)&GRB170524A&0.1&319.49&48.61&0.072&n/a&(12)(13)\\
        GRB161104A&0.1&77.89&-51.46&0.05&n/a&(12)(13)&GRB120305A&0.1&47.54&28.49&0.032&n/a&(12)(13)\\
        GRB160601A&0.12&234.94&64.54&0.052&n/a&(12)(13)&GRB100206A&0.12&47.16&13.16&0.055&0.4068&(12)(13)\\
        GRB140622A&0.13&317.17&-14.42&0.04&0.959&(12)(13)&GRB060502B&0.131&278.94&52.63&0.087&0.287&(12)(13)\\
        GRB220412B&0.14&320.758&-0.261&1.6&n/a&(12)(13)&GRB090621B&0.14&313.47&69.03&0.068&n/a&(12)(13)\\
        GRB150710A&0.15&194.47&14.32&0.047&n/a&(12)(13)&GRB210919A&0.16&80.25&1.31&0.067&n/a&(12)(13)\\
        GRB201221D&0.16&171.06&42.14&0.072&1.046&(12)(13)&GRB130626A&0.16&273.128&-9.525&1.8&n/a&(12)(13)\\
        GRB100702A&0.16&245.6969&-56.5316&0.04&n/a&(12)(13)&GRB180402A&0.18&251.93&-14.97&0.035&n/a&(12)(13)\\
        GRB130603B&0.18&172.2&17.07&0.023&0.3564&(12)(13)&GRB151127A&0.19&19.48&-82.77&0.032&n/a&(12)(13)\\
        GRB140516A&0.19&252.99&39.96&0.037&n/a&(12)(13)&GRB170428A&0.2&330.08&26.92&0.038&0.454&(12)(13)\\
        GRB160624A&0.2&330.19&29.64&0.028&0.483&(12)(13)&GRB101224A&0.2&285.92&45.71&0.053&0.4536&(12)(13)\\
        GRB081101&0.2&95.836&-0.112&1.7&n/a&(12)(13)&GRB061217&0.21&160.41&-21.12&0.1&0.827&(12)(13)\\
        GRB200411A&0.22&47.66&-52.32&0.023&n/a&(12)(13)&GRB150423A&0.22&221.58&12.28&0.027&1.394&(12)(13)\\
        GRB120229A&0.22&20.033&-35.796&1.9&n/a&(12)(13)&GRB050906&0.258&52.82&-14.61&4.0&n/a&(12)(13)\\
        GRB181123B&0.26&184.37&14.6&0.025&1.754&(12)(13)&GRB130313A&0.26&236.41&-0.37&0.08&n/a&(12)(13)\\
        GRB151228A&0.27&214.017&-17.665&1.8&n/a&(12)(13)&GRB050202&0.27&290.584&-38.73&2.3&n/a&(12)(13)\\
        GRB130912A&0.28&47.59&14&0.025&n/a&(12)(13)&GRB191031D&0.29&283.29&47.64&0.028&n/a&(12)(13)\\
        GRB160525B&0.29&149.38&51.21&0.025&n/a&(12)(13)&GRB130515A&0.29&283.44&-54.28&0.04&n/a&(12)(13)\\
        GRB190427A&0.3&280.217&40.304&2.4&n/a&(12)(13)&GRB170325A&0.3&127.483&20.526&2.0&n/a&(12)(13)\\
        GRB141212A&0.3&39.12&18.15&0.043&0.596&(12)(13)&GRB140903A&0.3&238.01&27.6&0.023&0.351&(12)(13)\\
        GRB100117A&0.3&11.27&-1.59&0.04&0.92&(12)(13)&GRB091109B&0.3&112.74&-54.09&0.04&n/a&(12)(13)\\
        GRB090510&0.3&333.55&-26.58&0.01&0.903&(12)(13)&GRB071112B&0.3&260.213&-80.884&2.2&n/a&(12)(13)\\
        GRB160408A&0.32&122.62&71.13&0.027&n/a&(12)(13)&GRB100625A&0.33&15.8&-39.09&0.03&0.452&(12)(13)\\
        GRB140606A&0.34&201.799&37.599&2.4&0.384&(12)(13)&GRB160714A&0.35&234.49&63.809&2.7&n/a&(12)(13)\\
        GRB101129A&0.35&155.921&-17.645&3.0&n/a&(12)(13)&GRB160411A&0.36&349.36&-40.24&0.037&n/a&(12)(13)\\
        GRB210726A&0.39&193.29&19.19&0.035&n/a&(12)(13)&GRB111020A&0.4&287.05&-38.01&0.027&n/a&(12)(13)\\
        GRB090305A&0.4&241.764&-31.572&2.3&n/a&(12)(13)&GRB081226A&0.4&120.527&-69.006&3.0&n/a&(12)(13)\\
        GRB070724&0.4&27.81&-18.59&0.028&0.457&(12)(13)&GRB210605B&0.448&15.7456&-6.429&0.095&n/a&(13)\\
        GRB140320A&0.45&281.86&-11.19&0.082&n/a&(12)(13)&GRB120521A&0.45&148.72&-49.42&0.03&n/a&(12)(13)\\
        GRB050813&0.45&241.99&11.25&0.048&n/a&(12)(13)&GRB111117A&0.47&12.7042&23.0167&3.0&2.211&(12)(13)\\
        GRB070429B&0.47&328.02&-38.83&0.01&0.904&(12)(13)&GRB160927A&0.48&256.24&17.33&0.028&n/a&(12)(13)\\
        GRB160821B&0.48&279.98&62.39&0.037&0.16&(12)(13)&GRB150301A&0.48&244.3&-48.71&0.083&n/a&(12)(13)\\
        GRB201006A&0.49&61.89&65.16&0.032&n/a&(12)(13)&GRB060801&0.49&213.01&16.98&0.025&1.131&(12)(13)\\
        GRB110112A&0.5&329.93&26.46&0.028&n/a&(12)(13)&GRB080702&0.5&313.05&72.31&0.03&n/a&(12)(13)\\
        GRB170127B&0.51&19.98&-30.36&0.033&n/a&(12)(13)&GRB120630A&0.6&353.29&42.55&0.008&n/a&(12)(13)\\
        GRB101219A&0.6&74.59&-2.54&0.028&0.718&(12)(13)&GRB090815C&0.6&64.49&-65.943&2.7&n/a&(12)(13)\\
        GRB080919&0.6&265.22&-42.37&0.027&n/a&(12)(13)&GRB200522A&0.62&5.68&-0.28&0.037&0.554&(12)(13)\\
        GRB190610A&0.62&46.244&-7.66&1.9&n/a&(12)(13)&GRB180715A&0.68&235.085&-0.899&2.0&n/a&(12)(13)\\
        GRB160726A&0.7&98.809&-6.617&1.3&n/a&(12)(13)&GRB140414A&0.7&195.31&56.902&4.0&n/a&(12)(13)\\
        GRB080121&0.7&137.235&41.841&3.0&n/a&(12)(13)&GRB060313&0.74&66.62&-10.84&0.023&n/a&(12)(13)\\
        GRB061201&0.76&332.13&-74.58&0.008&0.111&(12)(13)&GRB130716A&0.8&179.57&63.05&0.033&n/a&(12)(13)\\
        GRB111126A&0.8&276.057&51.461&3.0&n/a&(12)(13)&GRB120804A&0.81&233.95&-28.78&0.023&n/a&(12)(13)\\
        GRB200907B&0.83&89.03&6.91&0.028&n/a&(12)(13)&GRB150728A&0.83&292.23&33.92&0.067&n/a&(12)(13)\\
        GRB140930B&0.84&6.35&24.29&0.027&n/a&(12)(13)&GRB070729&0.9&56.32&-39.32&0.042&n/a&(12)(13)\\
        GRB200325B&0.96&167.55381&27.8196&0.082&n/a&(13)&GRB081211&0.97&328.12&-33.84&0.027&n/a&(12)(13)\\
        GRB210704A&1&159.08&57.31&0.002&n/a&(13)&GRB121226A&1&168.64&-30.41&0.035&n/a&(12)(13)\\
        GRB111222A&1&179.21983&69.07039&0.045&n/a&(13)&GRB080905&1&287.67&-18.88&0.027&0.1218&(12)(13)\\
        GRB210413B&1.088&182.558&55.965&3.0&n/a&(12)(13)&GRB200219A&1.1&342.65&-59.1&0.05&n/a&(12)(13)\\
        GRB180727A&1.1&346.67&-63.05&0.028&n/a&(12)(13)&GRB141205A&1.1&92.859&37.876&2.0&n/a&(12)(13)\\
        GRB070707S&1.1&267.74396&-68.92422&0.008&n/a&(13)&GRB210323A&1.12&317.95&25.37&0.037&n/a&(12)(13)\\
        GRB150831A&1.15&221.02&-25.63&0.027&n/a&(12)(13)&GRB180204A&1.16&330.13&30.84&0.023&n/a&(12)(13)\\
        GRB150120A&1.2&10.32&33.99&0.03&n/a&(12)(13)&GRB090426&1.2&189.08&32.99&0.008&2.609&(12)(13)\\
        GRB070406&1.2&198.956&16.53&3.3&n/a&(12)(13)&GRB040924&1.2&31.5792&16.0167&0.06&0.859&(13)\\
        GRB170728A&1.25&58.89&12.18&0.009&n/a&(12)(13)&GRB120403A&1.25&42.458&40.489&2.3&n/a&(12)(13)\\
        GRB211023B&1.3&170.31&39.14&0.009&n/a&(12)(13)&GRB161004A&1.3&263.15&-0.95&0.009&n/a&(12)(13)\\
        GRB070809&1.3&203.7833&-22.1333&3.0&0.2187&(12)(13)&GRB051210&1.3&330.17&-57.61&0.027&n/a&(12)(13)\\
        GRB140129B&1.36&326.76&26.21&0.008&n/a&(12)(13)&GRB151205B&1.4&41.19&-43.461&2.3&n/a&(12)(13)\\
        GRB100724A&1.4&194.54&-11.1&0.025&1.288&(12)(13)&GRB051221&1.4&328.7&16.89&0.023&0.5465&(12)(13)\\
        GRB131004A&1.54&296.11&-2.96&0.009&0.717&(12)(13)&GRB190627A&1.6&244.83&-5.29&0.008&1.942&(12)(13)\\
        GRB180805A&1.68&167.57&-45.33&0.033&n/a&(12)(13)&GRB220730A&1.7&225.014&59.495&0.11&n/a&(12)(13)\\
        GRB080426&1.7&26.5&69.47&0.025&n/a&(12)(13)&GRB211106A&1.75&343.59&-53.23&0.057&n/a&(12)(13)\\
        GRB151229A&1.78&329.37&-20.73&0.023&n/a&(12)(13)&GRB081024A&1.8&27.87&61.33&0.03&n/a&(12)(13)\\
        GRB071227&1.8&58.13&-55.98&0.01&0.383&(12)(13)&GRB140611A&1.9&349.917&40.104&0.917&n/a&(12)(13)\\
        GRB200826A&1.95&6.817&34.038&21.0&0.7481&(13)&GRB201008A&2&161.8588&46.1019&0.08&n/a&(13)\\
        GRB170817A&2&197.45&-23.38&0.022&0.01&(13)&GRB070714&2&42.93&30.24&0.03&1.58&(12)(13)\\
        GRB060121&2&137.46742&45.66039&0.005&n/a&(13)&GRB210618A&2.13&235.82&46.01&0.05&n/a&(8)(12)(13)\\
        GRB090927&2.2&343.97&-70.98&0.011&1.37&(11)(12)(13)&GRB090607&2.3&191.17&44.11&0.062&n/a&(2)(12)(13)\\
        GRB100213A&2.4&349.39&43.38&0.038&n/a&(12)(13)(18)&GRB100816A&2.9&351.74&26.58&0.008&0.8049&(12)(13)(17)\\
        GRB050724&3&246.19&-27.54&0.003&0.258&(12)(13)(14)&GRB210217A&4.22&97.61&68.72&0.05&n/a&(9)(12)(13)\\
        GRB080913A&8&65.73&-25.13&0.032&6.695&(12)(13)(19)&GRB160410A&8.2&150.69&3.48&0.008&1.717&(12)(13)(20)\\
        GRB070714B&64&57.84&28.3&0.008&0.92&(1)(12)(13)&GRB061210&85.3&144.52&15.62&0.03&0.4095&(6)(12)(13)\\
        GRB150424A&91&152.31&-26.63&0.008&0.3&(4)(12)(13)&GRB081211B&102&168.27&53.83&0.03&n/a&(10)(12)(13)\\
        GRB171007A&105&135.6&42.82&0.032&n/a&(3)(12)(13)&GRB080123&115&338.95&-64.9&0.027&0.495&(12)(13)(21)\\
        GRB111121A&119&154.76&-46.67&0.023&n/a&(7)(12)(13)&GRB061006&130&111.03&-79.2&0.008&0.4377&(12)(13)(15)\\
        GRB080503&170&286.62&68.79&0.004&n/a&(12)(13)(16)&GRB050709&220&345.36&-38.98&0.008&0.1606&(5)(13)\\

    \end{longtable}
    \footnotesize{\textbf{Note.}
    
    $^a$ The localization uncertainty in units of arcmin. 
    
    $^b$ The n/a means that this SGRB does not have a redshift measurement.}
    \begin{tablenotes}
	\item 
	   The references:
        (1)\citep{2007GCN..6623....1B},
        (2)\citep{2009GCN..9494....1B},
        (3)\citep{2017GCN.21981....1B},
        (4)\citep{2015GCN.17743....1B},
        (5)\citep{2005GCN..3570....1B},
        (6)\citep{2006GCN..5904....1C},
        (7)\citep{2011GCN.12578....1D},
        (8)\citep{2021GCN.30248....1F},
        (9)\citep{2021GCN.29536....1F}, 
        (10)\citep{2008GCN..8676....1G},
        (11)\citep{2009GCN..9945....1G},
        (12)\href{https://swift.gsfc.nasa.gov/archive/grb\_table.html/}{https://swift.gsfc.nasa.gov/archive/grb\_table.html/},
        (13)J-G source (\href{https://www.mpe.mpg.de/~jcg/grbgen.html\#userconsent\#}{https://www.mpe.mpg.de/$\sim$jcg/grbgen.html\#userconsent\#}),
        (14)050724\citep{2005GCN..3667....1K},
        (15)\citep{2006GCN..5704....1K},
        (16)\citep{2008GCN..7665....1M},
        (17)\citep{2010GCN.11113....1N},  
        (18)\citep{2010GCN.10427....1N},
        (19)\citep{2008GCN..8256....1P},
        (20)\citep{2016GCN.19276....1S} and
        (21)\citep{2008GCN..7203....1U}.
    \end{tablenotes}

    \clearpage
    \section{FRB and GRB 060502B host galaxy properties}
    \renewcommand\arraystretch{1.5}
    \setlength{\tabcolsep}{2pt}

	\begin{longtable}{ccccccccc}
    \caption{Host galaxy properties for FRBs and GRB 060502B.}
    \label{table1}\\
        
        \hline
        Name & Instrument & Offset (kpc) & SFR ($M_{\odot}\,{\rm yr}^{-1}$) & log($M_*/M_{\odot}$)& Repeater? & Ref\\
	    \hline
	    \endfirsthead
	    
        \hline
        Name & Instrument & Offset (kpc) & SFR ($M_{\odot}\,{\rm yr}^{-1}$) & log($M_*/M_{\odot}$)& Repeater? & Ref\\
	    \hline
	    \endhead
	    
	    \hline
	    \endfoot
	   GRB060502B & {Swift} & $73.0^{+19}_{-19}$& $0.6^{+0.2}_{-0.2}$ &11.85  & $-$ & (2) \\
    \hline
	   FRB121102A & Arecibo & 0.8$^{+0.1}_{-0.1}$ & 0.15$^{+0.04}_{-0.04}$  &8.16& y &(1)(3) \\
	   FRB171020A & ASKAP  & - &0.13 & 8.95 & n & (1) \\
	   FRB180301A & Parkes & 10.2$^{+3.0}_{-3.0}$ & 1.93$^{+0.58}_{-0.58}$  &9.36 & y &(1)(3)\\
	   FRB180916B & CHIME & 5.4$^{+0.0}_{-0.0}$ & 0.06$^{+0.02}_{-0.02}$ &9.33 & y & (1)(3)\\
	   FRB180924B & ASKAP & 3.4$^{+0.8}_{-0.8}$ & 0.88$^{+0.26}_{-0.26}$ & 10.12& n&(1)(3)(5)\\
	   FRB181030A & CHIME & 0 & 0.36$^{+0.10}_{-0.10}$  &9.76& y& (3) \\
	   FRB181112A & ASKAP & 3.1$^{+15.7}_{-3.1}$ & 0.37$^{+0.11}_{-0.11}$  &9.60 & n & (3)(5)\\
	   FRB190102C & ASKAP & 2.3$^{+4.2}_{-2.3}$ & 0.86$^{+0.26}_{-0.26}$ & 9.67  & n & (1)(3)\\
	   FRB190523A & DSA-10 & 27.2$^{+22.6}_{-22.6}$ & 0.09$^{+0.03}_{-0.03}$  &10.79& n & (1)(3)(5)\\
	   FRB190608B & ASKAP & 6.5$^{+0.8}_{-0.8}$ & 0.69$^{+0.21}_{-0.21}$  &10.06 & n &(1)(3)\\
	   FRB190611B & ASKAP & 11.7$^{+5.8}_{-5.8}$ & 0.27$^{+0.08}_{-0.08}$  &8.88& n & (1)(3)\\
	   FRB190711A & ASKAP & 1.6$^{+4.5}_{-4.5}$ & 0.42$^{+0.12}_{-0.12}$  &8.91& y &(1)(3)\\
	   FRB190714A & ASKAP & 2.7$^{+1.8}_{-1.8}$ & 0.65$^{+0.20}_{-0.20}$  &10.15& n &(1)(3)\\
	   FRB191001A & ASKAP & 11.1$^{+0.8}_{-0.8}$ & 8.06$^{+2.42}_{-2.42}$ &10.67& n &(1)(3)\\
	   FRB191228A & ASKAP & 5.7$^{+3.3}_{-3.3}$ & 0.50$^{+0.15}_{-0.15}$&9.73 & n &(1)(3) \\
	   FRB200120E & CHIME & 20.1$^{+3.0}_{-3.0}$ & 0.89$^{+0.27}_{-0.27}$ &10.86 & y & (1)(3)\\
	   FRB200430A & ASKAP & 1.7$^{+2.2}_{-1.7}$ & 0.26$^{+0.08}_{-0.08}$  &9.32& n & (1)(3)\\
	   FRB200906A & ASKAP & 5.9$^{+2.0}_{-2.0}$ & 0.48$^{+0.14}_{-0.14}$  &10.12& n & (1)(3)\\
	   FRB201124A & CHIME & 1.3$^{+0.1}_{-0.1}$ & 2.12$^{+0.69}_{-0.28}$  &10.20& y & (1)(3)\\

      FRB220207C & DSA-110 & 7.4 & 2.14$^{+0.26}_{-0.26}$ & 9.90 & n & (4)\\
      FRB220307B & DSA-110 & 6.0 & 3.52$^{+1.18}_{-0.60}$ & 11.73 & n & (4)\\
      FRB220310F & DSA-110 & 13.3 & 0.15$^{+0.01}_{-0.01}$ & 10.75 & n & (4)\\
      FRB220319D & DSA-110 & 2.8 & 1.17$^{+0.75}_{-0.72}$ & 9.93 & n & (4)\\
      FRB220418A & DSA-110 & 10.6 & 0.37$^{+0.40}_{-0.27}$ & 10.83 & n & (4)\\
      FRB220506D & DSA-110 & 12.0 & 7.01$^{+0.43}_{-0.64}$ & 10.45 & n & (4)\\
      FRB220509G & DSA-110 & 6.9 & 0.08$^{+0.06}_{-0.04}$ & 11.13 & n & (4)\\
      FRB220825A & DSA-110 & 8.1 & 1.34$^{+0.04}_{-0.04}$ & 11.14 & n & (4)\\
      FRB220914A & DSA-110 & 2.8 & 1.45$^{+1.05}_{-0.61}$ & 9.99 & n & (4)\\
      FRB220920A & DSA-110 & 6.2 & 0.39$^{+0.02}_{-0.02}$ & 10.85 & n & (4)\\
      FRB221012A & DSA-110 & 14.7 & 0.49$^{+0.44}_{-0.30}$ & 11.30 & n & (4)\\

        \hline
    \end{longtable}
	\begin{tablenotes}
	\item 
	   The references are as follows: 
          (1) \citet{2022AJ....163...69B},
	   (2) \citet{2007ApJ...654..878B}, 
	   (3) \href{http://frbhosts.org}{http://frbhosts.org},
          (4) \citet{2023arXiv230703344L}, and
	   (5) \citet{2020ApJ...899L...6L}.
	\end{tablenotes}

\end{appendix}

\end{document}